\documentclass[11pt,english,prb,reprint,nofootinbib]{revtex4-2}
\usepackage[T1]{fontenc}
\usepackage[latin9]{inputenc}
\usepackage{color}
\usepackage{amsmath}
\usepackage{amssymb}
\usepackage{graphicx}
\PassOptionsToPackage{normalem}{ulem}
\usepackage{ulem}
\usepackage{physics}
\usepackage{amsmath}
\usepackage{tikz}

\usepackage{soul}
\usepackage{mathdots}
\usepackage{yhmath}
\usepackage{cancel}
\usepackage{color}
\usepackage{siunitx}
\usepackage{array}
\usepackage{multirow}
\usepackage{amssymb}
\usepackage{gensymb}
\usepackage{tabularx}
\usepackage{extarrows}
\usepackage{booktabs}
\usetikzlibrary{fadings}
\usetikzlibrary{patterns}
\usetikzlibrary{shadows.blur}
\usetikzlibrary{shapes}
\makeatletter

\definecolor{caribbeangreen}{rgb}{0.0, 0.8, 0.6}

\newcommand{\eqnref}[1]{Eq.~\eqref{#1}}
\newcommand{\secref}[1]{Sec.~\ref{#1}}
\newcommand{\figref}[1]{Fig.~\ref{#1}}

\renewcommand{\braket}[1]{\langle{{#1}}\rangle}

\usepackage{hyperref}

\hypersetup{
    colorlinks=true,
    linkcolor=blue,
    filecolor=magenta,      
    urlcolor=cyan,
    pdftitle={Sharelatex Example},
    bookmarks=true,
    pdfpagemode=FullScreen,
}

\makeatother
\newcommand{\fr}{\mathfrak{r}}
\newcommand{\fb}{\mathfrak{b}}
\newcommand{\fp}{\mathfrak{p}}
\newcommand{\fm}{\mathfrak{m}}

\graphicspath{{../}}

\makeatother

\begin{document}
\title{Operator backflow and the classical simulation of quantum transport}

\author{Curt von Keyserlingk}

\affiliation{School of Physics \& Astronomy, University of Birmingham, Birmingham, B15 2TT,
UK}

\author{Frank Pollmann}

\affiliation{Department of Physics, T42, Technische Universit{\"a}t M{\"u}nchen, James-Franck-Stra{\ss}e 1, D-85748 Garching, Germany}
\affiliation{Munich Center for Quantum Science and Technology (MCQST), Schellingstr. 4, D-80799 M\"unchen, Germany }

\author{Tibor Rakovszky}

\affiliation{Department of Physics, Stanford University, Stanford, CA 94305, USA}

\begin{abstract}
Tensor product states have proved extremely powerful for simulating the area-law entangled states  of many-body systems, such as the ground states of gapped Hamiltonians in one dimension. The applicability of such methods to the \emph{dynamics} of many-body systems is less clear: the memory required  grows exponentially in time in most cases, quickly becoming unmanageable. New methods reduce the memory required by selectively discarding/dissipating parts of the many-body wavefunction which are expected to have little effect on the hydrodynamic observables typically of interest: for example, some methods discard fine-grained correlations associated with $n$-point functions, with $n$ exceeding some cutoff $\ell_*$. In this work, we present a theory for the sizes of `backflow corrections', i.e., systematic errors due to discarding this fine-grained information. In particular, we focus on their effect on transport coefficients. Our results suggest that backflow corrections are exponentially suppressed in the size of the cutoff $\ell_*$. Moreover, the backflow errors themselves have a hydrodynamical expansion, which we elucidate. We test our predictions against numerical simulations run on random unitary circuits and ergodic spin-chains. These results lead to the conjecture that transport coefficients in ergodic diffusive systems can be captured to a given precision $\epsilon$ with an amount of memory scaling as $\exp[\mathcal{O}(\log(\epsilon)^2)]$, significantly better than the naive estimate of memory $\exp[\mathcal{O}(\mathrm{poly}(\epsilon^{-1}))]$ required by more brute-force methods. 
\end{abstract}

\maketitle

\section{Introduction}

Transport properties, such as heat, charge and spin conductivities, are among the most experimentally accessible and practically important features of materials, and also serve to distinguish the plethora of phases of condensed matter. Yet, our ability to predict these properties from first principles in interacting quantum matter is limited. Improving existing methods has the potential to facilitate sharper predictions for existing experiments, aid in the exploration and discovery of new phenomena, and allow the design of future experimental systems with bespoke transport properties. 

Even in the linear response regime, the calculation of transport properties requires computing unequal time correlation functions, and thus depends on our ability to simulate the dynamics of quantum many-body systems. Unsurprisingly, existing methods like exact evolution \cite{SteinigewegReview}, and time-evolved block decimation (TEBD) \cite{MurgReview} are strongly limited for such \emph{ab initio} studies. Exact evolution is limited to small systems sizes, while estimates from TEBD are limited to just a few interaction times before the memory requirements become impractical/truncation errors become significant~\cite{SCHOLLWOCK201196,PAECKEL2019167998}.

Recently, a number of methods have been proposed for sidestepping
these issues  \cite{white_zaletel_1,parkerhypothesis,marko_method_1,jens_hydro_1,me_daoe_1,white2021effective,WURTZ2018341}. All of these methods (with the exception of Ref.
 \onlinecite{marko_method_1}) rely on the intuition that much of the information content
of a many-body density matrix -- particularly correlation functions involving products of many-operators~\cite{me_daoe_1}, or operators with large spatial support \cite{white_zaletel_1,white2021effective,jens_hydro_1,WURTZ2018341} --
can be thrown away without greatly affecting the calculation of transport
coefficients. This intuition has been articulated in various forms
for decades in (for example) the formulation of the BBGKY hierarchy, and the approximations employed in the mode-coupling/memory-matrix formalism and fluctuating hydrodynamics approaches to many-body systems \cite{mct_review,landau_lifshitz_fluids} . 

It is apparent from the references above that many investigators believe
that the general approach of discarding complicated correlations should give good approximate results. These expectations
have been largely supported by numerous (but by no means systematic)
numerical calculations which show impressive convergence requiring
surprisingly few computational resources  \cite{white_zaletel_1,parkerhypothesis,marko_method_1,jens_hydro_1,me_daoe_1}.
Here `surprisingly few' means the calculations produce  accurate estimates (seemingly to within a few percent) of transport coefficients using only modestly powered classical computers; one might
have expected that such many-body quantum problems generically require
the use of a quantum computer for accurate results.

A detailed understanding of the sizes of systematic errors for these
methods is lacking, however. In other words, if one does discard higher
order correlations, what is the size of the resulting systematic errors in the estimates of transport coefficients? Answering this question requires us to understand the size of \emph{backflow} processes, wherein information from higher order correlations `flows back' to the subspace of local observables.

The present paper aims to remedy this gap in the
literature. Although our results are relevant to various recently proposed  methods \cite{white2021effective,white_zaletel_1,jens_hydro_1}, our primary interest is in the dissipation assisted operator evolution (`DAOE') method \cite{me_daoe_1}. For DAOE, the correlation functions are approximated by modifying unitary Heisenberg dynamics so that operators with support on, say, more than $\ell_*$ sites are discarded or strongly dissipated. While we have highlighted the issue of estimating transport coefficients, our results are also relevant to the numerical simulation of hydrodynamical correlation functions (HCFs) which can shed light on more general non-equilibrium processes, such as quantum quenches.

%

We principally study ergodic systems in 1D with a single U$(1)$ conserved density undergoing diffusion, although we also expect our results to apply to systems for which the only conserved quantity is energy. In such a system, HCFs are expected to have an expansion in $1/t$ of general form
\begin{equation}\label{eq:corr}
\langle o(t)  o \rangle= c_1 t^{-a_1} + c_2 t^{-a_2}+\ldots
\end{equation}
For example in $d=1$ spatial dimensions, if $o$ is chosen to be the local charge density, then $a_1 = 1/2$ and $c_1 \propto D^{-1/2}$ for a diffusion constant $D$.
We develop a physical picture for how HCFS are modified when DAOE dissipation is applied (\eqnref{eq:trunc_picture_gen}). This physical picture leads to a prediction: The difference, or backflow, induced in \eqnref{eq:corr} by using DAOE also has a hydrodynamical expansion
\begin{equation}\label{eq:backflow}
 \langle o(t)  o \rangle - \langle o(t)  o \rangle_\mathrm{DAOE}  = c'_1 t^{-a'_1} + c'_2 t^{-a'_2}+\ldots
\end{equation}

Inspired by the study of random circuits, we argue that the leading order contributions to backflow are exponentially suppressed in $\ell_*$. In other words $c'_1=\exp[{-\mathcal{O}(\ell_*)}]$. Moreover, our hydrodynamical theory of backflow corrections also predicts the leading order exponent $a'_1$ for the time decay of corrections in \eqnref{eq:backflow}. It turns out that the predicted value depends on the hydrodynamical operator $o$, and whether or not the system has spatiotemporal randomness (like random circuits), but in all cases the backflow corrections decay at least as quickly as the original correlation function, i.e.,  $a'_1\geq a_1$. 

These results lead to the conclusion that the error induced in the diffusion constant $D$ due to the DAOE protocol is exponentially small in $\ell_*$
\begin{equation}\label{eq:backflowD}
D - D_{\mathrm{DAOE}}=\exp[{-\mathcal{O}(\ell_*)}].
\end{equation}

We support our theoretical predictions  with a numerical study of deterministic systems (\figref{fig:IsingandLadder}), although naturally this study is plagued by the usual memory limitations that impede our simulation of many-body systems, and so these results are not decisive. We supplement it with a study of backflow in the U$(1)$-symmetric random unitary circuit (RUC) (\figref{fig:total}). Specifically, we compute the average
effect of DAOE dissipation on the charge diffusion coefficient and
various correlation functions. The advantage of using random circuits
is that circuit-averaged quantities can be mapped to statistical mechanics
models, which tend to be easier to simulate to longer times and for larger systems sizes than would be available by simulating real-time dynamics directly   \cite{vonKeyserlingk2018_diffusive}. We also provide further consistency checks of our hydrodynamical picture for backflow corrections using a semi-analytical treatment of circuit averaged dynamics in a U(1) RUC. 




As we detail in \secref{sec:memory}, our backflow picture implies that accurate estimates of diffusion constants can be obtained by restricting dynamics to a smaller effective Hilbert space -- the space of short operators. This in turn implies that DAOE dramatically reduces the memory required to estimate transport coefficients. To be more precise, we reason that $\epsilon$-accurate approximations to transport coefficients can be obtained using DAOE in combination with matrix product operators with relatively small bond dimension $\chi \sim \exp [\mathcal{O}(\log(1/\epsilon)^2)]$. In contrast, a naive estimate suggests that the standard more brute-force approaches (TEBD, exact evolution) require substantially higher bond dimension $\chi \sim \exp [\mathrm{poly}(\epsilon^{-1})]$. Therefore, in the limit $\epsilon \rightarrow 0$, DAOE is far more memory efficient than these existing approaches.

This work is organized as follows.  \secref{sec:defconv} reviews conventions, and describes the DAOE protocol. \secref{ssec:hydro_sans_trunc} summarizes the expected behavior of hydrodynamical correlation functions without any truncation.  \secref{sec:theory} develops a theory for the form of backflow corrections due to DAOE, explaining their exponential suppression in $\ell_*$ and time dependence, and presents numerical results (\figref{fig:IsingandLadder}) supporting this picture. \secref{sec:U(1)random} then investigates backflow in a U$(1)$ random circuit, presenting  numerical results (\figref{fig:total}) which further support the picture in \secref{sec:theory}, as well as a semi-analytical calculation demonstrating that a subclass of backflow corrections are exponentially suppressed in $\ell_*$.  \secref{sec:memory} uses the theoretical picture in \secref{sec:theory} to argue that DAOE substantially reduces memory overheads compared to more brute-force approaches (asymptotically). We conclude and present future directions in \secref{sec:conclusion}.

\section{Setting and conventions}\label{sec:defconv}
We focus on 1D ergodic lattice systems at high temperature, with local
unitary dynamics and possessing a single conserved density which diffuses. For much of the discussion, and for concreteness, we will take our system to be a spin-1/2 chain, and for the conserved quantity to be the total  $z$-component of the spin, namely $S^z$ (which can equivalently be interpreted as the charge of hard-core bosons). We denote the local magnetization by $z_j=\sigma^z_j$, so that $S^z=\sum_j z_j$.  We expect our results to generalize to the study of energy diffusion in systems with a local Hamiltonian. 
The ergodicity requirement eliminates
certain finely-tuned models (like Bethe-Ansatz integrable models),
and localized systems. Our requirement that the conserved quantity diffuses is not very restrictive, given the ubiquity of systems exhibiting diffusion at high temperatures \cite{diffusion_ubiquitous_1,diffusion_ubiquitous_2,diffusion_ubiquitous_3}.

It is useful to define an inner product between operators
$\langle A | B\rangle\equiv\bigl\langle A^{\dagger}B\bigr\rangle$
; throughout this paper we work at infinite temperature, so that $\langle A | B\rangle=\text{tr}(A^{\dagger}B)/\text{tr}\left(I\right)$. This turns the space of operators in to a Hilbert space; as we shall see, it is natural to phrase various transport-related quantities in this language. For example, dynamical correlations are equivalent to an inner product between an operator at time $t$ and another at time $0$.

The unitary matrix $U(t_{2},t_{1})$ evolves states from time $t_{1}\rightarrow t_{2}$.
For energy conserving systems time is a continuous variable and $U(t_{2},t_{1})=e^{-\mathrm{i}(t_{2}-t_{1})\hat{H}}$
where $\hat{H}$ is the Hamiltonian. For Floquet/random circuit systems,
time is a discrete variable. Moreover, for Floquet systems $U(t_{2},t_{1})=U_{\text{f}}^{(t_{2}-t_{1})}$
for Floquet unitary $U_{\mathrm{f}}$. Unitary time evolution acts
on density matrices via a superoperator $\mathcal{U}(t_{2},t_{1})$, so that the expectation value of a Hermitian observable $A$ can be written

\begin{equation*}
 \langle A(t_2)\rangle = \langle A | \mathcal{U}(t_2,t_1) | \rho(t_1)\rangle
\end{equation*}
where $\rho(t_1)$ is the density matrix at some earlier time $t_1$. In this work, we focus on the structure of time evolved operators, which evolve according to the Heisenberg picture
\begin{equation*}
\langle A | \mathcal{U}(t_{2},t_{1})\equiv\langle U^{\dagger}(t_{2},t_{1})AU(t_{2},t_{1}) |.    
\end{equation*}

Our primary object of study is the diffusion constant. It can be written as $D\equiv \lim_{\tau\rightarrow \infty}\lim_{L\rightarrow \infty} D(\tau)$ where the time-dependent diffusion constant is defined as
\begin{equation}
D(\tau)\equiv\sum_{x}\frac{x^{2}}{2 \tau}\langle z_{x} | \mathcal{U}(\tau,0) |  z_{0}\rangle.\label{eq:D_1st}
\end{equation}
An exact calculation of $D$ is limited by the exponential growth of Hilbert space dimension with system size, with numerical studies limited to $\approx 20-30$ spins at most \cite{SteinigewegReview}.
One may instead attempt to evaluate Eq.~\ref{eq:D_1st} using TEBD, representing the time-evolved operator $q_x(t)$ as a matrix product operator (MPO). However, as $t$ increases, the bond dimension needed for a high fidelity MPO representation of $\langle q_{x}|\mathcal{U}(t)$ also grows. In particular, for the ergodic systems we consider, it is expected that the corresponding \emph{operator entanglement} grows linearly with $t$ \cite{jonay2018coarsegrained}, making the memory requirements grow \emph{exponentially}, as $\sim e^{\alpha t}$.Therefore one needs an  approximation scheme in order to be able to accurately evaluate $D(\tau)$ at long times. Recently, several such approximation schemes were proposed that rely on an idea of `operator truncation'~\cite{me_daoe_1,jens_hydro_1,white2021effective}. The shared intuition between these methods is that information regarding the component of $\langle q_{x}|\mathcal{U}(t)$ on operators that are highly non-local can be safely discarded, as these are are believed to have little quantitative effect on hydrodynamics.


In the following, we focus on one particular such approximation scheme, termed DAOE, that was introduced in Ref. \onlinecite{me_daoe_1} by the present authors. DAOE is implemented as follows. We estimate the diffusion
  constant or hydrodynamical correlation functions using a modified time evolution. In addition to the unitary Heisenberg evolution, we apply a superoperator $\mathcal{G}_{\ell_{*},\gamma}$ with time period $t_D$ that suppresses operators
  longer than some cutoff length $\ell_{*}$, i.e., in \eqnref{eq:D_1st},
  we replace $\mathcal{U}(\tau,0)$ with $\grave{\mathcal{U}}(\tau,0) \equiv \prod^{\tau/t_D  -1 }_{j=0}[\mathcal{U}(t_{j+1},t_j) \mathcal{G}_{\ell_{*},\gamma}]$ (assuming $t/t_\mathrm{D}$ is integer) and where $t_j = j t_D$.
  The original DAOE dissipator is most easily defined in terms of its action on \emph{Pauli strings} $\ket{\sigma^\mu}$, which form a basis in the space of all operators acting on the spin chain. The dissipator then takes the form
  \begin{equation}\label{eq:DAOEdefn}
  \mathcal{G}_{\ell_{*},\gamma}|\sigma^{\mu}\rangle=e^{-\gamma\max(0,\ell(\mu)-\ell_{*})}|\sigma^{\mu}\rangle,
  \end{equation}
  where $\ell(\mu)$ -- the `length' of operator $\sigma^\mu$ -- is the number of non-identity matrices in the string $\sigma^{\mu}$ \footnote{This quantity is known as the \emph{Pauli weight} in quantum computing literature.}. Using this nomenclature, note that the operator $\sigma^z_1 \sigma^x_{27}$ has length $\ell=2$, even though it is spread over a large spatial region (of diameter $27$). It will also be useful to define the superoperator 
  \begin{equation*}
    \mathcal{N}_{\ell_*}  |\sigma^{\mu}\rangle =\max(0,\ell(\mu)-\ell_{*}) |\sigma^{\mu}\rangle.
    \end{equation*}
  The DAOE dissipator may then be written {$\mathcal{G}_{\ell_{*},\gamma} =e^{-\gamma \mathcal{N}_{>\ell_*}}$}.  The DAOE dissipator has a compact MPO representation, which is important for the implementation of the DAOE procedure on tensor product representations of operators \cite{me_daoe_1}. 
  
  With dissipation
  present, we obtain in general a different estimate for the diffusion
  constant which we denote as 
  \begin{equation}
  D(\ell_{*},\gamma)=\lim_{n\rightarrow\infty}\sum_{x}\frac{x^{2}}{2 \tau}\langle q_{x}|\prod_{k=0}^{n}[\mathcal{U}(t_{k+1},t_{k})\mathcal{G}_{\ell_{*},\gamma}]| q_{0}\rangle,\label{eq:D_2nd}
  \end{equation}
  $t_{k} \equiv k t_\mathrm{D}$, and $\tau \equiv t_n$. Note that this expression is identically equal to
  $D$ if $\gamma=0$ or $\ell_{*}=\infty$, since the dissipation is switched off in both of these limits. We also note that while the method depends on the parameter $t_\text{D}$, one can show~\cite{me_daoe_1} that in the limit of small $\gamma$, it is only the ratio $\gamma/t_\text{D}$ that matters; we therefore fix $t_\text{D}$ to some $O(1)$ value and vary $\gamma$. We will assume that $\ell_{*}$
  exceeds the length of operators contained in $q_{x}$, which ensures
  that the DAOE dynamics at least conserve $q$; in most cases we consider, this
  simply requires $\ell_{*}\geq1$. Our goal will be to estimate the size
  of 
  \begin{equation}\label{eq:delta_D}
  \delta D(\ell_{*},\gamma)=D(\ell_{*},\gamma)-D    
  \end{equation}
  as $\ell_{*}$ is increased further.
  
\section{Summary of diffusive hydrodynamics}\label{ssec:hydro_sans_trunc}

  Before turning to our main topic of estimating backflow corrections, it is worth first clarifying how correlation functions behave in the absence of truncation, in the long-time regime where  hydrodynamics holds sway~\cite{forster}. While these results are well known, we are not aware of any reference that presents all the facts that we require for our analysis, so we will need to synthesize results from multiple works~\cite{forster,huse_tails,Khemani_2018,RoschTails,delacretaz2019,delacretazsu2} into a unified picture. 
 
 At long times, the correlation functions between local operators $o_{0,x}$ are dominated by the slowest operators $\eta_{0,x}$ with which  they can develop an overlap under time evolution. In other words, the correlation function can be approximated as
  \begin{equation}\label{eq:hydroheur}
      C(\tau,x) \equiv \langle o_0 (\tau)| o_x\rangle \sim \langle \eta_0 (\tau) | \eta_x\rangle 
  \end{equation}
  at leading order in $\tau^{-1}$, and up to multiplicative constants.
  
  The physical idea underpinning \eqnref{eq:hydroheur} is similar to that of the operator product expansion in quantum field theory \cite{delacretazOPE}. Using the language developed in the previous section, we can explain it as follows. We choose some appropriate basis of operators (not necessarily Pauli strings), which we denote $\ket{\eta^\mu}$, and write
  \begin{align}\label{eq:OPE}
      \braket{o_0|\mathcal{U}(\tau,0)|o_x} = \braket{o_0(\Delta)|\mathcal{U}(\tau-\Delta,\Delta)|o_x(-\Delta)} \nonumber \\  
      = \sum_{\mu,\nu} c^*_\mu(0,\Delta)c_\nu(x,-\Delta) \braket{\eta^\mu | \mathcal{U}(\tau-\Delta,\Delta) |\eta^\nu},
  \end{align}
  where we assume time-translation invariance and $c_\mu(x,\Delta) \equiv \braket{\eta^\mu|o_x(\Delta)}$. Here, $\Delta \ll \tau$ is some short time scale that depends on the size of $o$. The sum in \eqnref{eq:OPE} will be dominated by the basis operators with slowly decaying correlations. In particular, we expect $\braket{\eta^\mu | \mathcal{U}(\tau-\Delta,\Delta) |\eta^\nu} \approx \braket{\eta^\mu | \mathcal{U}(\tau,0) |\eta^\nu} = A_{\mu\nu} \tau^{-a_{\mu\nu}} + \ldots$ to leading order in $\tau$. At large $\tau$, the sum in \eqnref{eq:OPE} will be dominated by the smallest $a_{\mu\nu}$. The exception is if there is a cancellation between terms with the same $a_{\mu\nu}$; however, this can be avoided by making an appropriate choice for the basis $\eta^\mu$, which will be dictated by the choice of $o_{0,x}$.
  
  \eqnref{eq:hydroheur} can be illustrated with the following picture:
  \begin{equation}\label{eq:hydro_picture_gen}
\begin{tikzpicture}[x=0.75pt,y=0.75pt,yscale=-0.6,xscale=0.6]

\draw   (75,44.25) -- (32.5,79.25) -- (32.5,9.25) -- cycle ;
\draw  [fill={rgb, 255:red, 0; green, 0; blue, 0 }  ,fill opacity=0.45 ] (75,44.25) -- (32.5,79.25) -- (32.5,9.25) -- cycle ;

\draw   (468.5,60.83) -- (511,25.83) -- (511,95.83) -- cycle ;
\draw  [fill={rgb, 255:red, 0; green, 0; blue, 0 }  ,fill opacity=0.45 ] (468.5,60.83) -- (511,25.83) -- (511,95.83) -- cycle ;

\draw [line width=1.5]  [dash pattern={on 1.69pt off 2.76pt}]  (75,44.25) .. controls (115,14.25) and (179.5,115.67) .. (219.5,85.67) ;
\draw [line width=1.5]  [dash pattern={on 1.69pt off 2.76pt}]  (318,91) .. controls (358,61) and (428.5,90.83) .. (468.5,60.83) ;
\draw    (39,6) -- (506,6) ;
\draw [shift={(508,6)}, rotate = 180] [color={rgb, 255:red, 0; green, 0; blue, 0 }  ][line width=0.75]    (10.93,-3.29) .. controls (6.95,-1.4) and (3.31,-0.3) .. (0,0) .. controls (3.31,0.3) and (6.95,1.4) .. (10.93,3.29)   ;
\draw [shift={(37,6)}, rotate = 0] [color={rgb, 255:red, 0; green, 0; blue, 0 }  ][line width=0.75]    (10.93,-3.29) .. controls (6.95,-1.4) and (3.31,-0.3) .. (0,0) .. controls (3.31,0.3) and (6.95,1.4) .. (10.93,3.29)   ;
\draw [line width=1.5]  [dash pattern={on 1.69pt off 2.76pt}]  (219.5,85.67) .. controls (259.5,55.67) and (278,121) .. (318,91) ;

\draw (7,50.5) node [anchor=north west][inner sep=0.75pt]  [rotate=-270] [align=left] {$\displaystyle o_{0}$};
\draw (515.5,66.5) node [anchor=north west][inner sep=0.75pt]  [rotate=-270] [align=left] {$\displaystyle o_{x}$};
\draw (262,8.4) node [anchor=north west][inner sep=0.75pt]    {$\tau $};
\draw (430,44) node [anchor=north west][inner sep=0.75pt]    {$\eta _{x}$};
\draw (85,50) node [anchor=north west][inner sep=0.75pt]    {$\eta _{0}$};
\end{tikzpicture}
   \end{equation}
  Here $o_0$ contracts down to $\eta_0$ in some $\mathcal{O}(1)$ time, which then propagates until a time close to $\tau$, at which point the operator grows to form $o_x$. It will turn out that, modulo a few special cases and caveats we will shortly mention, the slowest variable is  the local density $\eta= z$ if $\langle o|S^z\rangle \neq 0$. Otherwise, it is the spatial gradient of the density $\eta= \partial z$. Readers willing to take this on trust may skip the remainder of this section, which serves to justify this result.
  
  In the systems studied here, $z$ operators are subject to diffusive hydrodynamics, i.e.,
  \begin{equation}\label{eq:diff_propagator}
      \langle z_0(\tau)| z_x(0)\rangle = K(\tau, x)
  \end{equation} 
  where $K(\tau,x)$ is a function whose width scales as $\sqrt{ D \tau}$, and takes approximate scaling form $K(\tau,x) = \tau^{-d/2} F(|x|/\sqrt{\tau})$ in the diffusive limit in $d$ spatial dimensions. It further obeys $\int_x K(\tau,x)=1$ due to the fact that $\int_x z_x$ is a conserved quantity \footnote{To simplify notation, we will occasionally lapse into using continuum notation $\int,\partial_x$ rather than their more cumbersome lattice analogues.}.

  What are the slowest hydrodynamical variables? Natural candidates involve products of $z$'s and spatial derivatives, which we denote as $\partial^n z^m$ (the notation suppresses the details of how the spatial derivatives are distributed among the $z$'s, and the regularization of the product $z^m$). According to naive dimension counting, this variable will have autocorrelation functions decaying as $\tau^{-n-m d /2}$. That is because each \emph{pair} of $z$'s is associated with a single factor of the propagator $K\sim \tau^{-d/2}$, while diffusive scaling implies $\partial_x \sim \tau^{-1/2}$. Thus, according to this naive dimension counting, the slowest variables are those with as few powers of $z,\partial$ as possible. In $d=1$ spatial dimensions (our main focus), the four most relevant variables are $z,z^2,z^3,\partial z$ which by dimension counting decay as $\tau^{-1/2,-1,-3/2,-3/2}$ respectively.

   $\partial z$ will play a special role in what follows. While $\partial z$ is not the slowest of all hydrodynamical variables, it turns out that many operators are prevented from developing an overlap with $z^{1,2,3}$ by symmetry constraints. These  constraints arise from the fact that the overlap of an operator with the total spin (or indeed any function $f$ of the spin) {$\langle o(\tau) | f(S^z) \rangle$} is independent of time because the operator $S^z$ is exactly conserved under time evolution. 
   
   We will illustrate this point with a few examples of operators $o_{0,x}$, applying \eqnref{eq:hydroheur} and symmetry constraints to predict the long-time behavior of their correlation function. If $o_x=z_x,o_0=z_0$, then the answer is already clear. $C(\tau,x)=K(\tau,x)$ by definition. As noted, this decays at late times as $\tau^{-1/2}$.

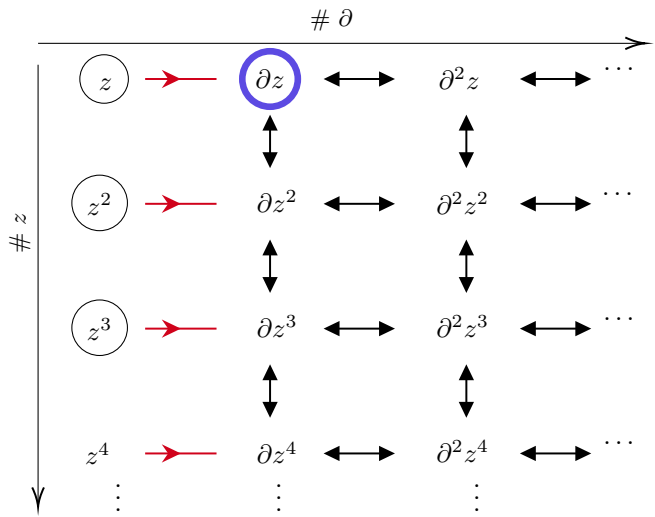
\begin{figure}
\begin{tikzpicture}[x=0.75pt,y=0.75pt,yscale=-0.9,xscale=0.9]
\draw [color={rgb, 255:red, 208; green, 2; blue, 27 }  ,draw opacity=1 ][line width=0.75]    (110,50) -- (150,50) ;
\draw [shift={(130,50)}, rotate = 180] [fill={rgb, 255:red, 208; green, 2; blue, 27 }  ,fill opacity=1 ][line width=0.08]  [draw opacity=0] (10.72,-5.15) -- (0,0) -- (10.72,5.15) -- (7.12,0) -- cycle    ;
\draw [line width=0.75]    (213,50) -- (247,50) ;
\draw [shift={(250,50)}, rotate = 180] [fill={rgb, 255:red, 0; green, 0; blue, 0 }  ][line width=0.08]  [draw opacity=0] (8.93,-4.29) -- (0,0) -- (8.93,4.29) -- cycle    ;
\draw [shift={(210,50)}, rotate = 0] [fill={rgb, 255:red, 0; green, 0; blue, 0 }  ][line width=0.08]  [draw opacity=0] (8.93,-4.29) -- (0,0) -- (8.93,4.29) -- cycle    ;
\draw [line width=0.75]    (323,50) -- (357,50) ;
\draw [shift={(360,50)}, rotate = 180] [fill={rgb, 255:red, 0; green, 0; blue, 0 }  ][line width=0.08]  [draw opacity=0] (8.93,-4.29) -- (0,0) -- (8.93,4.29) -- cycle    ;
\draw [shift={(320,50)}, rotate = 0] [fill={rgb, 255:red, 0; green, 0; blue, 0 }  ][line width=0.08]  [draw opacity=0] (8.93,-4.29) -- (0,0) -- (8.93,4.29) -- cycle    ;
\draw [color={rgb, 255:red, 208; green, 2; blue, 27 }  ,draw opacity=1 ][line width=0.75]    (110,120) -- (150,120) ;
\draw [shift={(130,120)}, rotate = 180] [fill={rgb, 255:red, 208; green, 2; blue, 27 }  ,fill opacity=1 ][line width=0.08]  [draw opacity=0] (10.72,-5.15) -- (0,0) -- (10.72,5.15) -- (7.12,0) -- cycle    ;
\draw [line width=0.75]    (213,120) -- (247,120) ;
\draw [shift={(250,120)}, rotate = 180] [fill={rgb, 255:red, 0; green, 0; blue, 0 }  ][line width=0.08]  [draw opacity=0] (8.93,-4.29) -- (0,0) -- (8.93,4.29) -- cycle    ;
\draw [shift={(210,120)}, rotate = 0] [fill={rgb, 255:red, 0; green, 0; blue, 0 }  ][line width=0.08]  [draw opacity=0] (8.93,-4.29) -- (0,0) -- (8.93,4.29) -- cycle    ;
\draw [line width=0.75]    (323,120) -- (357,120) ;
\draw [shift={(360,120)}, rotate = 180] [fill={rgb, 255:red, 0; green, 0; blue, 0 }  ][line width=0.08]  [draw opacity=0] (8.93,-4.29) -- (0,0) -- (8.93,4.29) -- cycle    ;
\draw [shift={(320,120)}, rotate = 0] [fill={rgb, 255:red, 0; green, 0; blue, 0 }  ][line width=0.08]  [draw opacity=0] (8.93,-4.29) -- (0,0) -- (8.93,4.29) -- cycle    ;
\draw [line width=0.75]    (180,73) -- (180,97) ;
\draw [shift={(180,100)}, rotate = 270] [fill={rgb, 255:red, 0; green, 0; blue, 0 }  ][line width=0.08]  [draw opacity=0] (8.93,-4.29) -- (0,0) -- (8.93,4.29) -- cycle    ;
\draw [shift={(180,70)}, rotate = 90] [fill={rgb, 255:red, 0; green, 0; blue, 0 }  ][line width=0.08]  [draw opacity=0] (8.93,-4.29) -- (0,0) -- (8.93,4.29) -- cycle    ;
\draw [line width=0.75]    (290,73) -- (290,97) ;
\draw [shift={(290,100)}, rotate = 270] [fill={rgb, 255:red, 0; green, 0; blue, 0 }  ][line width=0.08]  [draw opacity=0] (8.93,-4.29) -- (0,0) -- (8.93,4.29) -- cycle    ;
\draw [shift={(290,70)}, rotate = 90] [fill={rgb, 255:red, 0; green, 0; blue, 0 }  ][line width=0.08]  [draw opacity=0] (8.93,-4.29) -- (0,0) -- (8.93,4.29) -- cycle    ;
\draw [color={rgb, 255:red, 208; green, 2; blue, 27 }  ,draw opacity=1 ][line width=0.75]    (110,190) -- (150,190) ;
\draw [shift={(130,190)}, rotate = 180] [fill={rgb, 255:red, 208; green, 2; blue, 27 }  ,fill opacity=1 ][line width=0.08]  [draw opacity=0] (10.72,-5.15) -- (0,0) -- (10.72,5.15) -- (7.12,0) -- cycle    ;
\draw [line width=0.75]    (213,190) -- (247,190) ;
\draw [shift={(250,190)}, rotate = 180] [fill={rgb, 255:red, 0; green, 0; blue, 0 }  ][line width=0.08]  [draw opacity=0] (8.93,-4.29) -- (0,0) -- (8.93,4.29) -- cycle    ;
\draw [shift={(210,190)}, rotate = 0] [fill={rgb, 255:red, 0; green, 0; blue, 0 }  ][line width=0.08]  [draw opacity=0] (8.93,-4.29) -- (0,0) -- (8.93,4.29) -- cycle    ;
\draw [line width=0.75]    (323,190) -- (357,190) ;
\draw [shift={(360,190)}, rotate = 180] [fill={rgb, 255:red, 0; green, 0; blue, 0 }  ][line width=0.08]  [draw opacity=0] (8.93,-4.29) -- (0,0) -- (8.93,4.29) -- cycle    ;
\draw [shift={(320,190)}, rotate = 0] [fill={rgb, 255:red, 0; green, 0; blue, 0 }  ][line width=0.08]  [draw opacity=0] (8.93,-4.29) -- (0,0) -- (8.93,4.29) -- cycle    ;
\draw [line width=0.75]    (180,143) -- (180,167) ;
\draw [shift={(180,170)}, rotate = 270] [fill={rgb, 255:red, 0; green, 0; blue, 0 }  ][line width=0.08]  [draw opacity=0] (8.93,-4.29) -- (0,0) -- (8.93,4.29) -- cycle    ;
\draw [shift={(180,140)}, rotate = 90] [fill={rgb, 255:red, 0; green, 0; blue, 0 }  ][line width=0.08]  [draw opacity=0] (8.93,-4.29) -- (0,0) -- (8.93,4.29) -- cycle    ;
\draw [line width=0.75]    (290,143) -- (290,167) ;
\draw [shift={(290,170)}, rotate = 270] [fill={rgb, 255:red, 0; green, 0; blue, 0 }  ][line width=0.08]  [draw opacity=0] (8.93,-4.29) -- (0,0) -- (8.93,4.29) -- cycle    ;
\draw [shift={(290,140)}, rotate = 90] [fill={rgb, 255:red, 0; green, 0; blue, 0 }  ][line width=0.08]  [draw opacity=0] (8.93,-4.29) -- (0,0) -- (8.93,4.29) -- cycle    ;
\draw    (50,42) -- (50,288) ;
\draw [shift={(50,290)}, rotate = 270] [color={rgb, 255:red, 0; green, 0; blue, 0 }  ][line width=0.75]    (10.93,-3.29) .. controls (6.95,-1.4) and (3.31,-0.3) .. (0,0) .. controls (3.31,0.3) and (6.95,1.4) .. (10.93,3.29)   ;
\draw    (50,30) -- (388,30) ;
\draw [shift={(390,30)}, rotate = 180] [color={rgb, 255:red, 0; green, 0; blue, 0 }  ][line width=0.75]    (10.93,-3.29) .. controls (6.95,-1.4) and (3.31,-0.3) .. (0,0) .. controls (3.31,0.3) and (6.95,1.4) .. (10.93,3.29)   ;
\draw [color={rgb, 255:red, 208; green, 2; blue, 27 }  ,draw opacity=1 ][line width=0.75]    (110,260) -- (150,260) ;
\draw [shift={(130,260)}, rotate = 180] [fill={rgb, 255:red, 208; green, 2; blue, 27 }  ,fill opacity=1 ][line width=0.08]  [draw opacity=0] (10.72,-5.15) -- (0,0) -- (10.72,5.15) -- (7.12,0) -- cycle    ;
\draw [line width=0.75]    (213,260) -- (247,260) ;
\draw [shift={(250,260)}, rotate = 180] [fill={rgb, 255:red, 0; green, 0; blue, 0 }  ][line width=0.08]  [draw opacity=0] (8.93,-4.29) -- (0,0) -- (8.93,4.29) -- cycle    ;
\draw [shift={(210,260)}, rotate = 0] [fill={rgb, 255:red, 0; green, 0; blue, 0 }  ][line width=0.08]  [draw opacity=0] (8.93,-4.29) -- (0,0) -- (8.93,4.29) -- cycle    ;
\draw [line width=0.75]    (323,260) -- (357,260) ;
\draw [shift={(360,260)}, rotate = 180] [fill={rgb, 255:red, 0; green, 0; blue, 0 }  ][line width=0.08]  [draw opacity=0] (8.93,-4.29) -- (0,0) -- (8.93,4.29) -- cycle    ;
\draw [shift={(320,260)}, rotate = 0] [fill={rgb, 255:red, 0; green, 0; blue, 0 }  ][line width=0.08]  [draw opacity=0] (8.93,-4.29) -- (0,0) -- (8.93,4.29) -- cycle    ;
\draw [line width=0.75]    (180,213) -- (180,237) ;
\draw [shift={(180,240)}, rotate = 270] [fill={rgb, 255:red, 0; green, 0; blue, 0 }  ][line width=0.08]  [draw opacity=0] (8.93,-4.29) -- (0,0) -- (8.93,4.29) -- cycle    ;
\draw [shift={(180,210)}, rotate = 90] [fill={rgb, 255:red, 0; green, 0; blue, 0 }  ][line width=0.08]  [draw opacity=0] (8.93,-4.29) -- (0,0) -- (8.93,4.29) -- cycle    ;
\draw [line width=0.75]    (290,213) -- (290,237) ;
\draw [shift={(290,240)}, rotate = 270] [fill={rgb, 255:red, 0; green, 0; blue, 0 }  ][line width=0.08]  [draw opacity=0] (8.93,-4.29) -- (0,0) -- (8.93,4.29) -- cycle    ;
\draw [shift={(290,210)}, rotate = 90] [fill={rgb, 255:red, 0; green, 0; blue, 0 }  ][line width=0.08]  [draw opacity=0] (8.93,-4.29) -- (0,0) -- (8.93,4.29) -- cycle    ;

\draw    (87, 50) circle [x radius= 13.6, y radius= 13.6]   ;
\draw (93,57.6) node [anchor=south east] [inner sep=0.75pt]  [font=\normalsize]  {$z$};
\draw  [color={rgb, 255:red, 92; green, 74; blue, 226 }  ,draw opacity=1 ][line width=2.5]   (180, 50) circle [x radius= 15.56, y radius= 15.56]   ;
\draw (180,50) node  [font=\normalsize]  {$\partial z$};
\draw (286,49.5) node  [font=\normalsize]  {$\partial ^{2} z$};
\draw (376.67,44.17) node  [font=\normalsize]  {$\dotsc $};
\draw    (84.5, 119.5) circle [x radius= 15.57, y radius= 15.57]   ;
\draw (93,127.6) node [anchor=south east] [inner sep=0.75pt]  [font=\normalsize]  {$z^{2}$};
\draw (184,118.5) node  [font=\normalsize]  {$\partial z^{2}$};
\draw (288,119.5) node  [font=\normalsize]  {$\partial ^{2} z^{2}$};
\draw (376,114) node  [font=\normalsize]  {$\dotsc $};
\draw    (84.5, 189.5) circle [x radius= 15.57, y radius= 15.57]   ;
\draw (93,197.6) node [anchor=south east] [inner sep=0.75pt]  [font=\normalsize]  {$z^{3}$};
\draw (184,188.5) node  [font=\normalsize]  {$\partial z^{3}$};
\draw (288,188.5) node  [font=\normalsize]  {$\partial ^{2} z^{3}$};
\draw (376.67,184.17) node  [font=\normalsize]  {$\dotsc $};
\draw (32.4,149) node [anchor=north west][inner sep=0.75pt]  [rotate=-270]  {$\#\ z\ $};
\draw (201,9.4) node [anchor=north west][inner sep=0.75pt]    {$\# \ \partial \ $};
\draw (92,267.6) node [anchor=south east] [inner sep=0.75pt]  [font=\normalsize]  {$z^{4}$};
\draw (184,259.5) node  [font=\normalsize]  {$\partial z^{4}$};
\draw (288,258.5) node  [font=\normalsize]  {$\partial ^{2} z^{4}$};
\draw (376.67,254.17) node  [font=\normalsize]  {$\dotsc $};
\draw (95,286) node  [font=\normalsize,rotate=-90]  {$\dotsc $};
\draw (296,286) node  [font=\normalsize,rotate=-90]  {$\dotsc $};
\draw (185,286) node  [font=\normalsize,rotate=-90]  {$\dotsc $};

\end{tikzpicture}

\caption{Slowest modes in diffusive hydrodynamics. 
We consider operators of form $\partial^n z^m$, where $z$ is a local conserved density; note that this notation suppresses details of how the spatial derivatives are distributed among the $z$'s, and the spatial regularization of the product $z^m$. Arrows indicate symmetry-allowed transitions between operators. The red arrows are one-way, indicating a transition whose reverse is not symmetry allowed. The four slowest operators in $d=1$ spatial dimensions are circled. $\partial z$ plays a special role among these, because any local operator can (considering symmetry alone) develop an overlap with it. By way of comparison, $z^2$ is a slower operator in $d=1$, but many operators are excluded from developing an overlap with it because they are orthogonal to $(S^z)^2$. \label{fig:hydro}}

\end{figure}

 If on the other hand $o_x=\sigma^+_{x} \sigma^-_{x+1},o_0=\sigma^+_{0} \sigma^-_{1}$, then the answer is more interesting. The component of $o_{x} (\tau)$ on single $z$ operators can be written as a superposition $\int_y g(y,\tau) z_y$ for some function $g$. However, as $o_{x}$ is orthogonal to the total spin, we have the constraint that $\int_y g=0$, which implies that this can be rewritten as $\int_y f(y,\tau) \partial_y z_y$ with $f(y)=\int_y^{\infty} \mathrm{d}y' g(y')$ supported in the light-cone.
 The conclusion is that $o_x$ cannot develop a component of $z$ that survives coarse-graining. Similarly, $o_{x}$ is orthogonal to $(S^z)^{2}$, $(S^z)^{3}$ which (after coarse-graining) precludes developing an overlap with the next slowest operators $z^2,z^3$. 

 These considerations show that the slowest hydrodynamical operator with which $o_{x,0}$ can develop an overlap is $\partial z$. Thus we expect $C(\tau,x)\sim \langle \partial z_x(\tau) | \partial z_0 \rangle = \partial^2_{x} K(\tau,x)\sim \tau^{-3/2}$. If $o_x=\sigma^+_{x} \sigma^-_{x+1},o_0=z_0$, the same style of argument shows that $C(\tau,x)\sim \langle \partial z_x(\tau) | z_0 \rangle \sim \partial_x K(\tau,x)$. 
    
Returning to the case of operators of form $o = \partial^n z^m$, recall that their autocorrelations decay as $\tau^{-n-m d /2}$ according to naive dimension counting. However, this does not always correctly predict the decay of correlations for operators involving many $z$'s. That is because such operators can (in accordance with \eqnref{eq:hydroheur}) develop an overlap with slower hydrodynamical variables involving a product of fewer $z$'s, and as a result decay slower than would be expected from dimension counting. However, the question of which hydrodynamical variables can develop an overlap is once again constrained by symmetry: The situation for $\partial^n z^m$ is summarized in \figref{fig:hydro}.

The same reasoning leads to a more general principle, useful for later discussions in this work. If $o_{x,0}$ are operator strings orthogonal to $S^z,(S^z)^2,(S^z)^3$, and both are local and charge neutral (i.e., commute with $S^z$), then $C(\tau,x)\sim \langle \partial z_x(\tau) | \partial z_0 \rangle \sim \partial^2_x K(\tau,x)$ should hold at large $\tau$, up to multiplicative constants. In other words, $\partial z$ is the slowest hydrodynamical variable with which $o_{x,0}$ can develop overlap~\footnote{  In $d=2$, the same statement holds but one can drop the requirement of orthogonality with $(S^z)^3$. In $d\geq 3$ one may additionally drop the requirement of orthogonality with $(S^z)^2$.}.

Finally, we note that if an operator $o$ has $[S^z,o]=\lambda o$ with $\lambda\neq 0$ (i.e., the operator creates/destroys charge), then $o$ can never have an overlap with any hydrodynamical variable. That is because all the hydrodynamical variables mentioned above commute with $S^z$, and 
  \begin{equation}
      \langle [S^z,a](\tau) | b \rangle =  \langle a(\tau) | [S^z,b] \rangle 
  \end{equation}
  for any operators $A,B$. In words, the adjoint action of $S^z$ commutes with time evolution. In the absence of additional slow modes in the system, such a variable is expected to have exponentially decaying correlations.

This section has focused on the case where the sole conserved quantity is a U$(1)$ charge. While we think our main conclusions generalize to systems with energy conservation and to finite temperature, some of the details (particularly those involving operator overlaps) may change\footnote{We thank Luca Delacretaz for a related discussion.}. 

\section{Theoretical picture for backflow contributions}\label{sec:theory}

 In this section, we present a theoretical justification for our central claim: systematic errors (e.g., \eqnref{eq:delta_D}) due to DAOE are exponentially small in $\ell_*$. Our strategy is as follows. We first consider a stripped-down version of the question: What is the effect on a hydrodynamic correlation function $C(\tau,x)$ of discarding all operators above some length $\ell_*$ half-way through the evolution (at time $t=\tau/2$)? We will call this quantity $C_{>\ell_*}(\tau,x)$ and we shall argue that it is exponentially suppressed in $\ell_*$. Then, in \secref{ss:periodic_diss}, we argue that this result extends to the case of periodically applied dissipation. 
 
 
 \subsection{A single dissipation event}\label{ssec:first bound}
 
Our first aim is to bound the quantities
\begin{subequations}
\label{eq:stripped}
    \begin{align}
    C_{>\ell_*}(\tau,x) &\equiv \langle o_0| \mathcal{U}(\tau=2t,t) \mathcal{P}_{>\ell_*} \mathcal{U}(t,0)| o_x \rangle\\
    C_{\ell}(\tau,x) &\equiv \langle o_0| \mathcal{U}(\tau=2t,t) \mathcal{P}_{\ell} \,\mathcal{U}(t,0)| o_x \rangle{,}
\end{align}
\end{subequations}
where $o_{0,x}$ are local operators based around positions $0,x$ respectively. Here $ \mathcal{P}_{\ell}$ is a projector onto all operators of length  $\ell$. Similarly $\mathcal{P}_{>\ell_*}=\sum_{\ell>\ell_*} \mathcal{P}_{\ell}$ is a projector onto all operators greater than length $\ell_*$; these quantities measure the contributions to $C(\tau,x)$ from `backflow' type processes where the operator grows to a significant size ($\ell>\ell_*$) and then shrinks back to $o$.

  The hydrodynamical picture of Eqs.~\eqref{eq:hydroheur} and~\eqref{eq:hydro_picture_gen} suggests how to bound \eqnref{eq:stripped} at long times. We contend that the leading contributions take the form
\begin{equation}\label{eq:trunc_picture_gen}
\begin{tikzpicture}[x=0.75pt,y=0.75pt,yscale=-0.6,xscale=0.6]

\draw   (75,44.25) -- (32.5,79.25) -- (32.5,9.25) -- cycle ;
\draw  [fill={rgb, 255:red, 0; green, 0; blue, 0 }  ,fill opacity=0.45 ] (75,44.25) -- (32.5,79.25) -- (32.5,9.25) -- cycle ;

\draw   (468.5,60.83) -- (511,25.83) -- (511,95.83) -- cycle ;
\draw  [fill={rgb, 255:red, 0; green, 0; blue, 0 }  ,fill opacity=0.45 ] (468.5,60.83) -- (511,25.83) -- (511,95.83) -- cycle ;

\draw [line width=1.5]  [dash pattern={on 1.69pt off 2.76pt}]  (75,44.25) .. controls (115,14.25) and (179.5,115.67) .. (219.5,85.67) ;
\draw [line width=1.5]  [dash pattern={on 1.69pt off 2.76pt}]  (322.36,85.67) .. controls (362.36,55.67) and (428.5,90.83) .. (468.5,60.83) ;
\draw    (216,29) -- (324,29) ;
\draw [shift={(326,29)}, rotate = 180] [color={rgb, 255:red, 0; green, 0; blue, 0 }  ][line width=0.75]    (10.93,-3.29) .. controls (6.95,-1.4) and (3.31,-0.3) .. (0,0) .. controls (3.31,0.3) and (6.95,1.4) .. (10.93,3.29)   ;
\draw [shift={(214,29)}, rotate = 0] [color={rgb, 255:red, 0; green, 0; blue, 0 }  ][line width=0.75]    (10.93,-3.29) .. controls (6.95,-1.4) and (3.31,-0.3) .. (0,0) .. controls (3.31,0.3) and (6.95,1.4) .. (10.93,3.29)   ;
\draw  [fill={rgb, 255:red, 155; green, 155; blue, 155 }  ,fill opacity=0.34 ] (270.93,34.24) -- (322.36,85.67) -- (270.93,137.09) -- (219.5,85.67) -- cycle ;
\draw    (330.93,40.67) -- (330.93,138) ;
\draw [shift={(330.93,140)}, rotate = 270] [color={rgb, 255:red, 0; green, 0; blue, 0 }  ][line width=0.75]    (10.93,-3.29) .. controls (6.95,-1.4) and (3.31,-0.3) .. (0,0) .. controls (3.31,0.3) and (6.95,1.4) .. (10.93,3.29)   ;
\draw [shift={(330.93,38.67)}, rotate = 90] [color={rgb, 255:red, 0; green, 0; blue, 0 }  ][line width=0.75]    (10.93,-3.29) .. controls (6.95,-1.4) and (3.31,-0.3) .. (0,0) .. controls (3.31,0.3) and (6.95,1.4) .. (10.93,3.29)   ;

\draw (7,50.5) node [anchor=north west][inner sep=0.75pt]  [rotate=-270] [align=left] {$\displaystyle o_{0}$};
\draw (515.5,66.5) node [anchor=north west][inner sep=0.75pt]  [rotate=-270] [align=left] {$\displaystyle o_{x}$};
\draw (264,76.4) node [anchor=north west][inner sep=0.75pt]    {$\Gamma $};
\draw (233,80) node [anchor=north west][inner sep=0.75pt]    {$y$};
\draw (295,77.07) node [anchor=north west][inner sep=0.75pt]    {$y'$};
\draw (262,9.4) node [anchor=north west][inner sep=0.75pt]    {$\Delta$};
\draw (338,140) node [anchor=north west][inner sep=0.75pt]  [rotate=-270]  {$\ell \geq \ell _{*}$};
\end{tikzpicture}
  \end{equation}
  In words, $o_0$ decays into a hydrodynamical operator $\eta$, which propagates until a time near $t=\tau/2$. It then grows so as to satisfy the length constraint imposed by the projector $\mathcal{P}_{\ell}$ with $\ell>\ell_*$, before contracting back to a hydrodynamical operator, which propagates to $o_x$ as before. It will be sufficient for us to consider the case where $o_{0,x}$ are both orthogonal to $z^{2,3}$, so that the dotted line corresponds to the propagation of a single $z$, with  $\eta_{0,x} \in \{z,\partial z\}$. 
  
 The growth/contraction process is represented by the grey box (which we refer to as $\Gamma_\ell(\Delta,y,y')$) and occurs over a time-scale $\Delta$ which is in turn set by $\ell$; note that this time scale will need to be at least as long as $\mathcal{O}(\ell)$ to give the hydrodynamic operator sufficient time (set by a Lieb-Robinson bound) to grow to length $\ell$. The initial and final spatial locations $y,y'$ will need to be summed over, but they are forced by locality and the relative smallness of $\Delta$ to be close to one another.

  The resulting process will decay with a power law in $t$, coming from the diffusive propagators (dotted lines) making up \eqnref{eq:trunc_picture_gen}. The precise exponent of this power law can be  determined using the hydrodynamical rules discussed in \secref{ssec:hydro_sans_trunc}; we will do this in the next section. 

 However, the key point is that the process will be additionally suppressed by the amplitude associated with the grey box. It is this suppression that allows us to isolate the $\ell$ dependence of \eqnref{eq:stripped}. Our claim, which we support with both analytical arguments and numerical results below,  is that such amplitudes are exponentially suppressed in $\ell$. After summing the contributions over all $\ell>\ell_*$, it follows that $C_{>\ell_*}$ is exponentially small in $\ell_*$ as required. 
 
 There are various assumptions made in the picture presented above. The key idea is that the grow-shrink process represents an excursion out of the space of small operators, and the longer the excursion, the harder it is for the operator to find its way back to a small hydrodynamical operator. This idea leads to the statement above that $\Delta$ is set by $\ell_*$ rather than $\tau$ (which is equivalent to the fast decay of $\Gamma_\ell$ with $\Delta$), and also responsible the exponential suppression of $\Gamma_\ell$ in $\ell$.

\subsection{Estimating $\Gamma_\ell$}\label{ssec:estgamma}

We now present a quantitative  justification of this idea. To see why $\Gamma_\ell$ is exponentially suppressed in $\ell$, introduce a resolution of the identity at time $t$:
   \begin{equation}\label{eq:Gamma}
      \Gamma_{\ell}(\Delta,y,y')= \sum_{a: \ell(a)=\ell} T_{\eta_y \rightarrow a}T_{a \rightarrow \eta_{y'}},
  \end{equation}
  where we sum over all operators of length $\ell$, and the $T$'s are  transition amplitudes. For example, {$T_{\eta_y \rightarrow a}= \langle \eta_y|\mathcal{U}(\Delta,\Delta/2)| a\rangle$} is the amplitude for $\eta_y$ to evolve to operator $a$ over time $\Delta/2$; we have dropped the explicit $\Delta$ dependence of $T_{\eta_y \rightarrow a}$ to simplify the notation.
  
  \subsubsection{Dynamics without conserved quantities}

  We first note that there are good theoretical reasons to expect amplitudes of form \eqnref{eq:Gamma} to be small in a system \emph{without} any conservation laws (including conservation of energy). Operator spreading in such models have been studied in Refs. \onlinecite{nahum_no_symm,vonKeyserlingk2017_no_symm} (see also Ref. \onlinecite{nahumfatop}). In this case, a natural expectation is that the summand in \eqnref{eq:Gamma} has a randomly fluctuating sign as a function of $a$. That allows us to make the approximation
  \begin{align}
  &|\Gamma_{\ell}(\Delta,y,y')|^2 \nonumber \\
  &= \sum_{a,a': \ell(a),\ell(a')=\ell} T_{\eta_y \rightarrow a} T_{\eta_y \rightarrow a'}^* T_{a \rightarrow \eta_{y'}}  T^*_{a' \rightarrow \eta_{y'}} \label{eq:GammawRUCline1}\\
  &\approx \sum_{a: \ell(a)=\ell} |T_{\eta_y \rightarrow a}|^2 |T_{a \rightarrow \eta_{y'}}|^2. \label{eq:GammawRUCline2}
  \end{align}
  Using the sum rule $\sum_{a} |T_{o\rightarrow a}|^2 = 1$ for a normalized operator $o$,  we can further bound this by 
  \begin{equation}\label{eq:RUCsanssymmfinal}
  |\Gamma_{\ell}(\Delta,y,y')|^2 \lesssim \sup_{a:\ell(a)=\ell} |T_{\eta_y\rightarrow a}|^2.
  \end{equation}
  One can further justify this equation by an analytical calculation in Haar random unitary circuits of the kind employed in Refs. \onlinecite{vonKeyserlingk2017_no_symm,nahum_no_symm}, where taking the average of \eqnref{eq:GammawRUCline1} leads to $\overline{|\Gamma_{\ell}|^2} \leq \sup_{a:\ell(a)=\ell} \overline{|T_{\eta_y\rightarrow a}|^2}$ exactly.
  
  Considering the RHS of \eqnref{eq:RUCsanssymmfinal}, we note that there are exponentially many operators $a$ of length $\ell$ in the forward light cone of $\eta_y$. Since we assumed no conservation laws, one expects that these all occur with roughly equal amplitudes, which readily gives the required exponential suppression of $\Gamma_\ell$. Once again, this can be verified explicitly in random circuits (see e.g., Eq.~(C2) in ~\cite{vonKeyserlingk2017_no_symm}). In that case, one finds that the result is exponentially decaying both in $\ell$ and in $\Delta$.

\subsubsection{Systems with U$(1)$ symmetry}\label{sssec:u1}

The preceding argument is well controlled (e.g. it can be made exact in random circuits), but flawed because it ignores the role of continuous symmetries in the grow/shrink processes. To address that problem, we want to consider a system which has a single conservation law associated to a U$(1)$ symmetry (we still exclude energy conservation). Paradigmatic models of this kind can be constructed as random unitary circuits where the structure of the local gates is restricted by the symmetry.~\cite{Khemani_2018,vonKeyserlingk2018_diffusive}. We shall study such a circuit model in detail in \secref{sec:U(1)random}; here we give a more heuristic argument. 
  
  When continuous global symmetries are present, the path from \eqnref{eq:GammawRUCline1} to \eqnref{eq:GammawRUCline2} becomes less controlled, (even in the aforementioned U$(1)$ circuits, we do not have exact expressions for the circuit-averaged operator spreading coefficients). Nevertheless, we now argue (non-rigorously) that the final result (exponential suppression in $\ell$) continues to hold. Intuitively, the symmetry qualitatively changes the behavior of processes that involve relatively short operators, leading to the diffusive (rather than exponential) decay of correlations. However, despite the underlying symmetry, it still seems reasonable (and in-line with existing works \cite{vonKeyserlingk2018_diffusive,Khemani_2018}) to assume that the amplitudes governing the ballistic growth of operators are somewhat random, so that an argument analogous to that presented above still holds.

  An operator $a$ appearing in \eqnref{eq:Gamma} can be expressed as a tensor product of on-site operators of form $I,z,\sigma^{+},\sigma^{-}$. Separate out the $\sigma^{\pm}$ part of $a$ and call it $a_\perp$. Likewise separate out the $z$ components (which can only have support on sites not already occupied by $a_\perp$), calling it $a_z$. Then $a=a_z a_\perp$. For example, the string $a= z_1 I_2 \sigma^+_3 \sigma^-_4$ has $a_z=z_1$ and $a_\perp=\sigma^+_3 \sigma^-_4$. 
  Within the random circuit calculation, $T_{\eta \rightarrow a }T^*_{\eta \rightarrow a' }$ can be non-vanishing (on average) only if $a_\perp=a'_\perp$, while $a_z,a'_z$ are allowed to differ. This gives a generalization of the random-phase approximation above, and one which we expect to hold approximately for more general U$(1)$-symmetric models as well. Another constraint is that, as $\eta$ is hydrodynamical it is also charge neutral. That is equivalent to requiring the number of $\sigma^+$'s is equal to the number of $\sigma^-$'s in $a_\perp$. 
  
  For simplicity of presentation, we restrict ourselves to estimating $\Gamma_\ell$ with $y=y'$. Once more we isolate those backflow corrections due to operators of some size $\ell$ (we will eventually sum over $\ell>\ell_*$). We will assume that, as in the previous case without conservation laws, $\Delta = \Delta_\ell$ is some function of $\ell$ alone, rather than the long time scale $\tau$ in \eqnref{eq:stripped}. We define the size of the effective light cone $d_\ell \equiv v_{\mathrm{LR}} \Delta_\ell$ where $v_{\mathrm{LR}}$ is a Lieb-Robinson velocity. The processes contributing to $\Gamma(\Delta)$ require that $d_\ell\geq \ell$, because the grow/shrink process should produces operators of length at least $\ell$ part-way through the evolution.

  Starting from \eqnref{eq:GammawRUCline1}, and using $a_\perp = a'_\perp$ gives
  \begin{align}
    &|\Gamma_{\ell}(\Delta)|^2 \nonumber\\
    &= {\sum^{ \ell \,\,\,\prime}_{m=0}}
      \sum_{\substack{a_z,a'_z\\a_\perp: \ell(a_\perp)=m }} \delta_{\ell(a_\perp a_z),\ell(a_\perp a'_z)=\ell}  T_{\eta \rightarrow a} T_{\eta \rightarrow a'}^* T_{a \rightarrow \eta}  T^*_{a' \rightarrow \eta}\label{eq:u1dephasing}
  \end{align}
  where $m$ is the length of $a_\perp=a'_\perp$. The dash on the sum over $m$ indicates that $m$ must be even (so that $a$ is charge neutral). We now make a further  approximation: the coefficients $T$ have roughly the same size $|T|$ for all $a$ that are supported within a ball of diameter $d_\ell$ of $y$. In other words, for given length $\ell$, the operator spread coefficients of operators supported in $d_\ell$ are to a good approximation independent of the detailed content of the operator e.g., independent of $m=\ell(a_\perp)$ and the particular locations of operators. This approximation refines an existing picture \cite{abaninho} so as to include some of the effects of charge conservation.
  
  We can now perform the $a_z,a'_z$ sums, picking out combinatorial factor $\binom{d_\ell-m}{\ell-m}^2$. After that, we perform the $a_\perp$ sum, giving factor $\binom{d_\ell}{m}\binom{m}{m/2}$. The result is
  \begin{equation}\label{eq:boundU1_pt1}
    |\Gamma_{\ell}(\Delta)|^2 \lesssim \sum^{ \ell \,\,\,\prime}_{m=0} N(\ell,m) \binom{d_\ell-m}{\ell-m} |T|^4
  \end{equation}
  where $N(\ell,m) = \binom{d_\ell}{m}\binom{m}{m/2}\binom{d_\ell-m}{\ell-m} $ . To  estimate the RHS of this equation, we note the sum rule $1\geq \sum_{a} |T_{\eta\rightarrow a}|^2$. Using the same approximations leading to \eqnref{eq:boundU1_pt1}, this bound may be expressed as
  \begin{equation}\label{eq:boundU1_pt2}
    1\gtrsim  \sum^{ \ell \,\,\,\prime}_{m=0} N(\ell,m) |T|^2\sim |T|^{2}3^{\ell}\binom{d_\ell}{\ell}
  \end{equation}
  The sum on the RHS of \eqnref{eq:boundU1_pt2} has been estimated using a saddle point approximation and the assumption of large $\ell$. The upshot is that $|T|^{2}$ is bounded above by the inverse of the number of operators of length $\ell$ within the region of width $d_\ell$. We can now bound \eqnref{eq:boundU1_pt1} using \eqnref{eq:boundU1_pt2} to give
  \begin{equation}\label{eq:finalbound}
    |\Gamma_{\ell}(\Delta)|^2 \lesssim |T|^2 \sup_{0\leq m\leq \ell} \binom{d_\ell-m}{\ell-m} \sim 3^{-\ell}.
  \end{equation}
  The supremum is achieved for $m=0$. Thus, if we assume operators of length $\ell$ have somewhat uniform operator spreading coefficients for large $\ell$, it is natural that backflow processes are exponentially suppressed in $\ell$. The final step of this argument involves summing \eqnref{eq:finalbound} over contributions for all $\ell>\ell_*$, with the end result result is that the backflow due to a single dissipation event is exponentially suppressed in $\ell_*$

The above argument makes the assumption that $\Delta$ is set by $\ell$. It also assumes that the size of operator spread coefficients to operators of fixed length, within a forward causal cone, are to good approximation uniform, which is certainly not true in general. Therefore, it behooves us to perform additional consistency checks on the results in this section.
We will perform multiple checks. First, we investigate numerically our backflow hypothesis in some ergodic (non-random) spin chains (see Fig.~\ref{fig:IsingandLadder}). Later, in \secref{sec:U(1)random} we will give a more detailed analysis of the U$(1)$-symmetric random circuit model. In that case, we can evaluate backflow contributions at long times numerically (see Fig.~\ref{fig:total}); all of these checks agree with our prediction of exponentially suppressed backflow effects. Moreover, the numerical checks in Fig.~\ref{fig:IsingandLadder} suggest that our arguments apply to systems for which the conserved quantity is energy rather than the U$(1)$ charge assumed in this section. Finally, we will lend further support to our hypothesis by evaluating analytically certain sub-classes of contributions to $|\Gamma_{\ell}|^2$ in the random circuit model and showing that they are consistent with that quantities' exponential suppression in $\ell$ and the conjecture that $\Delta$ is set only by $\ell$.

\subsubsection{Numerical checks in deterministic spin chains}\label{ssec:IsingLadder}

\begin{figure}
\centering
\includegraphics[width=1.\linewidth]{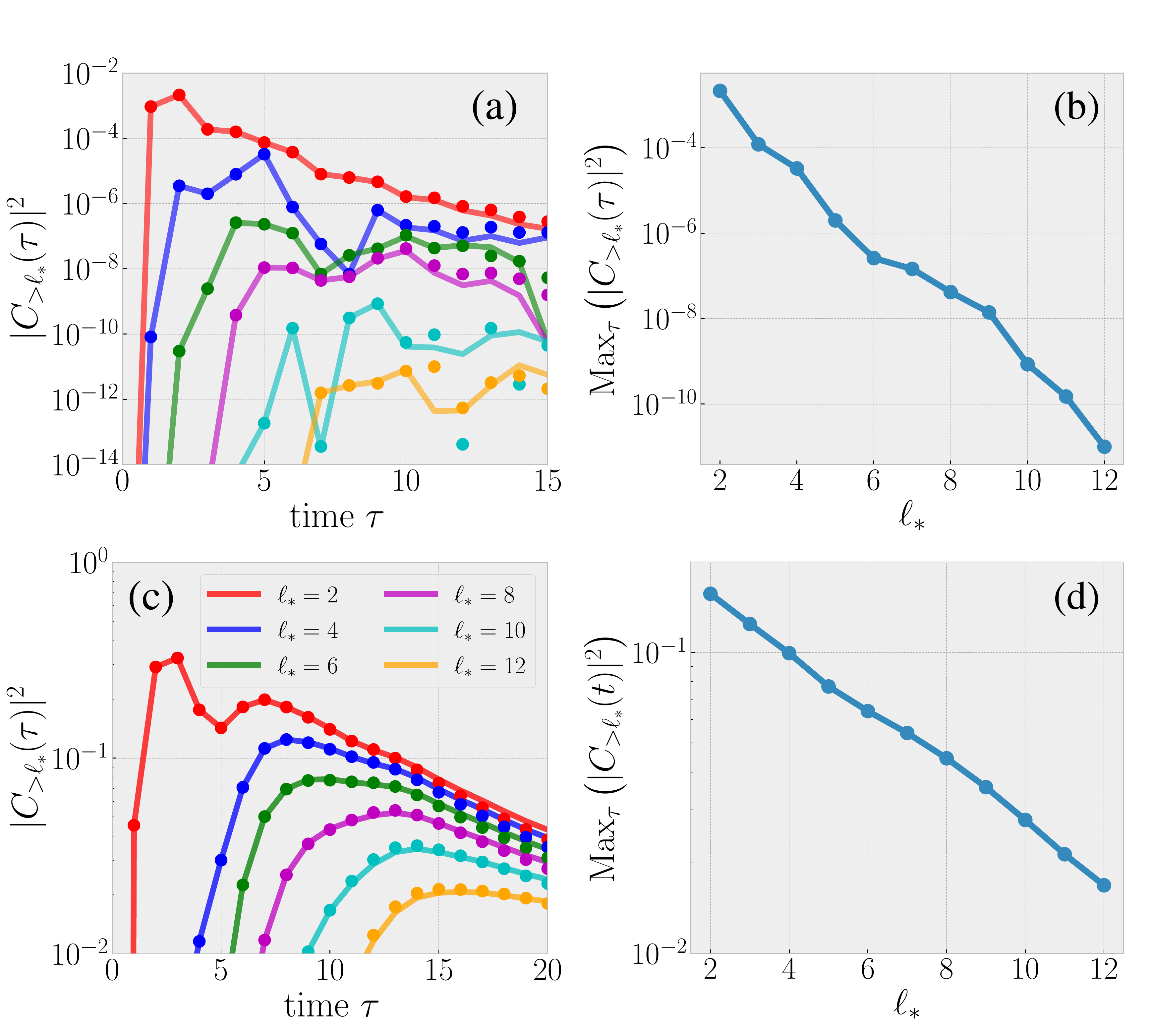}
\caption{Numerical results for backflow correction to autocorrelations due to a single dissipation event. We plot $C_{>\ell_*}(0,\tau)$ as defined in \eqnref{eq:stripped} for the conserved density $o=h_j$ ($=z_j$) in the Ising ($XX$ ladder) model in a,b (c,d). a and c show the dependence of the backflow correction against time; the lines and dots correspond to two different bond dimension $\chi$ ($\chi=512, 1024$ for Ising and $\chi=256, 512$ for the ladder) and allow for estimating truncation effects. b and d show the maximum backflow over all times as a function of $\ell_*$. The data are broadly consistent with the prediction that backflow is exponentially suppressed in $\ell_*$. \label{fig:IsingandLadder}}
\end{figure}

In order to support our argument, here we investigate backflow in two deterministic models (tilted-field Ising and XX ladder) studied using DAOE in our previous work \cite{me_daoe_1}, as well as in numerous other works using alternative methods~\cite{Leviatan2017,jens_hydro_1,Steinigeweg2014_2,Kloss2018}. We investigate the quantity $C_{>\ell_*}(\tau)$, as applied to the autocorrelation function of a conserved density (energy in the Ising case, and $z$-magnetization for the ladder). Both cases are consistent (\figref{fig:IsingandLadder}) with our conjecture that $C_{>\ell_*}(\tau)$ decays exponentially in $\ell_*$, although the data are certainly less clean and decisive than those calculated using random circuits. 
 
We first consider the tilted-field Ising model
\begin{align}\label{eq:IsingDef}
H = \sum_j h_j \equiv \sum_j \left( g_x X_j + g_z Z_j + \frac{Z_{j-1} Z_{j} + Z_j Z_{j+1}}{2} \right),
\end{align}
fixing $g_x = 1.4$ and $g_z = 0.9045$. The model is expected to be strongly chaotic~\cite{Karthik2007,KimHuse2013}. $h_j$ is the energy associated to site $j$. This is the only local conserved density in the model, and its correlations capture energy (or heat) transport~\cite{KimHuse2013}. We represent $h_j(t)$ as an MPO and evolve it using the TEBD method~\cite{VidalTEBD}. We calculate the quantity $C_{>\ell_*}(\tau) = \braket{h_j(-\tau/2)|\mathcal{P}_{>\ell_*}|h_j(\tau/2)}$ which is the $x=0$ version of the quantity defined in \eqnref{eq:stripped}, where we used that the dynamics generated by \eqnref{eq:IsingDef} is time-reversal invariant.

The results are shown in \figref{fig:IsingandLadder}a,b. First, we perform the calculation for various values of $\ell_*$. $C_{>\ell_*}(\tau)$ is suppressed at early times, $\tau < \ell_* / v_\text{LR}$, by the Lieb-Robinson bound. At that time, it grows and reaches a peak value; at late times $\tau \gg \ell_* / v_\text{LR}$ it decays again. While the data is rather noisy, we clearly observe that the curves corresponding to large $\ell_*$ ($\ell_* = 12$ is the largest we consider) are strongly suppressed compared to ones with smaller $\ell_*$. To quantify this, we consider the maximum of $C_{>\ell_*}(\tau)$ over the range of $\tau$ we access and plot it as a function of $\ell_*$ in \figref{fig:IsingandLadder}b. The results are well fit by the exponential decay we predict theoretically; there is a difference of about 7 orders of magnitude in going from $\ell_*=2$ to $\ell_*=12$. 

 We also consider a spin-$1/2$ model on a two-leg ladder~\cite{Steinigeweg2014_2,Karrasch2015,Kloss2018}. We denote by $j=1,\ldots, L$ the rungs of the ladder, and use $a=1,2$ for the two legs. Pauli operators on a given site are specified as $X_{j,a}$, etc. The Hamiltonian then reads
\begin{align}\label{eq:XXladder}
    H = &\sum_{j=1}^L \sum_{a=1,2}  \left( X_{j,a} X_{j+1,a} + Y_{j,a} Y_{j+1,a} \right) \nonumber \\
    +&\sum_{j=1}^L \left( X_{j,1} X_{j,2} + Y_{j,1} Y_{j,2} \right).
\end{align}
Besides energy, this model also conserves the spin $z$ component, $\sum_{j,a} Z_{j,a}$. We calculate $C_{>\ell_*}(\tau)$ as above, with $h_j$ replaced by the average magnetization on a rung, $z_j \equiv (Z_{j,1} + Z_{j,2})/2$. We plot the results in \figref{fig:IsingandLadder}c,d. We find that the trends with $\ell_*$ are cleaner than in the Ising model, although their magnitude is much smaller. In particular, we find that $\ell_*=10$ results are only suppressed by about a factor of $10$ compared to $\ell_*=2$; nonetheless, they are very well fit by an exponential decay. 
 
\subsection{Hydrodynamical expansion for backflow corrections}\label{ss:backflow_hydro_structure}

  We have argued that backflow processes are exponentially suppressed in $\ell_*$. We now determine the dependence of quantities like \eqnref{eq:trunc_picture_gen} on time. The answer will depend both on the structure of $o_0,o_x$ and on whether or not the system is noisy (e.g., a random circuit system). We will verify our predictions using simulations of backflow in the U(1) RUC (\figref{fig:total}); our deterministic data (\figref{fig:IsingandLadder}) do not go to sufficiently long times to test our predictions for time dependence.
  
  As discussed in \secref{ssec:hydro_sans_trunc}, without any truncation the correlator $\langle o_x (\tau)| o_0 \rangle \sim \langle \eta_x (\tau)| \eta _0 \rangle \sim \tau^{-\alpha}$ decays with some power $\alpha$ determined by the slowest hydrodynamical variables $\eta_x,\eta_0$ with which $o_x,o_0$ (respectively) have overlap. 
  When we examine the effects of a single truncation event \eqnref{eq:stripped}, we force the operators to be of size $\ell>\ell_*$ part way through their evolution. Thus, at least for the purposes of understanding time dependence, we expect the leading power in $t$ to be determined by a correlator of form 
  
  \begin{equation}\label{eq:half_way_contribution}
    \langle \eta_x (2t)| a_y(t) \rangle \langle a_y(t) | \eta_0\rangle \Gamma_{y,t},
  \end{equation}
  where $a_y$ is some operator of length  $\ell$, localized near both $y,y'$. $\Gamma_{y,t}$ is the `grey box' amplitude associated with the grow/shrink process. As long as $\ell>2$, $\partial z$ is the  slowest hydro operator with which $a_y$ can develop overlap. Therefore we expect backflow corrections \eqnref{eq:trunc_picture_gen}, fixing $y,y'$, to have the same temporal decay as $\langle \eta_x (\tau)| \partial z_y(t) \rangle \langle \partial z_y(t)  | \eta_0\rangle$. Simple power counting, or substituting in an explicit form for the hydrodynamical correlators \eqnref{eq:diff_propagator}, shows that this decays as $t^{-\alpha - \frac{3}{2}}$. [Note from $\langle z_0(t)| z_0 (0)\rangle \sim t^{-1/2}$ that $z$ scales as $t^{-1/4}$. Furthermore, in diffusive systems $\partial \sim t^{-1/2}$].

  However, to fully account for the backflow processes contributing to \eqnref{eq:stripped}, we will need to integrate over intermediate positions $y\sim y'$. Here, the results will strongly depend on whether the system has randomness, in other words whether the grow/shrink amplitude represented by the grey box in \eqnref{eq:trunc_picture_gen} depends on $y,y'$ individually, rather than just their difference. 
 
  In systems without such randomness, the integration is straightforward, and gives rise to an additional power of $t^{1/2}$, so that \eqnref{eq:stripped} scales as $|\Gamma| \tau^{-\alpha - 1}$, which decays faster than the original correlation function by a power of $\tau$. The exponential suppression in $\ell$ is encoded in $\Gamma$. 
  
  In systems with randomness, relevant to the random circuits we study numerically, the grow/shrink amplitude will have a random sign dependent on position $y$. On average, the contribution to \eqnref{eq:half_way_contribution} will be zero, but its typical modulus will scale as the standard deviation 
  \begin{equation}\label{eq:backflow_1event}
    \sqrt{|\Gamma|^2 \int_y  |\langle \eta_x (2t)| \partial z_y(t) \rangle|^2 |\langle \partial z_y(t) | \eta_0\rangle|^2}.
  \end{equation}

   The scaling in time of \eqnref{eq:backflow_1event} follows immediately from dimension counting, and takes form $\Gamma \tau^{-\alpha - 5/4}$, which decays a factor of $\tau^{-1/4}$ faster than in the case without noise. Once again, the exponential suppression in $\ell$ is provided by the $\Gamma$ factor. 
  
  Combining our results so far, we can evaluate the backflow correction to a diffusion constant \eqnref{eq:D_1st} due to a single dissipation event:
  \begin{equation}\label{eq:d_1 D}
    \delta_1 D(\ell_*,\tau)\equiv \sum_{x} \frac{x^2}{2 \tau} \langle z_x | \mathcal{U}(\tau,\tau/2)  P_{>\ell_*} \mathcal{U}(\tau/2,0) | z_0\rangle.
  \end{equation}
  We thus have to integrate our results over $x$. Estimating the result is straightforward: When there is no noise, $\delta_1 D \sim \Gamma/\tau$. When there is noise, the result show scale as $\delta_1 D \sim \Gamma/\tau^{5/4}$. We test this latter prediction in \figref{fig:total}(a,b) below. 
  
 
  \subsection{Periodic dissipation}\label{ss:periodic_diss}
  
  So far we examined the effect of a single dissipation event on a correlation function, with the conclusion that the error introduced by dissipation is exponentially suppressed in $\ell_*$. What effect do multiple applications of a dissipator have on correlators (e.g., as would be required to calculate \eqnref{eq:D_2nd})? In other words, we want to estimate the error
  \begin{equation}\label{eq:error_order_gamma_corr}
  \delta C(\tau,\gamma)\equiv
    \langle o_x |  [\prod^{n-1}_{j=0}  \mathcal{U}(t_{j+1},t_j) \mathcal{G}(\ell_*,\gamma)] -  \mathcal{U}(t,0) | o_0 \rangle
  \end{equation}
  where $t_j = j t_\text{D}$ and $t_{n}=\tau$. 
  %
  %
  We will use our hydrodynamical picture of backflow processes to argue that  $\partial_\gamma \delta C(\tau,\gamma)$ is exponentially suppressed in $\ell_*$ for all $\gamma$. From this it follows (integrating over $\gamma$) that $\delta C(\tau,\gamma)=\exp[{-\mathcal{O}(\ell_*)}]$. For simplicity, we will be satisfied with showing that $\partial_\gamma \delta C(\tau,\gamma)|_{\gamma=0}$ is exponentially suppressed; the computation for finite $\gamma$ goes through in the same way provided one assumes that DAOE dissipation does not change the fact that the system undergoes diffusion (our previous work \cite{me_daoe_1} supports this assertion). 
  
  The derivative with respect to $\gamma$ is
  \begin{equation}\label{eq:error}
  \partial_\gamma \delta C(\tau,\gamma)=  -\sum^{n}_{j=0}\langle o_x |   \mathcal{U}(\tau,t_j) \mathcal{N}_{>\ell_*} \mathcal{U}(t_{j-1},0)  | o_0 \rangle
  \end{equation}
  The summand here is similar to \eqnref{eq:stripped}, and for the same reasons suppressed exponentially in $\ell_*$. The novelty in the present calculation is that we must sum over times at which we insert the projector $\mathcal{N}$ \footnote{Note $\mathcal{N}_{>\ell_*} = (\ell - \ell_*) \mathcal{P}_{>\ell_*}$ are the same up to a polynomial factor in $\ell$; hence whether we apply $\mathcal{N}$ or $\mathcal{P}$ part way through the evolution will not affect whether the contribution decays exponentially in $\ell_*$}. 
  
  Using the hydrodynamical assumptions above, we represent the derivative as
   \begin{equation}\label{eq:error_schematic_corr}
    - \int^\tau_0 dt \int dy \langle \eta_x (\tau) | \partial z_y(t)\rangle \langle \partial z_y(t) | \eta_0 \rangle \Gamma_{y,t}.
  \end{equation}

  A result from a previous section is that for each fixed $t$, \eqnref{eq:error_schematic_corr} is exponentially decaying in $\ell_*$ and also suppressed by a further factor $\mathcal{O}(t^{-\alpha-1})$ when there is no spatial randomness, and by a factor $\mathcal{O}(t^{-\alpha-5/4})$ when there is. 

  When there is no randomness in space or time, we therefore expect the backflow correction due to be of order $\int_t \mathcal{O}(t^{-\alpha-1}) \sim \mathcal{O}(t^{-\alpha})$; so the correction due to backflow is exponentially decaying in $\ell_*$ but has the same power of $t$ as the original correlation function. 

  When there is randomness in space and time, we need to integrate $\mathcal{O}(t^{-\alpha-5/4})$ over time, taking into account the fact that the sign of the integrand will fluctuate randomly in time ($\Gamma_{y,t}$ now depends randomly on $t$ as well as $y$). The result of that integral is an $\tau^{1/2}$ enhancement, giving a result $\tau^{-\alpha-3/4}$.

  The same reasoning can be applied to quantities like the diffusion constant.  
  \begin{equation}\label{eq:diff_const_full_correction}
  \partial_\gamma \delta D(\ell_*,\tau)\sim  \int_{x,y}  \int^\tau_0 dt \frac{x^2}{2\tau}\langle z_x(\tau) | \partial z_y(t)\rangle \langle \partial z_y(t) | z_0\rangle \Gamma_{t,y}.
  \end{equation}
  
  When there is no noise, simple dimension counting indicates that this term goes as $ \Gamma$; i.e., it represents a correction to the diffusion constant which decays exponentially with $\ell_*$ because $\Gamma$ does. Integrating over $\gamma$ leads to the statement that $\delta D =\exp[{-\mathcal{O}(\ell)}]$.

  When there is noise, we must account for the fluctuating signs of $\Gamma_{t,y}$. In this limit, \eqnref{eq:diff_const_full_correction} may be approximated as 
  \begin{equation}\label{eq:diff_const_full_correction_noise}
     \int dx \sqrt{ |\Gamma|^2\int^\tau_0 dt \int dy ( \frac{x^2}{2\tau})^2 |\langle z_x(\tau) | \partial z_y(t)\rangle \langle z_y(t) | z_0\rangle|^2 }.
  \end{equation}
  Dimension counting, and integrating over $\gamma$, predicts that $ \delta D(\ell_*,\tau) \sim \Gamma \tau^{-3/4}$ which is indeed exponentially decaying in $\ell$. We test this hypothesis numerically in \figref{fig:total}(c,d), for the U$(1)$-symmetric random circuit model.

In summary, we have shown that the backflow corrections to hydrodynamical quantities are accompanied by an exponential suppression in $\ell_*$, and a power law which depends on whether or not the system has randomness, and which hydrodynamical correlation is being considered. In all cases, however, the backflow correction decays in time at least as quickly as the original hydrodynamical correlation function.

\section{U$(1)$ random circuit analysis}\label{sec:U(1)random}
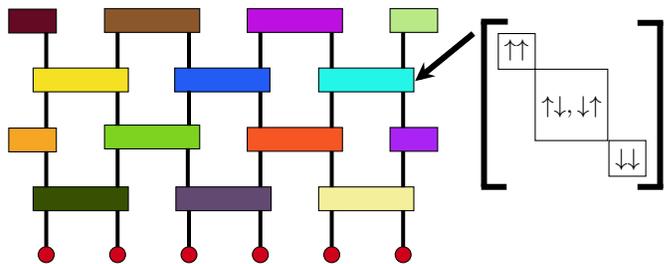
\begin{figure}
\begin{tikzpicture}[x=0.75pt,y=0.75pt,yscale=-0.6,xscale=0.6]

\draw  [fill={rgb, 255:red, 139; green, 87; blue, 42 }  ,fill opacity=1 ] (100,10) -- (180,10) -- (180,30) -- (100,30) -- cycle ;
\draw  [fill={rgb, 255:red, 189; green, 16; blue, 224 }  ,fill opacity=1 ] (220,10) -- (300,10) -- (300,30) -- (220,30) -- cycle ;
\draw  [fill={rgb, 255:red, 100; green, 10; blue, 35 }  ,fill opacity=1 ] (19.5,10.5) -- (59.5,10.5) -- (59.5,30.5) -- (19.5,30.5) -- cycle ;
\draw  [fill={rgb, 255:red, 184; green, 233; blue, 134 }  ,fill opacity=1 ] (340,10) -- (380,10) -- (380,30) -- (340,30) -- cycle ;
\draw  [fill={rgb, 255:red, 126; green, 211; blue, 33 }  ,fill opacity=1 ] (100,108) -- (180,108) -- (180,128) -- (100,128) -- cycle ;
\draw  [fill={rgb, 255:red, 245; green, 86; blue, 35 }  ,fill opacity=1 ] (220,110) -- (300,110) -- (300,130) -- (220,130) -- cycle ;
\draw  [fill={rgb, 255:red, 245; green, 166; blue, 35 }  ,fill opacity=1 ] (19.5,110.5) -- (59.5,110.5) -- (59.5,130.5) -- (19.5,130.5) -- cycle ;
\draw  [fill={rgb, 255:red, 169; green, 35; blue, 245 }  ,fill opacity=1 ] (340,110) -- (380,110) -- (380,130) -- (340,130) -- cycle ;
\draw  [fill={rgb, 255:red, 245; green, 225; blue, 35 }  ,fill opacity=1 ] (40,60) -- (120,60) -- (120,80) -- (40,80) -- cycle ;
\draw  [fill={rgb, 255:red, 35; green, 91; blue, 245 }  ,fill opacity=1 ] (159,60) -- (239,60) -- (239,80) -- (159,80) -- cycle ;
\draw  [fill={rgb, 255:red, 35; green, 245; blue, 230 }  ,fill opacity=1 ] (280,60) -- (360,60) -- (360,80) -- (280,80) -- cycle ;
\draw  [fill={rgb, 255:red, 55; green, 80; blue, 1 }  ,fill opacity=1 ] (40,160) -- (120,160) -- (120,180) -- (40,180) -- cycle ;
\draw  [fill={rgb, 255:red, 97; green, 73; blue, 113 }  ,fill opacity=1 ] (160,160) -- (240,160) -- (240,180) -- (160,180) -- cycle ;
\draw  [fill={rgb, 255:red, 244; green, 239; blue, 154 }  ,fill opacity=1 ] (280,160) -- (360,160) -- (360,180) -- (280,180) -- cycle ;
\draw [line width=1.5]    (111,30) -- (111,60.33) ;
\draw [line width=1.5]    (51,30) -- (51,60.33) ;
\draw [line width=1.5]    (171,30) -- (171,60.33) ;
\draw [line width=1.5]    (291,30) -- (291,60.33) ;
\draw [line width=1.5]    (231,30) -- (231,60.33) ;
\draw [line width=1.5]    (351,30) -- (351,60.33) ;
\draw [line width=1.5]    (111,80) -- (111,108.5) ;
\draw [line width=1.5]    (51,80) -- (51,110.33) ;
\draw [line width=1.5]    (171,79) -- (171,108.5) ;
\draw [line width=1.5]    (291,80) -- (291,110.33) ;
\draw [line width=1.5]    (231,80) -- (231,110.33) ;
\draw [line width=1.5]    (351,80) -- (351,110.33) ;
\draw [line width=1.5]    (51,130) -- (51,160.33) ;
\draw [line width=1.5]    (170,127.5) -- (170,160.33) ;
\draw [line width=1.5]    (291,130) -- (291,160.33) ;
\draw [line width=1.5]    (231,130) -- (231,160.33) ;
\draw [line width=1.5]    (351,130) -- (351,160.33) ;
\draw [line width=1.5]    (111,180) -- (111,210.33) ;
\draw [line width=1.5]    (51,180) -- (51,210.33) ;
\draw [line width=1.5]    (171,180) -- (171,210.33) ;
\draw [line width=1.5]    (291,180) -- (291,210.33) ;
\draw [line width=1.5]    (231,180) -- (231,210.33) ;
\draw [line width=1.5]    (351,180) -- (351,210.33) ;
\draw  [fill={rgb, 255:red, 208; green, 2; blue, 27 }  ,fill opacity=1 ] (44.33,217) .. controls (44.33,213.32) and (47.32,210.33) .. (51,210.33) .. controls (54.68,210.33) and (57.67,213.32) .. (57.67,217) .. controls (57.67,220.68) and (54.68,223.67) .. (51,223.67) .. controls (47.32,223.67) and (44.33,220.68) .. (44.33,217) -- cycle ;
\draw  [fill={rgb, 255:red, 208; green, 2; blue, 27 }  ,fill opacity=1 ] (104.33,217) .. controls (104.33,213.32) and (107.32,210.33) .. (111,210.33) .. controls (114.68,210.33) and (117.67,213.32) .. (117.67,217) .. controls (117.67,220.68) and (114.68,223.67) .. (111,223.67) .. controls (107.32,223.67) and (104.33,220.68) .. (104.33,217) -- cycle ;
\draw  [fill={rgb, 255:red, 208; green, 2; blue, 27 }  ,fill opacity=1 ] (164.33,217) .. controls (164.33,213.32) and (167.32,210.33) .. (171,210.33) .. controls (174.68,210.33) and (177.67,213.32) .. (177.67,217) .. controls (177.67,220.68) and (174.68,223.67) .. (171,223.67) .. controls (167.32,223.67) and (164.33,220.68) .. (164.33,217) -- cycle ;
\draw  [fill={rgb, 255:red, 208; green, 2; blue, 27 }  ,fill opacity=1 ] (224.33,217) .. controls (224.33,213.32) and (227.32,210.33) .. (231,210.33) .. controls (234.68,210.33) and (237.67,213.32) .. (237.67,217) .. controls (237.67,220.68) and (234.68,223.67) .. (231,223.67) .. controls (227.32,223.67) and (224.33,220.68) .. (224.33,217) -- cycle ;
\draw  [fill={rgb, 255:red, 208; green, 2; blue, 27 }  ,fill opacity=1 ] (284.33,217) .. controls (284.33,213.32) and (287.32,210.33) .. (291,210.33) .. controls (294.68,210.33) and (297.67,213.32) .. (297.67,217) .. controls (297.67,220.68) and (294.68,223.67) .. (291,223.67) .. controls (287.32,223.67) and (284.33,220.68) .. (284.33,217) -- cycle ;
\draw  [fill={rgb, 255:red, 208; green, 2; blue, 27 }  ,fill opacity=1 ] (344.33,217) .. controls (344.33,213.32) and (347.32,210.33) .. (351,210.33) .. controls (354.68,210.33) and (357.67,213.32) .. (357.67,217) .. controls (357.67,220.68) and (354.68,223.67) .. (351,223.67) .. controls (347.32,223.67) and (344.33,220.68) .. (344.33,217) -- cycle ;
\draw [line width=1.5]    (111,128) -- (111,160.5) ;
\draw [line width=2.25]    (410,31) -- (364.87,67.84) ;
\draw [shift={(361,71)}, rotate = 320.77] [fill={rgb, 255:red, 0; green, 0; blue, 0 }  ][line width=0.08]  [draw opacity=0] (16.07,-7.72) -- (0,0) -- (16.07,7.72) -- (10.67,0) -- cycle    ;
\draw [line width=2.25]    (419,21) -- (419,160) ;
\draw [line width=2.25]    (416.5,21) -- (439,21) ;
\draw [line width=2.25]    (416.65,160) -- (438,160) ;
\draw [line width=2.25]    (567,21) -- (567,161) ;
\draw [line width=2.25]    (548,22) -- (569.5,22) ;
\draw [line width=2.25]    (548,160) -- (569.5,160) ;
\draw   (431,31) -- (462,31) -- (462,61) -- (431,61) -- cycle ;
\draw   (462,61) -- (523,61) -- (523,121) -- (462,121) -- cycle ;
\draw   (524,121) -- (555,121) -- (555,151) -- (524,151) -- cycle ;

\draw (434,34) node [anchor=north west][inner sep=0.75pt]    {$\uparrow \uparrow $};
\draw (465.5,81.4) node [anchor=north west][inner sep=0.75pt]    {$\uparrow \downarrow ,\downarrow \uparrow $};
\draw (527.3,125.5) node [anchor=north west][inner sep=0.75pt]    {$\downarrow \downarrow $};
\end{tikzpicture}
    \caption{\label{fig:The-random-circuit} The random circuit brick-work geometry
    used in \eqnref{eq:RC}. Each circuit element is chosen to be a
    different random unitary subject to U$(1)$ ($S^{z}$) symmetry. Each
    unitary therefore has the block structure depicted in the figure,
    where the blocks are independent $1\times1,2\times2$ and $1\times1$
    Haar random unitaries respectively. }
  
    \end{figure}

  \begin{figure*}[t]
    \renewcommand{\arraystretch}{2.0}
      \begin{tabular}{|c|c|c|}
          \hline 
          Rule & Input example & Rule Output\tabularnewline
          \hline 
         \hline 
          0 & $\mathfrak{1}\otimes\mathfrak{1}$ & $\mathfrak{1}\otimes\mathfrak{1}$\tabularnewline
          \hline 
          1 & $\mathfrak{r}\otimes\mathfrak{1}$ & $\frac{1}{2}(\mathfrak{r}\otimes\mathfrak{1}+\mathfrak{1}\otimes\mathfrak{r})$\tabularnewline
          \hline 
          2 & $\mathfrak{r}\otimes\mathfrak{r}$ & $\mathfrak{r}\otimes\mathfrak{r}$\tabularnewline
          \hline 
          3 & $\mathfrak{rb}\otimes\mathfrak{1},\mathfrak{r}\otimes\mathfrak{b}$ & $\frac{1}{4}(\mathfrak{r}\otimes\mathfrak{1}+\mathfrak{1}\otimes\mathfrak{r})(\mathfrak{b}\otimes\mathfrak{1}+\mathfrak{1}\otimes\mathfrak{b})$\tabularnewline
          \hline 
          4 & $\mathfrak{m}\otimes\mathfrak{1},\mathfrak{m}\otimes\mathfrak{\mathfrak{rb}}$ & \multirow{1}{*}{$\frac{1}{4}\left[\mathfrak{m}\otimes\mathfrak{1}+\mathfrak{1}\otimes\mathfrak{m}+\mathfrak{m}\otimes\mathfrak{\mathfrak{rb}}+\mathfrak{\mathfrak{rb}}\otimes\mathfrak{m}\right]$}\tabularnewline
          \hline 
          5 & $\mathfrak{m}\otimes\mathfrak{b},\mathfrak{m}\otimes\mathfrak{r}$ & \multirow{1}{*}{$\frac{1}{4}\left[\mathfrak{m}\otimes\mathfrak{b}+\mathfrak{b}\otimes\mathfrak{m}+\mathfrak{m}\otimes\mathfrak{r}+\mathfrak{r}\otimes\mathfrak{m}\right]$}\tabularnewline
          \hline 
          6 & $\mathfrak{m}\otimes\mathfrak{m}$ & $\mathfrak{m}\otimes\mathfrak{m}$\tabularnewline
          \hline 
          7 & $\frac{1}{2}\partial\mathfrak{r}\partial\mathfrak{b},\mathfrak{m}\otimes\mathfrak{p},\mathfrak{p}\otimes\mathfrak{m}$ & \multirow{1}{*}{$\frac{1}{3}\left(\frac{1}{2}\partial\mathfrak{r}\partial\mathfrak{b}+\mathfrak{m}\otimes\mathfrak{p}+\mathfrak{p}\otimes\mathfrak{m}\right)$}\tabularnewline
          \hline 
          \end{tabular}\hspace{5mm}
          \centering
          \begin{tikzpicture}[x=0.75pt,y=0.75pt,yscale=-1,xscale=1,baseline=(current bounding box.center)]
      \draw [color={rgb, 255:red, 155; green, 155; blue, 155 }  ,draw opacity=0.33 ][fill={rgb, 255:red, 128; green, 128; blue, 128 }  ,fill opacity=1 ][line width=2.25]  [dash pattern={on 2.53pt off 3.02pt}]  (122,21) -- (199.67,21) ;
      \draw [color={rgb, 255:red, 74; green, 144; blue, 226 }  ,draw opacity=1 ][fill={rgb, 255:red, 74; green, 144; blue, 226 }  ,fill opacity=1 ][line width=1.5]    (120,60.33) -- (158,60.33) -- (199.67,60.33) ;
      \draw [color={rgb, 255:red, 208; green, 2; blue, 27 }  ,draw opacity=1 ][fill={rgb, 255:red, 74; green, 144; blue, 226 }  ,fill opacity=1 ][line width=1.5]    (120,100.33) -- (174,100.33) -- (200.33,100.33) ;
      \draw [color={rgb, 255:red, 0; green, 0; blue, 0 }  ,draw opacity=1 ][fill={rgb, 255:red, 74; green, 144; blue, 226 }  ,fill opacity=1 ][line width=1.5]    (119.33,140.33) -- (199.67,140.33) ;
      \draw [shift={(159.5,140.33)}, rotate = 180] [fill={rgb, 255:red, 0; green, 0; blue, 0 }  ,fill opacity=1 ][line width=0.08]  [draw opacity=0] (13.4,-6.43) -- (0,0) -- (13.4,6.44) -- (8.9,0) -- cycle    ;
      \draw [color={rgb, 255:red, 0; green, 0; blue, 0 }  ,draw opacity=1 ][fill={rgb, 255:red, 74; green, 144; blue, 226 }  ,fill opacity=1 ][line width=1.5]    (119.33,180.33) -- (199.67,180.33) ;
      \draw [shift={(159.5,180.33)}, rotate = 0] [fill={rgb, 255:red, 0; green, 0; blue, 0 }  ,fill opacity=1 ][line width=0.08]  [draw opacity=0] (13.4,-6.43) -- (0,0) -- (13.4,6.44) -- (8.9,0) -- cycle    ;
      
      \draw (54,52) node [anchor=north west][inner sep=0.75pt]  [font=\small] [align=left] {$\displaystyle \mathfrak{b} =z\boxtimes I$};
      \draw (54,92) node [anchor=north west][inner sep=0.75pt]  [font=\small] [align=left] {$\displaystyle \mathfrak{r} =I\boxtimes z$};
      \draw (54,12) node [anchor=north west][inner sep=0.75pt]  [font=\small] [align=left] {$\displaystyle \mathfrak{1} =I\boxtimes I$};
      \draw (34,132) node [anchor=north west][inner sep=0.75pt]  [font=\small] [align=left] {$\displaystyle \mathfrak{p} =2\sigma ^{+} \boxtimes \sigma ^{-}$};
      \draw (32,173) node [anchor=north west][inner sep=0.75pt]  [font=\small] [align=left] {$\displaystyle \mathfrak{m} =2\sigma ^{-} \boxtimes \sigma ^{+}$};
      \draw (136.67,11) node [anchor=north west][inner sep=0.75pt]   [align=left] {no line};
      
      \end{tikzpicture}
  \caption{The circuit averaged evolution is a sum of projectors onto the states in the Rule Output columns and their symmetry related analogues. The symmetry related rules (not shown for brevity) are obtained by swapping sites (across the tensor product $\otimes$) in both the input and output states. Similarly, swapping $\mathfrak{r}\leftrightarrow\mathfrak{b}$ or $\mathfrak{m}\leftrightarrow\mathfrak{p}$ in both the input and output states gives rise to new valid rules. We use shorthand notation $\partial\mathfrak{o}\equiv\mathfrak{o}\otimes\mathfrak{1}-\mathfrak{1\otimes o}$ for a discrete spatial derivative. The input example column lists examples of operators which are affected by the rule in question; the rule output column gives the contribution to the output of that particular operator due to the rule.}
  \label{tab:q1_feynman}
  \end{figure*}

In this section, we supplement our argument above with a more thorough analysis of a U$(1)$ symmetric random circuit model and use it to estimate Eq.~\ref{eq:stripped} both numerically and analytically. Along the way, we will see how the dynamics of U$(1)$ RUCs agrees with the hydrodynamical picture just explained. 
 
The random circuit model is defined as follows. The random unitary circuit has a brick-work structure (see also
 \figref{fig:The-random-circuit})
    \begin{equation}
    U(t)=\prod_{\tau=1}^{t}\Bigl[\prod_{k\text{ odd}}U_{j,j+1}^{(\tau)}\Bigr]\Bigl[\prod_{k\text{ even}}U_{k,k+1}^{(\tau)}\Bigr].\label{eq:RC}
    \end{equation}
  
    Each $U_{j,j+1}^{(\tau)}$ acts exclusively on sites $j,j+1$. Moreover,
    for each $j,\tau$, $U_{j,j+1}^{(\tau)}$ is a matrix drawn independently
    from some ensemble. We will consider the case where
    each $U_{j,j+1}^{(\tau)}$ is Haar random, subject to the constraint
    that it commutes with the global U$(1)$ symmetry $S^{z}=\sum_{j}\sigma_{j}^{z}$,
    endowing it with the block structure shown in \figref{fig:The-random-circuit}. This ensemble has
    been useful in capturing the OTOC/operator spreading \cite{Khemani_2018,vonKeyserlingk2018_diffusive}
    and entanglement growth properties of systems with continuous conservation
    laws, capturing novel behaviors that appear to be present even in the absence
    of randomness~\cite{Khemani_2018,vonKeyserlingk2018_diffusive,Rakovszky_2019}. In this specific random circuit model, $\sigma_{j}^{z}=z_j$  plays the role of our conserved density, thus we use the two interchangeably in our discussions of numerical results. 
  
  \subsection{Circuit-averaged dynamics}

  We now turn to the task of estimating backflow. As was shown in Refs. \onlinecite{Khemani_2018,vonKeyserlingk2018_diffusive}, {\it on average} there is no backflow in the random circuit: $C_{>\ell_*}$ averages to zero. This is due to a cancellation between different circuit realizations. Instead, we want to understand the size of backflow contributions in {\it typical} realizations. To this end, we examine $|C_{>\ell_*}(\tau,x)|^2$, which can be recast as
  \begin{equation}
      \label{eq:RUCstrippedsq}
      |C_{>\ell_*}(2 t,x)|^2=\langle o_0^{\boxtimes 2} | \mathcal{U}_2(2t ,t) P^{\boxtimes 2}_{>\ell_*} \mathcal{U}_2(t,0) | o^{\boxtimes 2}_x \rangle,
  \end{equation}
  where we use compact notation $a^{\boxtimes 2} = a\boxtimes a^\dagger$. The $\boxtimes$ denotes a tensor product between the two replica copies needed to calculate the squared quantity, while the $\otimes$ denotes a tensor product between different sites. Here
  $\mathcal{U}_2=\mathcal{U}\boxtimes\mathcal{U}$, and recall that $\mathcal{U}$ represents the adjoint action of the unitary on an operator. Recall that our circuits consist of two-site gates chosen independently for each position and time-step. Since all gates in the circuit are chosen independently, we can perform averages over them separately. After averaging, we are left with an effective Markovian evolution, which is generated by a circuit averaged super-operator $\overline{\mathcal{U}_2}$. This object is a large ($256\times 256$) matrix; it is a linear map on tensor product pairs of operators. However, it is a projector and has a small (13-dimensional) support spanned by the states in the third column of \figref{tab:q1_feynman} and their symmetry related partners (and $\mathfrak{1}\otimes\mathfrak{1}$), and thus admits a relatively simple description.

  In particular,  while the on-site Hilbert space in this replica-doubled problem is $4^2=16$ dimensional, states in the support of $\overline{\mathcal{U}_2}$ can be entirely described by the following $6$ operator pairs 
  \begin{align*}
      \mathfrak{1} &=I\boxtimes I &\mathfrak{b} &=z\boxtimes I\\
      \mathfrak{r}&=I\boxtimes z &\mathfrak{rb}&=z\boxtimes z\\
      \mathfrak{p}&=2\sigma^+\boxtimes\sigma^-  &\mathfrak{m} &=2\sigma^-\boxtimes\sigma^+.
  \end{align*}
  Note that operators of form $\sigma^{+}\boxtimes \sigma^{+},\sigma^{-}\boxtimes \sigma^{-}$ as well as $\sigma^{\pm}\boxtimes I,\sigma^{\pm}\boxtimes z$ (and their two additional replica swapped versions) do not appear here: they are annihilated by the evolution operator. The intuitive explanation for this result is that operator pairs that locally change charge are associated with rapidly fluctuating phases whose contributions tend to vanish under averaging.  This sort of phenomenon is well-known in ergodic non-random systems where, for example, particle propagators develop a finite lifetime (e.g., exponential decay of $\langle b^\dagger (\tau,x) b(0,0)   \rangle$ with time in a high temperature Bose-Hubbard model). This idea was used above to arrive at \eqnref{eq:u1dephasing} and consequently to argue for exponential suppression of backflow.

  The transition amplitudes induced by a single two-site averaged gate $\overline{\mathcal{U}_2}$ have been calculated previously \cite{Khemani_2018,vonKeyserlingk2018_diffusive} and are summarized in \figref{tab:q1_feynman}. The evolution is simply a sum of projectors onto the states in the rule output column and their symmetry related states, which can be obtained by globally swapping $x \leftrightarrow x+1$ or $\mathfrak{p}\leftrightarrow \mathfrak{m}$ or $\mathfrak{r}\leftrightarrow \mathfrak{b}$. These three symmetries reflect the inversion symmetry of the random ensemble, and (for the final two rules) the fact $\mathcal{U}_2$ is invariant under swapping replicas. 

Let us list some other important observations concerning the rules 0-6:

  \begin{itemize}
      \item Isolated $\mathfrak{r}$'s undergo diffusion (rule 1); but they cannot diffuse past one another (rule 2); in other words they undergo single file diffusion. Similarly for $\mathfrak{b}$'s
      \item  Except when they meet, $\mathfrak{p},\mathfrak{m}$  tend to random walk (rules 4,5 and their symmetry related versions). 
      \item As $\mathfrak{p},\mathfrak{m}$  random walk, they are equally likely as not to emit $\mathfrak{rb}$ pairs.
      
      \item Both of $\mathfrak{p},\mathfrak{m}$ can absorb $\mathfrak{rb}$ pairs or inter-convert $\mathfrak{r} \leftrightarrow\mathfrak{b}$. $\mathfrak{p},\mathfrak{m}$ play a crucial role in processes that change the total number of operators: they constantly attempt to produce $\mathfrak{rb}$ pairs, and they are the only operators that can absorb other operators. 
  \end{itemize}

  It is useful to note that rules 0-6 define a probability conserving classical stochastic process, on a state space where each site takes one of six values. Only rule 7 breaks this pattern, because some of its transition amplitudes are negative. Rule 7  dictates what happens when $\mathfrak{p},\mathfrak{m}$ meet, and determines (in combination with rule 3) what happens when $\mathfrak{r},\mathfrak{b}$ meet:

  \begin{subequations}
    \label{eq:all}
  \begin{align}
      \mathfrak{rb}\otimes\mathfrak{1} & \rightarrow\frac{1}{6}(2\mathfrak{rb}\otimes\mathfrak{1}+2\mathfrak{1}\otimes\mathfrak{rb}+\mathfrak{r\otimes b}+\mathfrak{b\otimes r})\nonumber\\
      & +\frac{1}{6}(\mathfrak{p}\otimes\mathfrak{m}+\mathfrak{m}\otimes\mathfrak{p})\label{rbrule1}\\
      \mathfrak{r}\otimes\mathfrak{b} & \rightarrow\frac{1}{6}(\mathfrak{rb}\otimes\mathfrak{1}+\mathfrak{1}\otimes\mathfrak{rb}+2\mathfrak{r\otimes b}+2\mathfrak{b\otimes r}\nonumber)\\
      & -\frac{1}{6}(\mathfrak{p}\otimes\mathfrak{m}+\mathfrak{m}\otimes\mathfrak{p})\label{rbrule2}\\      
      \mathfrak{p}\otimes\mathfrak{m} & \rightarrow\frac{1}{6}(2\mathfrak{p}\otimes\mathfrak{m}+2\mathfrak{m}\otimes\mathfrak{p}+\mathfrak{rb\otimes1}+\mathfrak{1\otimes rb})\nonumber\\
      & -\frac{1}{6}(\mathfrak{r}\otimes\mathfrak{b}+\mathfrak{b}\otimes\mathfrak{r})\label{pmrule}\nonumber\\ \end{align}
    \end{subequations}
      Note that the right hand sides of \eqnref{rbrule1} and \eqnref{rbrule2}  involve a sum of contributions from two of the rules (3 and 7) in \figref{tab:q1_feynman}, because the left hand sides have overlap with the corresponding two states in the output column.
      
      Due to the non-probability conserving nature of rule 7, Eq.~\ref{eq:all} have some negative amplitudes and the weights on the RHS of the equations need not sum up to $1$ as they do for rules 0-6. Nevertheless, these rules show that $\mathfrak{rb}$ pairs can fuse to form a $\mathfrak{pm}$ pair. Similarly, when $\mathfrak{m},\mathfrak{p}$ meet, they have some chance of forming an $\mathfrak{rb}$ pair. 

      We can now summarize and synthesize these observations together into a simpler story. For each time step that an $\mathfrak{m},\mathfrak{p}$ are present in an operator, that operator will tend to grow in support because $\mathfrak{m},\mathfrak{p}$ constantly tend to emit $\mathfrak{rb}$ pairs. This process is difficult to reverse because the $\mathfrak{r},\mathfrak{b}$ produced can only be absorbed by $\mathfrak{m}/\mathfrak{p}$, but are most likely to diffuse away from their parent $\mathfrak{m}/\mathfrak{p}$. Thus $\mathfrak{m}/\mathfrak{p}$ tend to lead to further operator growth. This manifests itself in a finite lifetime/exponentially decaying in time return probability for a $\mathfrak{mp}$ pair; we verify this story in more detail later (\secref{sec:self-energy}). 

      
    \subsection{Numerical results}
    
    Using the update rules summarized in \figref{tab:q1_feynman} for every gate in the circuit we arrive at a process that is formally analogous to a classical stochastic many-body evolution, albeit with some negative transition `probabilities'. One can then use this representation to evaluate quantities like the average squared correlator $\overline{|C(\tau,x)|^2}$ numerically. For $o=z$ this involves taking an initial configuration with $\mathfrak{rb}$ on site $0$ and $\mathfrak{1}$ everywhere else, evolving under the rules of \figref{tab:q1_feynman}, and then considering the `probability' (overlap) associated to a configuration with a single $\mathfrak{rb}$ on site $x$. Following previous works~\cite{vonKeyserlingk2018_diffusive,Rakovszky_2019}, one can do this in a tensor network language, representing the instantaneous distribution as a MPS and updating it using TEBD; it was found previously that the effective bond dimension in this problem does not grow very rapidly, allowing one to get reliable results for times and system sizes well beyond the reach of naive exact evolution. 
    
      \begin{figure}
    \includegraphics[width=1.0\columnwidth]{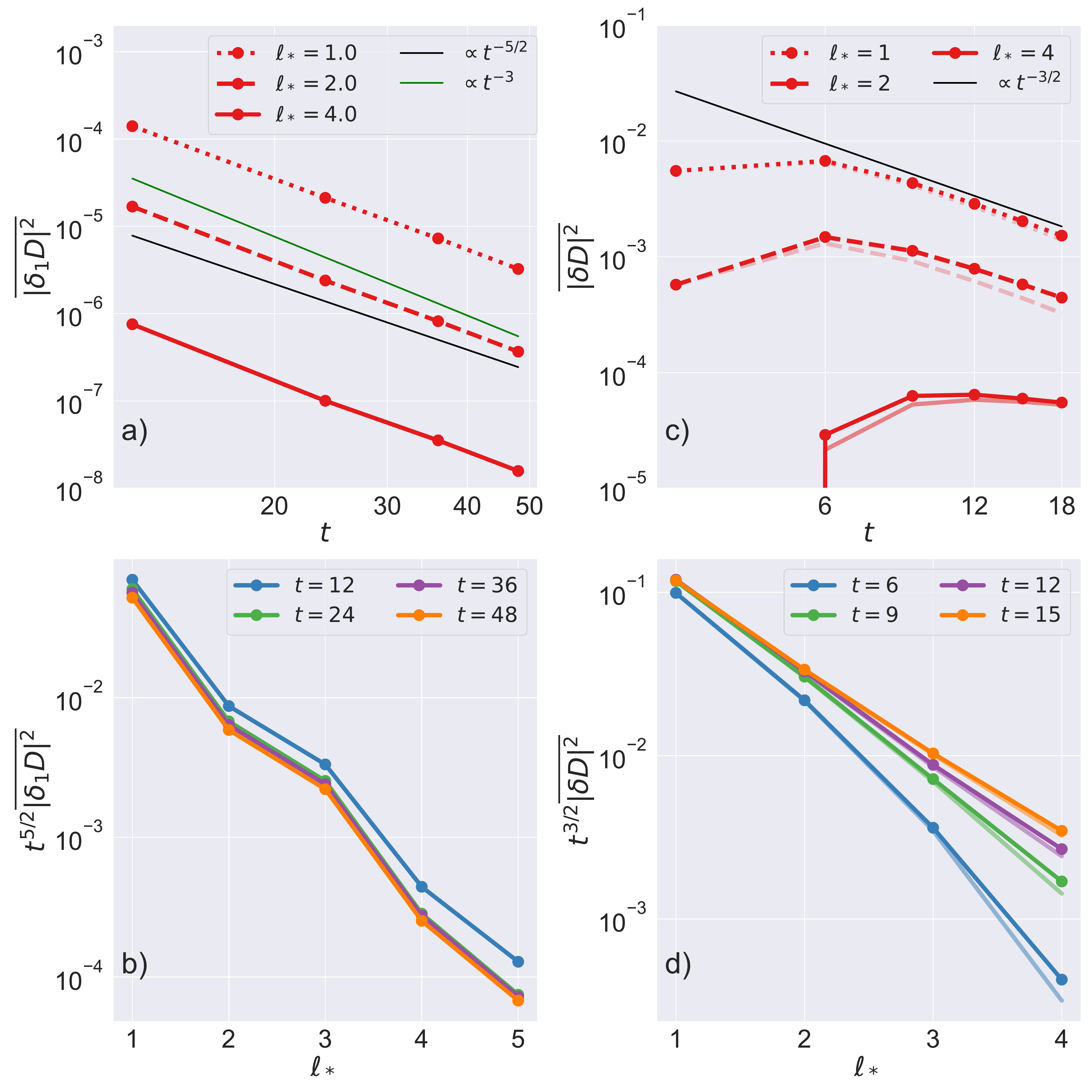}
    \caption{All simulations are carried out for system size $L=50$, except for the $\ell_*=1$ series in (a), where $L=80$ is needed to go to sufficiently long times to resolve the power law decay. All data shown appear to be converged in system size. (a,b) Show the effect on the diffusion constant due to a single application of DAOE with $\gamma=\infty$ at time $t=\tau/2$. The data are well converged with bond dimension ($\chi=400,768$ are both plotted). (c,d) Shows the error in the diffusion constant induced by acting with DAOE after every circuit layer. We achieve reasonable convergence in bond dimension: $\chi=100,300,500$ are plotted in increasingly bold hues.\label{fig:total}}
  \end{figure}
  
  One can easily extend this calculation to the backflow contribution, $\overline{|C_{>\ell_*}(\tau,x)|^2}$. In order to do so, one only has to insert $\mathcal{P}_{>\ell_*}^{\boxtimes 2}$ at time $\tau/2$ into the evolution. This allows us to evaluate the size of backflow corrections in the random circuit more reliably than we were able in the deterministic spin chains studied in \secref{ssec:IsingLadder}. This is further helped by the fact that the randomness of the circuit makes diffusive behavior set in immediately, while for a fixed Hamiltonian there is some short local thermalization time before which hydrodynamics does not apply. 
  
   In \figref{fig:total}a,b we plot the U$(1)$ circuit average of $|\delta_1 D_{>\ell_*}(t,x)|^2$, i.e., the error suffered by the diffusion constant due to a single dissipation event, as defined in \eqnref{eq:d_1 D}. In the top panel, we plot this quantity against time for different $\ell_*$. The data are consistent with a power law with decay exponent close to our prediction of $\tau^{-5/2}$; the fit is not decisive, although it can be much improved by allowing for a sum of power laws ($\tau^{-5/2}$ and $\tau^{-7/2}$) which can be easily accounted for by our hydrodynamical formalism. An alternative hypothesis, which the numerical simulations do not rule out but which is inconsistent with theory, is that there is power law decay with an exponent between $\tau^{-5/2}$ and $\tau^{-3}$. In the bottom panel (b), we plot the same quantity as a function of $\ell_*$ for various values of $t$, rescaling all results by  $t^{5/2}$. The results appear consistent with exponential decay in $\ell_*$.
   
   In \figref{fig:total}c,d we plot the average value of $|\delta D|^2$, the error in the diffusion constant when the dissipation $\mathcal{P}_{>\ell_*}$ is applied after every two layers of the circuit. We again find that the error decays algebraically in time. Here the fit is of a better quality and we extract an exponent of decay from the $\ell_*=1$ data that is $\tau^{-1.55(8)}$, in good agreement with our prediction of $\tau^{-3/2}$. Panel (d) also shows evidence of exponential decay in $\ell_*$, although we were not able to test larger $\ell_*$ in this case for want of computational resources. 

  \subsection{Consistency with hydrodynamical picture}\label{ssec:feynman_to_hydro}
  The update rules above also allow us to estimate some of the backflow contributions analytically. But first, to develop some intuition, we will show how to recover from them the hydrodynamical picture presented in \secref{ssec:hydro_sans_trunc}.

  Let us examine the circuit averaged behavior of 
  \begin{equation}\label{eq:C^2}
      |C(\tau,x)|^2 = \langle o_0^{\boxtimes 2} | \mathcal{U}_2(\tau,0) | o^{\boxtimes 2}_x \rangle,
  \end{equation}
a correlator in the absence of any dissipation events.
  Take $o_{0,x}$ to be local strings of $1,z,\sigma^{\pm}$ that are charge neutral (i.e., each has the same number of $\sigma^+$ and $\sigma^{-}$). Consider the limit of very large $\tau$ (in comparison to the lengths of $o_{0,x}$). In this limit,  $|C(\tau,x)|^2$ represents the amplitude for a local operator $o_0$ to evolve to another local operator $o_x$ over a long period of time. 

  We have already mentioned  $\mathfrak{p},\mathfrak{m}$ tend to lead to further operator growth. On the other hand, $\mathfrak{r,b}$  undergo single-file diffusion, which is a slow/hydrodynamical process. Thus, we expect the leading order behavior of \eqnref{eq:C^2} in $\tau$ to involve the propagation of these slow $\mathfrak{r,b}$ operators. 

  \begin{equation}\label{eq:C^2_figure}
      \begin{tikzpicture}[x=0.55pt,y=0.75pt,yscale=-1,xscale=1]
          \draw   (75,44.25) -- (32.5,79.25) -- (32.5,9.25) -- cycle ;
          \draw  [fill={rgb, 255:red, 0; green, 0; blue, 0 }  ,fill opacity=0.45 ] (75,44.25) -- (32.5,79.25) -- (32.5,9.25) -- cycle ;
          
          \draw   (378.5,60.83) -- (421,25.83) -- (421,95.83) -- cycle ;
          \draw  [fill={rgb, 255:red, 0; green, 0; blue, 0 }  ,fill opacity=0.45 ] (378.5,60.83) -- (421,25.83) -- (421,95.83) -- cycle ;
          
          \draw [color={rgb, 255:red, 74; green, 144; blue, 226 }  ,draw opacity=1 ][line width=0.75]    (74,44.25) .. controls (114,14.25) and (134,74.25) .. (174,44.25) ;
          \draw [color={rgb, 255:red, 74; green, 144; blue, 226 }  ,draw opacity=1 ][line width=0.75]    (174,44.25) .. controls (214,14.25) and (232,88) .. (272,58) ;
          \draw [color={rgb, 255:red, 74; green, 144; blue, 226 }  ,draw opacity=1 ][line width=0.75]    (272,58) .. controls (312,28) and (337.5,90.83) .. (377.5,60.83) ;
          \draw [color={rgb, 255:red, 208; green, 2; blue, 27 }  ,draw opacity=1 ][line width=0.75]    (378.5,60.28) .. controls (338.5,90.28) and (292,0) .. (252,30) ;
          \draw [color={rgb, 255:red, 208; green, 2; blue, 27 }  ,draw opacity=1 ][line width=0.75]    (252,30) .. controls (212,60) and (218.5,30.28) .. (178.5,60.28) ;
          \draw [color={rgb, 255:red, 208; green, 2; blue, 27 }  ,draw opacity=1 ][line width=0.75]    (178.5,60.28) .. controls (138.5,90.28) and (115,13.7) .. (75,43.7) ;
          
          \draw (2,66.5) node [anchor=north west][inner sep=0.75pt]  [rotate=-270] [align=left] {$\displaystyle o_{0} \boxtimes o_{o}^{\dagger }$};
          \draw (423,85.5) node [anchor=north west][inner sep=0.75pt]  [rotate=-270] [align=left] {$\displaystyle o_{x} \boxtimes o_{x}^{\dagger }$};        
          \end{tikzpicture}
  \end{equation}

  The grey triangles in \eqnref{eq:C^2_figure} represent processes whereby the initial/final operators $o$ contract into $\mathfrak{r},\mathfrak{b}$. Unless the $o$ is already a single $z$, said processes necessarily involve the propagation and annihilation of $\mathfrak{pm}$ pairs and so are short-lived (in comparison to $O(\tau)$). For example, if we choose $o_0 = z_{y_1} z_{y_2} z_{y_3} z_{y_4} $, a $\mathfrak{pm}$ will need to be generated (using rule 7) in order to on net remove three of the $\fr \fb$ pairs (using rule 4) 
  
  \begin{equation}
\begin{tikzpicture}[x=0.75pt,y=0.75pt,yscale=-1,xscale=1]
\draw [color={rgb, 255:red, 74; green, 144; blue, 226 }  ,draw opacity=1 ][line width=0.75]    (159.5,54) .. controls (173.05,57.53) and (186.05,56.17) .. (197,58.17) .. controls (207.95,60.17) and (228.04,67.08) .. (244.33,69.83) ;
\draw [color={rgb, 255:red, 208; green, 2; blue, 27 }  ,draw opacity=1 ][line width=0.75]    (244.33,69.83) .. controls (220.67,68.62) and (173.67,70.01) .. (159.5,54) ;
\draw [color={rgb, 255:red, 74; green, 144; blue, 226 }  ,draw opacity=1 ][line width=0.75]    (157,119.83) .. controls (170.92,118) and (185.67,108.83) .. (199.67,113.51) .. controls (213.67,118.2) and (271.14,93.93) .. (290,82.75) ;
\draw [color={rgb, 255:red, 208; green, 2; blue, 27 }  ,draw opacity=1 ][line width=0.75]    (290,82.75) .. controls (281.5,96.75) and (197,146.58) .. (157,119.83) ;
\draw    (225.5,83.25) .. controls (233.33,77.46) and (259.58,60.01) .. (276.93,54.3) .. controls (294.29,48.59) and (311.64,47.03) .. (325.67,55.83) ;
\draw [shift={(250.59,67.01)}, rotate = 329.22] [fill={rgb, 255:red, 0; green, 0; blue, 0 }  ][line width=0.08]  [draw opacity=0] (10.72,-5.15) -- (0,0) -- (10.72,5.15) -- (7.12,0) -- cycle    ;
\draw [shift={(301.8,49.63)}, rotate = 356.4] [fill={rgb, 255:red, 0; green, 0; blue, 0 }  ][line width=0.08]  [draw opacity=0] (10.72,-5.15) -- (0,0) -- (10.72,5.15) -- (7.12,0) -- cycle    ;
\draw    (225.5,83.25) .. controls (232.83,96.92) and (241,96.25) .. (290,82.75) ;
\draw [shift={(256.27,91.31)}, rotate = 529.65] [fill={rgb, 255:red, 0; green, 0; blue, 0 }  ][line width=0.08]  [draw opacity=0] (10.72,-5.15) -- (0,0) -- (10.72,5.15) -- (7.12,0) -- cycle    ;
\draw    (290,82.75) .. controls (301.58,80.4) and (321.67,62.17) .. (325.67,55.83) ;
\draw [shift={(309.62,71.63)}, rotate = 503.15] [fill={rgb, 255:red, 0; green, 0; blue, 0 }  ][line width=0.08]  [draw opacity=0] (10.72,-5.15) -- (0,0) -- (10.72,5.15) -- (7.12,0) -- cycle    ;
\draw [color={rgb, 255:red, 74; green, 144; blue, 226 }  ,draw opacity=1 ][line width=0.75]    (325.17,56.17) .. controls (336.5,51.07) and (336.5,56.27) .. (358.5,59.07) ;
\draw [color={rgb, 255:red, 208; green, 2; blue, 27 }  ,draw opacity=1 ][line width=0.75]    (358.1,63.47) .. controls (334.43,61.13) and (346.1,65.07) .. (325.17,56.17) ;
\draw [color={rgb, 255:red, 208; green, 2; blue, 27 }  ,draw opacity=1 ][line width=0.75]    (276.93,54.3) .. controls (264.93,53.68) and (245.11,41.85) .. (220,38.5) .. controls (195.59,35.24) and (165.82,40.56) .. (158.83,32.67) ;
\draw [color={rgb, 255:red, 74; green, 144; blue, 226 }  ,draw opacity=1 ][line width=0.75]    (156,89.83) .. controls (196,59.83) and (187.5,120.25) .. (225.5,83.25) ;
\draw [color={rgb, 255:red, 208; green, 2; blue, 27 }  ,draw opacity=1 ][line width=0.75]    (225.5,83.25) .. controls (185.5,80.25) and (196,116.58) .. (156,89.83) ;
\draw [color={rgb, 255:red, 74; green, 144; blue, 226 }  ,draw opacity=1 ][line width=0.75]    (158.83,32.67) .. controls (172.39,36.19) and (186,28.5) .. (196.33,36.83) .. controls (206.67,45.17) and (260.64,51.54) .. (276.93,54.3) ;

\draw (141.83,110.33) node [anchor=north west][inner sep=0.75pt]   [align=left] {$\displaystyle y_{1}$};
\draw (140.83,44.67) node [anchor=north west][inner sep=0.75pt]   [align=left] {$\displaystyle y_{3}$};
\draw (140.5,16.67) node [anchor=north west][inner sep=0.75pt]   [align=left] {$\displaystyle y_{4}$};
\draw (140.83,80.33) node [anchor=north west][inner sep=0.75pt]   [align=left] {$\displaystyle y_{2}$};
\draw (321.83,64.33) node [anchor=north west][inner sep=0.75pt]  [font=\small] [align=left] {$\displaystyle 7$};
\draw (219.83,87.33) node [anchor=north west][inner sep=0.75pt]  [font=\small] [align=left] {$\displaystyle 7$};
\draw (274.33,38.83) node [anchor=north west][inner sep=0.75pt]  [font=\small] [align=left] {$\displaystyle 4$};
\draw (244.83,68.83) node [anchor=north west][inner sep=0.75pt]  [font=\small] [align=left] {$\displaystyle 4$};
\draw (282.33,65.33) node [anchor=north west][inner sep=0.75pt]  [font=\small] [align=left] {4};
\end{tikzpicture}
  \end{equation}
  On the other hand if we choose $o_0 = \sigma^+_0 \sigma^-_y $ then the operator begins with a $\mathfrak{p},\mathfrak{m}$ pair; these will annihilate quickly (compared to $O(\tau)$) through rule 7:
  \begin{equation}
      \begin{tikzpicture}[x=0.75pt,y=0.75pt,yscale=-1,xscale=1]
      
      \draw    (159.33,42.83) .. controls (181,68) and (197.33,60) .. (226,78) ;
      \draw [shift={(190.98,63.41)}, rotate = 19.04] [fill={rgb, 255:red, 0; green, 0; blue, 0 }  ][line width=0.08]  [draw opacity=0] (10.72,-5.15) -- (0,0) -- (10.72,5.15) -- (7.12,0) -- cycle    ;
      \draw    (160.33,100.83) .. controls (172.33,83.33) and (207.58,69.31) .. (226,78) ;
      \draw [shift={(190.38,79.74)}, rotate = 517.53] [fill={rgb, 255:red, 0; green, 0; blue, 0 }  ][line width=0.08]  [draw opacity=0] (10.72,-5.15) -- (0,0) -- (10.72,5.15) -- (7.12,0) -- cycle    ;
      \draw [color={rgb, 255:red, 74; green, 144; blue, 226 }  ,draw opacity=1 ][line width=0.75]    (224.67,77.5) .. controls (236,72.4) and (236,77.6) .. (258,80.4) ;
      \draw [color={rgb, 255:red, 208; green, 2; blue, 27 }  ,draw opacity=1 ][line width=0.75]    (257.6,84.8) .. controls (233.93,82.47) and (245.6,86.4) .. (224.67,77.5) ;
      
      \draw (145.83,90.33) node [anchor=north west][inner sep=0.75pt]   [align=left] {$\displaystyle 0$};
      \draw (146.83,29.33) node [anchor=north west][inner sep=0.75pt]   [align=left] {$\displaystyle y$};
      \draw (220.83,54.33) node [anchor=north west][inner sep=0.75pt]   [align=left] {$\displaystyle 7$};
      
      \end{tikzpicture}    
  \end{equation}

  These examples illustrate a general point; unless $o_x,o_0$ have overlap with $S^z$, one needs to use rule 7 in order to reduce the operator down to an $\mathfrak{rb}$ pair  \footnote{We ignore the special situation where $o_{0,x}$ have overlap with $(S^z)^{2,3}$, in which case it may not be favorable to reduce the operator to an $\mathfrak{rb}$ pair.}. However, note that rule 7 produces $\mathfrak{rb}$ in the combination $\partial\mathfrak{r}\partial\mathfrak{b}$, which includes two spatial derivatives. This is consistent with the finding in \secref{ssec:hydro_sans_trunc}, namely that unless $o_{0,x}$ overlap with  $(S^z)^{1,2,3}$, then the slowest variable with which they can develop overlap is $\partial z$.



  \subsection{$\mathfrak{pm}$ pairs are short-lived}\label{sec:self-energy}
\begin{figure}
  \includegraphics[width=\columnwidth]{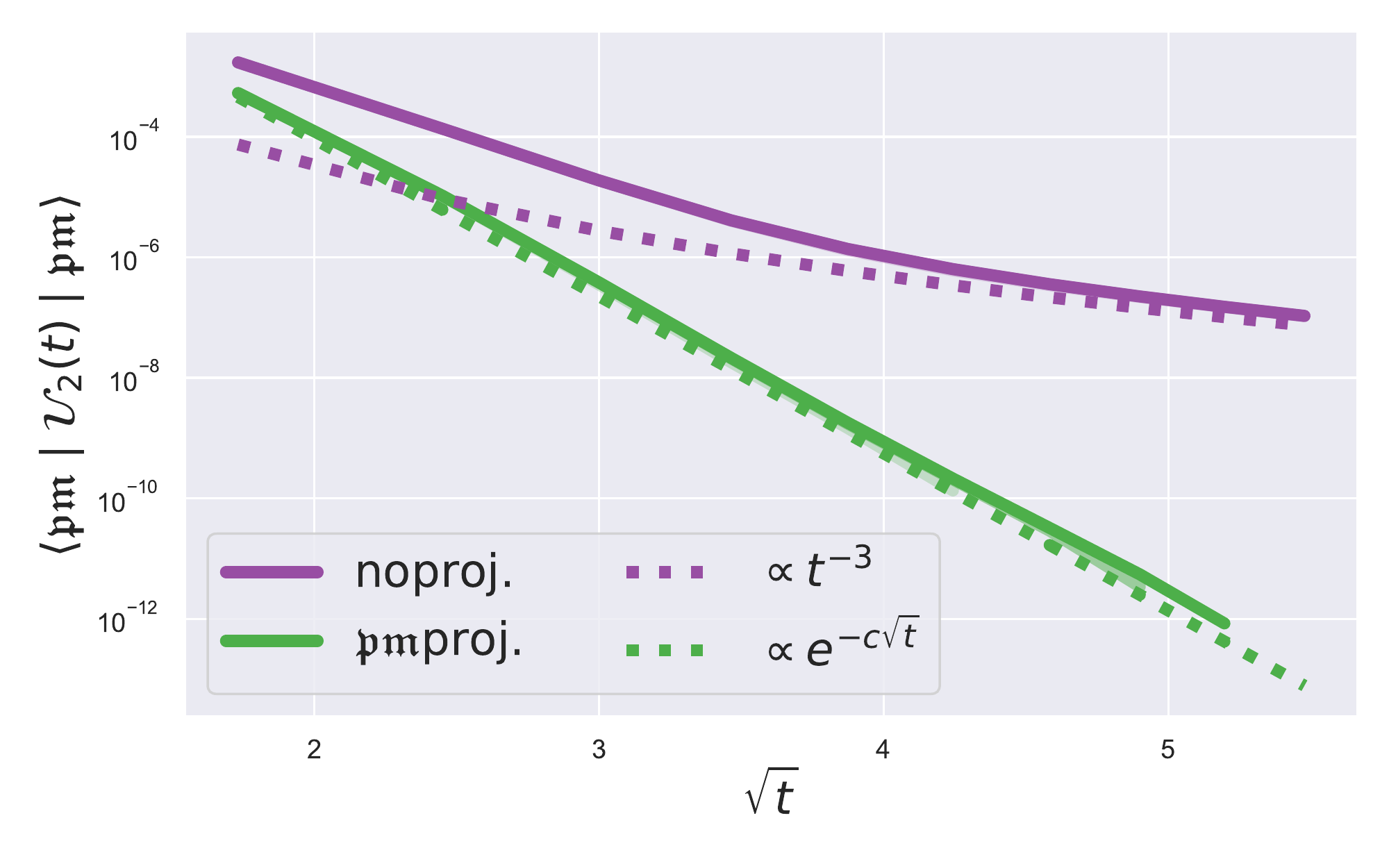}
  \caption{This figure shows a numerical simulation of the return amplitude \eqnref{eq:pm_prop}, in both the case where the $\fp,\fm$ pairs are prevented from recombining (`projected evolution' with $\overline{\mathcal{U}}_{2,\mathrm{f}}$) and under the usual circuit averaged evolution using $\overline{\mathcal{U}}_2$ (`unprojected evolution'). We utilize a TEBD algorithm to implement the circuit averaged dynamics. We use system size  $L=50$ in this plot, and runs with $\chi=100,200,300$ are shown in increasing order of opacity. \label{fig:pm_lifetime}}
  \end{figure}

    In this section we argue that the survival probability of a pair of $\fp,\fm$'s decays rapidly with time. Concretely, we conjecture that the amplitude
    \begin{equation}\label{eq:pm_prop}
        \langle \fp_0 \fm_1| \overline{\mathcal{U}_{2,\mathrm{f}}(t,0)} | \fp_0 \fm_1 \rangle
    \end{equation}
    decays faster than any power law, and we provide an argument for a stretched exponential decay $e^{-\mathcal{O}(\sqrt{t})}$. Our numerical simulation of the circuit average of \eqnref{eq:pm_prop} (see \figref{fig:pm_lifetime}) is consistent with this. The object  $\overline{\mathcal{U}_{2,\mathrm{f}}(t,0)}$ is the same as $ \overline{\mathcal{U}_{2}}$ except we set to zero all matrix elements which would allow a $\fp,\fm$ pair to convert into $\partial \fr \partial \fb$; this is a modification to rule 7. Thus \eqnref{eq:pm_prop} represents the amplitude that the $\langle \fp_0 \fm_1|$ pair (initially on sites $0,1$) persist for time $t$, and propagate back to their initial positions. 

    It is worth reminding the reader how \eqnref{eq:pm_prop} behaves if we do not insist the $\fp\fm$ pair persist. In this case, we are examining
    \begin{equation}\label{eq:pm_prop_noproj}
        \langle \fp_0 \fm_1| \overline{\mathcal{U}_{2}(t,0)} | \fp_0 \fm_1 \rangle,
    \end{equation}
     which decays as $t^{-3}$. That is because \eqnref{eq:pm_prop_noproj} is just the square of $\langle\sigma^+_0 \sigma^-_1 (t)|\sigma^+_0 \sigma^-_1 (t)\rangle$, which decays as $t^{-3/2}$ according to the hydrodynamical recipe in \secref{ssec:hydro_sans_trunc}. In the related construction \secref{ssec:feynman_to_hydro}, the $t^{-3}$ power law comes from processes were the $\fp,\fm$ pair combine to $\partial \fr \partial \fb$, which then propagate diffusively. A circuit averaged simulation of \eqnref{eq:pm_prop_noproj} (\figref{fig:pm_lifetime}) is consistent with the predicted power law.
  
 We now give a theoretical picture explaining why \eqnref{eq:pm_prop} decays as a stretched exponential.  As $\overline{\mathcal{U}_{2,\mathrm{f}}}$ evolves the $\fp,\fm$ pair, they will diffuse, and in the process source $\fr \fb$ pairs (according to rule 3). The number of $\fr, \fb$ that can be created is diffusion-limited because the $\fr, \fb$ diffuse, and at most one $\fr$ and at most one $\fb$ an occupy a given site. As a result, by time $t$ we expect $\langle \fp_0 \fm_1 | \overline{\mathcal{U}_{2,\mathrm{f}}}$ to have $\mathcal{O}(\sqrt{t})$ number of $\fr, \fb$, distributed in a region of size $\mathcal{O}(\sqrt{t})$ centered on the origin. The equal probability of  removing versus adding a pair of $\fr, \fb$ from a $\fm$ or $\fp$ (rule 3) suggests that the density $\nu$ of $\fr,\fb$ in this diffusive cone approaches a steady state limit $\nu\sim 1/2$\footnote{ Our argument only requires that the density approaches a limit strictly between $0< \nu < 1$.}. We will assume that, to a good approximation, each site within the diffusive cone has independent probability $\nu$ of being occupied by $\fr$ and $\fb$.

We are studying the overlap $\langle \fp_0 \fm_1 | \overline{\mathcal{U}_{2,\mathrm{f}}} |  \fp_0 \fm_1 \rangle$ which represents the rare event where at the last time step no $\fr,\fb$ are present. The probability that no $\fr,\fb$ are present are, from the reasoning in the previous paragraph, expected to scale as $\nu^{\mathcal{O}(\sqrt{t})}$, which is the stretched exponential result we wished to prove.  Note we have thus far ignored the requirement that the $\fp,\fm$ pair return to the origin. This will further suppress the amplitude by an unimportant power law in $t$.

Perhaps the least controlled assumption in the argument above is that we have ignored processes where $\fr\fb$ can fuse and form $\fp\fm$ pairs (additional to the original pair). We cannot rigorously argue that such processes are irrelevant for obtaining the stretched exponential, however it is at least self-consistent to ignore such processes: they give rise to $\fp\fm$ pairs which we are arguing (and have demonstrated numerically) are short-lived. As an aside, we note that the argument for stretched-exponential decay is almost identical to the one presented in our previous work \cite{Rakovszky_2019} showing that higher R\'enyi entropies grow diffusively, a result that can be proved independently under mild assumptions~\cite{Huang2020}. 

\subsection{A subset of backflow corrections}

  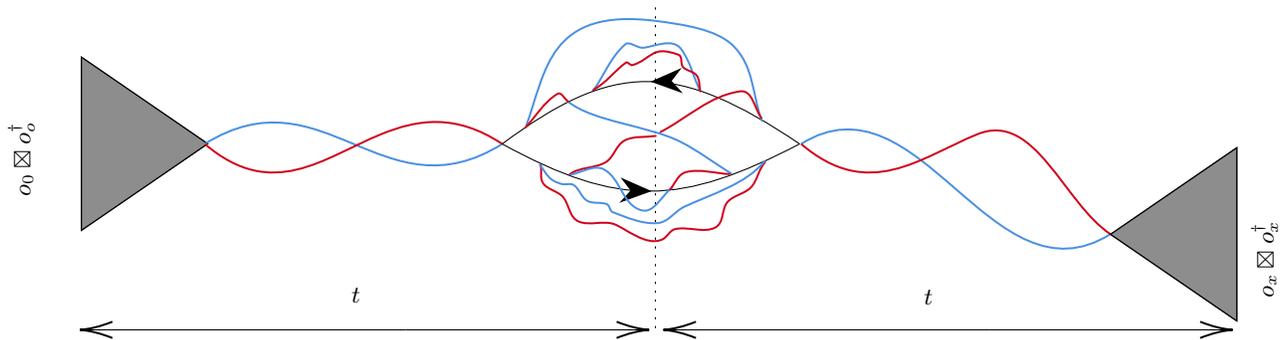
\begin{figure*}[t] 
   \begin{tikzpicture}[x=0.75pt,y=0.75pt,yscale=-1.25,xscale=1.5]
      
      \draw   (84,55.25) -- (41.5,90.25) -- (41.5,20.25) -- cycle ;
      \draw  [fill={rgb, 255:red, 0; green, 0; blue, 0 }  ,fill opacity=0.45 ] (84,55.25) -- (41.5,90.25) -- (41.5,20.25) -- cycle ;
      
      \draw [color={rgb, 255:red, 74; green, 144; blue, 226 }  ,draw opacity=1 ][line width=0.75]    (83,55.33) .. controls (123,25.33) and (143,85.33) .. (183,55.33) ;
      \draw [color={rgb, 255:red, 208; green, 2; blue, 27 }  ,draw opacity=1 ][line width=0.75]    (83,55.33) .. controls (105.82,78.72) and (122.93,60.87) .. (142.31,51.39) .. controls (154.36,45.5) and (167.28,42.84) .. (183,55.33) ;
      \draw    (183,55.33) .. controls (220.5,21.75) and (241.5,21.75) .. (283,55.33) ;
      \draw [shift={(232.86,30.17)}, rotate = 357.77] [fill={rgb, 255:red, 0; green, 0; blue, 0 }  ][line width=0.08]  [draw opacity=0] (10.72,-5.15) -- (0,0) -- (10.72,5.15) -- (7.12,0) -- cycle    ;
      \draw    (183,55.33) .. controls (222.5,81.25) and (243,80.25) .. (283,55.33) ;
      \draw [shift={(233.26,74.39)}, rotate = 182.19] [fill={rgb, 255:red, 0; green, 0; blue, 0 }  ][line width=0.08]  [draw opacity=0] (10.72,-5.15) -- (0,0) -- (10.72,5.15) -- (7.12,0) -- cycle    ;
      \draw [color={rgb, 255:red, 74; green, 144; blue, 226 }  ,draw opacity=1 ][line width=0.75]    (283.5,55.33) .. controls (323.5,25.33) and (347.5,121.75) .. (387.5,91.75) ;
      \draw [color={rgb, 255:red, 208; green, 2; blue, 27 }  ,draw opacity=1 ][line width=0.75]    (283.5,55.33) .. controls (306.32,78.72) and (323.43,60.87) .. (342.81,51.39) .. controls (362.2,41.91) and (371.78,79.26) .. (387.5,91.75) ;
      \draw   (387.5,91.83) -- (430,56.83) -- (430,126.83) -- cycle ;
      \draw  [fill={rgb, 255:red, 0; green, 0; blue, 0 }  ,fill opacity=0.45 ] (387.5,91.83) -- (430,56.83) -- (430,126.83) -- cycle ;
      
      \draw  [dash pattern={on 0.84pt off 2.51pt}]  (234.5,0.2) -- (234.5,131.4) ;
      \draw  [color={rgb, 255:red, 74; green, 144; blue, 226 }  ,draw opacity=1 ][line width=0.75] [line join = round][line cap = round] (191.07,48) .. controls (197.7,26.07) and (198.5,0.06) .. (234.8,5.69) .. controls (243.99,7.12) and (254.81,9.05) .. (260.37,15.37) .. controls (267.87,23.88) and (267,35.81) .. (270.32,46.03) ;
      \draw  [color={rgb, 255:red, 74; green, 144; blue, 226 }  ,draw opacity=1 ][line width=0.75] [line join = round][line cap = round] (260,67.3) .. controls (254.33,63.03) and (249.04,58.21) .. (243,54.5) .. controls (232.34,47.95) and (213.77,46.47) .. (205.2,38.3) ;
      \draw  [color={rgb, 255:red, 208; green, 2; blue, 27 }  ,draw opacity=1 ][line width=0.75] [line join = round][line cap = round] (190.8,48.7) .. controls (194.4,44.17) and (196.91,38.49) .. (201.6,35.1) .. controls (202.9,34.16) and (204,37.23) .. (205.2,38.3) ;
      \draw  [color={rgb, 255:red, 208; green, 2; blue, 27 }  ,draw opacity=1 ][line width=0.75] [line join = round][line cap = round] (205.6,67.7) .. controls (208.71,64.59) and (214.25,65.29) .. (217.8,62.7) .. controls (220.73,60.56) and (221.13,55.84) .. (224.2,53.9) .. controls (227.18,52.02) and (231.13,52.7) .. (234.6,52.1) ;
      \draw  [color={rgb, 255:red, 208; green, 2; blue, 27 }  ,draw opacity=1 ][line width=0.75] [line join = round][line cap = round] (236.2,50.5) .. controls (245.13,45.03) and (252.67,35.8) .. (263,34.1) .. controls (267.27,33.4) and (267.8,41.3) .. (270.2,44.9) ;
      \draw  [color={rgb, 255:red, 74; green, 144; blue, 226 }  ,draw opacity=1 ][line width=0.75] [line join = round][line cap = round] (206,67.7) .. controls (209.67,66.97) and (213.8,63.56) .. (217,65.5) .. controls (222.98,69.12) and (223.17,79.91) .. (229.8,82.1) .. controls (234.11,83.52) and (236.6,76.1) .. (240,73.1) ;
      \draw  [color={rgb, 255:red, 208; green, 2; blue, 27 }  ,draw opacity=1 ][line width=0.75] [line join = round][line cap = round] (239.4,73.7) .. controls (241.2,71.37) and (241.99,67.57) .. (244.8,66.7) .. controls (249.46,65.26) and (254.53,67.23) .. (259.4,67.5) ;
      \draw  [color={rgb, 255:red, 74; green, 144; blue, 226 }  ,draw opacity=1 ][line width=0.75] [line join = round][line cap = round] (249.4,33.7) .. controls (247.33,28.17) and (246.24,22.16) .. (243.2,17.1) .. controls (240.17,12.06) and (233.44,16.4) .. (230.2,16.5) .. controls (228.21,16.56) and (225.79,13.68) .. (224.4,15.1) .. controls (219.56,20.06) and (217.47,27.1) .. (214,33.1) ;
      \draw  [color={rgb, 255:red, 208; green, 2; blue, 27 }  ,draw opacity=1 ][line width=0.75] [line join = round][line cap = round] (249.6,34.1) .. controls (249.2,31.97) and (249.51,29.57) .. (248.4,27.7) .. controls (247.43,26.06) and (244.93,25.83) .. (243.8,24.3) .. controls (242.76,22.89) and (243.8,20) .. (242.2,19.3) .. controls (232.92,15.25) and (233.47,21.06) .. (226.8,23.5) .. controls (226.04,23.78) and (225.21,21.97) .. (224.6,22.5) .. controls (220.67,25.88) and (217.98,30.73) .. (213.6,33.5) ;
      \draw  [color={rgb, 255:red, 208; green, 2; blue, 27 }  ,draw opacity=1 ][line width=0.75] [line join = round][line cap = round] (196,63.9) .. controls (196.27,66.37) and (195.56,69.15) .. (196.8,71.3) .. controls (198.22,73.76) and (201.73,74.4) .. (203.4,76.7) .. controls (205.62,79.77) and (205.41,85.01) .. (208.8,86.7) .. controls (212.33,88.47) and (216.98,86.14) .. (220.6,87.7) .. controls (224.48,89.37) and (231.38,96.89) .. (236.8,93.9) .. controls (238.57,92.92) and (239.24,90.23) .. (241.2,89.7) .. controls (245.52,88.54) and (250.81,91.47) .. (254.6,89.1) .. controls (257.77,87.12) and (256.57,81.7) .. (258.8,78.7) .. controls (260.73,76.1) and (264.59,75.44) .. (266.6,72.9) .. controls (268.93,69.95) and (269.08,65.66) .. (271.4,62.7) ;
      \draw  [color={rgb, 255:red, 74; green, 144; blue, 226 }  ,draw opacity=1 ][line width=0.75] [line join = round][line cap = round] (195.6,63.7) .. controls (198.25,67.82) and (198.43,69.98) .. (203,71.5) .. controls (204.27,71.92) and (205.95,71.07) .. (207,71.9) .. controls (209.06,73.53) and (209.44,76.8) .. (211.6,78.3) .. controls (213.45,79.59) and (216.22,78.64) .. (218.2,79.7) .. controls (219.31,80.29) and (219.05,82.39) .. (220.2,82.9) .. controls (224.96,85.03) and (230.01,87.78) .. (235.2,87.3) .. controls (238.9,86.96) and (240.95,82.51) .. (244.2,80.7) .. controls (253.27,75.63) and (265.06,71.52) .. (271.2,62.3) ;
      \draw    (150.6,130.4) -- (230.4,130.4) ;
      \draw [shift={(232.4,130.4)}, rotate = 180] [color={rgb, 255:red, 0; green, 0; blue, 0 }  ][line width=0.75]    (10.93,-3.29) .. controls (6.95,-1.4) and (3.31,-0.3) .. (0,0) .. controls (3.31,0.3) and (6.95,1.4) .. (10.93,3.29)   ;
      \draw    (150.6,130.4) -- (43,130.4) ;
      \draw [shift={(41,130.4)}, rotate = 360] [color={rgb, 255:red, 0; green, 0; blue, 0 }  ][line width=0.75]    (10.93,-3.29) .. controls (6.95,-1.4) and (3.31,-0.3) .. (0,0) .. controls (3.31,0.3) and (6.95,1.4) .. (10.93,3.29)   ;
      \draw    (346.8,130.4) -- (426.6,130.4) ;
      \draw [shift={(428.6,130.4)}, rotate = 180] [color={rgb, 255:red, 0; green, 0; blue, 0 }  ][line width=0.75]    (10.93,-3.29) .. controls (6.95,-1.4) and (3.31,-0.3) .. (0,0) .. controls (3.31,0.3) and (6.95,1.4) .. (10.93,3.29)   ;
      \draw    (346.8,130.4) -- (239.2,130.4) ;
      \draw [shift={(237.2,130.4)}, rotate = 360] [color={rgb, 255:red, 0; green, 0; blue, 0 }  ][line width=0.75]    (10.93,-3.29) .. controls (6.95,-1.4) and (3.31,-0.3) .. (0,0) .. controls (3.31,0.3) and (6.95,1.4) .. (10.93,3.29)   ;
      
      \draw (16,77.5) node [anchor=north west][inner sep=0.75pt]  [rotate=-270] [align=left] {$\displaystyle o_{0} \boxtimes o_{o}^{\dagger }$};
      \draw (434,118.5) node [anchor=north west][inner sep=0.75pt]  [rotate=-270] [align=left] {$\displaystyle o_{x} \boxtimes o_{x}^{\dagger }$};
      \draw (131.5,112.4) node [anchor=north west][inner sep=0.75pt]    {$t $};
      \draw (324,113.4) node [anchor=north west][inner sep=0.75pt]    {$t $};
      
      \end{tikzpicture}
      \caption{A typical process contributing to backflow in the U(1) RUC calculation. \label{fig:typical}}
      \end{figure*}

  Our discussion in \secref{ssec:first bound} identifies the exponential suppression of growth/shrinkage processes as being the source of exponential suppression in backflow due to DAOE. This suppression is easy to prove in the RUC without symmetry. We are unable to rigorously prove exponential suppression in the U(1) RUC. We can, however, examine the contributions of certain sub-classes of process contributing to backflow. In other words, our aim  here is to use the U(1) RUC model to estimate a subset of the contributions to $\Gamma_\ell$ as defined in \eqnref{eq:trunc_picture_gen}. 

  Specifically, we examine the contributions of growth/shrinkage processes where the control parameter is the number of applications of Rule 7. This choice is made primarily because it affords  control over the diagrammatic expansion, but also because the processes contributing are irreducibly rely on the diffusive modes that distinguish the U(1) RUC from the RUC without symmetry. 

  Rule 7 must be used twice in a growth process (it is required both at the beginning and end of the grow/shrink process). An example of such an amplitude (which would contribute for $\ell_*< 5$) is

  \begin{equation}\label{eq:type1rule7twice}
    \begin{tikzpicture}[x=0.75pt,y=0.75pt,yscale=-1,xscale=1]
    
    \draw    (7,59.33) .. controls (44.5,25.75) and (65.5,25.75) .. (107,59.33) ;
    \draw [shift={(56.86,34.17)}, rotate = 357.77] [fill={rgb, 255:red, 0; green, 0; blue, 0 }  ][line width=0.08]  [draw opacity=0] (10.72,-5.15) -- (0,0) -- (10.72,5.15) -- (7.12,0) -- cycle    ;
    \draw    (7,59.33) .. controls (46.5,85.25) and (67,84.25) .. (107,59.33) ;
    \draw [shift={(57.26,78.39)}, rotate = 182.19] [fill={rgb, 255:red, 0; green, 0; blue, 0 }  ][line width=0.08]  [draw opacity=0] (10.72,-5.15) -- (0,0) -- (10.72,5.15) -- (7.12,0) -- cycle    ;
    \draw  [color={rgb, 255:red, 74; green, 144; blue, 226 }  ,draw opacity=1 ][line width=0.75] [line join = round][line cap = round] (15.07,52) .. controls (21.7,30.07) and (22.5,4.06) .. (58.8,9.69) .. controls (67.99,11.12) and (78.81,13.05) .. (84.37,19.37) .. controls (91.87,27.88) and (91,39.81) .. (94.32,50.03) ;
    \draw  [color={rgb, 255:red, 74; green, 144; blue, 226 }  ,draw opacity=1 ][line width=0.75] [line join = round][line cap = round] (84,71.3) .. controls (78.33,67.03) and (73.04,62.21) .. (67,58.5) .. controls (56.34,51.95) and (37.77,50.47) .. (29.2,42.3) ;
    \draw  [color={rgb, 255:red, 208; green, 2; blue, 27 }  ,draw opacity=1 ][line width=0.75] [line join = round][line cap = round] (14.8,52.7) .. controls (18.4,48.17) and (20.91,42.49) .. (25.6,39.1) .. controls (26.9,38.16) and (28,41.23) .. (29.2,42.3) ;
    \draw  [color={rgb, 255:red, 208; green, 2; blue, 27 }  ,draw opacity=1 ][line width=0.75] [line join = round][line cap = round] (29.6,71.7) .. controls (32.71,68.59) and (38.25,69.29) .. (41.8,66.7) .. controls (44.73,64.56) and (45.13,59.84) .. (48.2,57.9) .. controls (51.18,56.02) and (55.13,56.7) .. (58.6,56.1) ;
    \draw  [color={rgb, 255:red, 208; green, 2; blue, 27 }  ,draw opacity=1 ][line width=0.75] [line join = round][line cap = round] (60.2,54.5) .. controls (69.13,49.03) and (76.67,39.8) .. (87,38.1) .. controls (91.27,37.4) and (91.8,45.3) .. (94.2,48.9) ;
    \draw  [color={rgb, 255:red, 74; green, 144; blue, 226 }  ,draw opacity=1 ][line width=0.75] [line join = round][line cap = round] (30,71.7) .. controls (33.67,70.97) and (37.8,67.56) .. (41,69.5) .. controls (46.98,73.12) and (47.17,83.91) .. (53.8,86.1) .. controls (58.11,87.52) and (60.6,80.1) .. (64,77.1) ;
    \draw  [color={rgb, 255:red, 208; green, 2; blue, 27 }  ,draw opacity=1 ][line width=0.75] [line join = round][line cap = round] (63.4,77.7) .. controls (65.2,75.37) and (65.99,71.57) .. (68.8,70.7) .. controls (73.46,69.26) and (78.53,71.23) .. (83.4,71.5) ;
    \draw  [color={rgb, 255:red, 74; green, 144; blue, 226 }  ,draw opacity=1 ][line width=0.75] [line join = round][line cap = round] (73.4,37.7) .. controls (71.33,32.17) and (70.24,26.16) .. (67.2,21.1) .. controls (64.17,16.06) and (57.44,20.4) .. (54.2,20.5) .. controls (52.21,20.56) and (49.79,17.68) .. (48.4,19.1) .. controls (43.56,24.06) and (41.47,31.1) .. (38,37.1) ;
    \draw  [color={rgb, 255:red, 208; green, 2; blue, 27 }  ,draw opacity=1 ][line width=0.75] [line join = round][line cap = round] (73.6,38.1) .. controls (73.2,35.97) and (73.51,33.57) .. (72.4,31.7) .. controls (71.43,30.06) and (68.93,29.83) .. (67.8,28.3) .. controls (66.76,26.89) and (67.8,24) .. (66.2,23.3) .. controls (56.92,19.25) and (57.47,25.06) .. (50.8,27.5) .. controls (50.04,27.78) and (49.21,25.97) .. (48.6,26.5) .. controls (44.67,29.88) and (41.98,34.73) .. (37.6,37.5) ;
    \draw  [color={rgb, 255:red, 208; green, 2; blue, 27 }  ,draw opacity=1 ][line width=0.75] [line join = round][line cap = round] (20,67.9) .. controls (20.27,70.37) and (19.56,73.15) .. (20.8,75.3) .. controls (22.22,77.76) and (25.73,78.4) .. (27.4,80.7) .. controls (29.62,83.77) and (29.41,89.01) .. (32.8,90.7) .. controls (36.33,92.47) and (40.98,90.14) .. (44.6,91.7) .. controls (48.48,93.37) and (55.38,100.89) .. (60.8,97.9) .. controls (62.57,96.92) and (63.24,94.23) .. (65.2,93.7) .. controls (69.52,92.54) and (74.81,95.47) .. (78.6,93.1) .. controls (81.77,91.12) and (80.57,85.7) .. (82.8,82.7) .. controls (84.73,80.1) and (88.59,79.44) .. (90.6,76.9) .. controls (92.93,73.95) and (93.08,69.66) .. (95.4,66.7) ;
    \draw  [color={rgb, 255:red, 74; green, 144; blue, 226 }  ,draw opacity=1 ][line width=0.75] [line join = round][line cap = round] (19.6,67.7) .. controls (22.25,71.82) and (22.43,73.98) .. (27,75.5) .. controls (28.27,75.92) and (29.95,75.07) .. (31,75.9) .. controls (33.06,77.53) and (33.44,80.8) .. (35.6,82.3) .. controls (37.45,83.59) and (40.22,82.64) .. (42.2,83.7) .. controls (43.31,84.29) and (43.05,86.39) .. (44.2,86.9) .. controls (48.96,89.03) and (54.01,91.78) .. (59.2,91.3) .. controls (62.9,90.96) and (64.95,86.51) .. (68.2,84.7) .. controls (77.27,79.63) and (89.06,75.52) .. (95.2,66.3) ;
    \draw    (113,12) -- (113,96) ;
    \draw [shift={(113,98)}, rotate = 270] [color={rgb, 255:red, 0; green, 0; blue, 0 }  ][line width=0.75]    (10.93,-3.29) .. controls (6.95,-1.4) and (3.31,-0.3) .. (0,0) .. controls (3.31,0.3) and (6.95,1.4) .. (10.93,3.29)   ;
    \draw [shift={(113,10)}, rotate = 90] [color={rgb, 255:red, 0; green, 0; blue, 0 }  ][line width=0.75]    (10.93,-3.29) .. controls (6.95,-1.4) and (3.31,-0.3) .. (0,0) .. controls (3.31,0.3) and (6.95,1.4) .. (10.93,3.29)   ;
    \draw    (106,108) -- (11,108) ;
    \draw [shift={(9,108)}, rotate = 360] [color={rgb, 255:red, 0; green, 0; blue, 0 }  ][line width=0.75]    (10.93,-3.29) .. controls (6.95,-1.4) and (3.31,-0.3) .. (0,0) .. controls (3.31,0.3) and (6.95,1.4) .. (10.93,3.29)   ;
    \draw [shift={(108,108)}, rotate = 180] [color={rgb, 255:red, 0; green, 0; blue, 0 }  ][line width=0.75]    (10.93,-3.29) .. controls (6.95,-1.4) and (3.31,-0.3) .. (0,0) .. controls (3.31,0.3) and (6.95,1.4) .. (10.93,3.29)   ;
    
    \draw (116.4,67) node [anchor=north west][inner sep=0.75pt]  [rotate=-270]  {$\ell >\ell _{*}$};
    \draw (48,111.4) node [anchor=north west][inner sep=0.75pt]    {$\Delta$};

    \end{tikzpicture}
  \end{equation}

  Amplitudes of this type require a single $\fp,\fm$ pair to persist for time $\Delta$. However, we have argued in \secref{sec:self-energy} that such processes are suppressed as $e^{-\mathcal{O}(\sqrt{\Delta})}$. Moreover, in the same section, we argued that the operator support tends to increase diffusively, so that $\Delta = \mathcal{O}(\ell^2)$. The net conclusion is that such processes are suppressed as $e^{-\mathcal{O}(\ell)}$ (and  $e^{-\mathcal{O}(\ell_*)}$ after we sum over $\ell \geq \ell_*$). As a result, the subset of processes of form \eqnref{eq:type1rule7twice} are consistent with our main assumption in previous sections, namely that grow-shrink processes contributing to $|\Gamma_\ell|^2$ are exponentially suppressed in $\ell$.

 We can also estimate the contributions of diagrams involving a few more applications of rule 7. For example the following amplitudes has four such applications
  \begin{equation}
    \begin{tikzpicture}[x=0.75pt,y=0.75pt,yscale=-1,xscale=1]

    \draw    (7,59.33) .. controls (44.5,25.75) and (17.4,57) .. (66.2,56.6) ;
    \draw [shift={(34.47,47.5)}, rotate = 37.25] [fill={rgb, 255:red, 0; green, 0; blue, 0 }  ][line width=0.08]  [draw opacity=0] (10.72,-5.15) -- (0,0) -- (10.72,5.15) -- (7.12,0) -- cycle    ;
    \draw    (7,59.33) .. controls (40.2,69.4) and (26.2,81.52) .. (66.2,56.6) ;
    \draw [shift={(37.26,71.27)}, rotate = 190.16] [fill={rgb, 255:red, 0; green, 0; blue, 0 }  ][line width=0.08]  [draw opacity=0] (10.72,-5.15) -- (0,0) -- (10.72,5.15) -- (7.12,0) -- cycle    ;
    \draw    (84.6,42.13) .. controls (114.2,35) and (127,11) .. (143.8,39.4) ;
    \draw [shift={(115.97,28.69)}, rotate = 334.03] [fill={rgb, 255:red, 0; green, 0; blue, 0 }  ][line width=0.08]  [draw opacity=0] (10.72,-5.15) -- (0,0) -- (10.72,5.15) -- (7.12,0) -- cycle    ;
    \draw    (84.6,42.13) .. controls (117.8,52.2) and (103.8,64.32) .. (143.8,39.4) ;
    \draw [shift={(114.86,54.07)}, rotate = 190.16] [fill={rgb, 255:red, 0; green, 0; blue, 0 }  ][line width=0.08]  [draw opacity=0] (10.72,-5.15) -- (0,0) -- (10.72,5.15) -- (7.12,0) -- cycle    ;
    \draw  [color={rgb, 255:red, 208; green, 2; blue, 27 }  ,draw opacity=1 ][line width=0.75] [line join = round][line cap = round] (66.14,56.43) .. controls (66.46,55.24) and (70.87,41.55) .. (73.57,40.71) .. controls (76.96,39.67) and (81.38,44.07) .. (84.14,41.86) ;
    \draw  [color={rgb, 255:red, 74; green, 144; blue, 226 }  ,draw opacity=1 ][line width=0.75] [line join = round][line cap = round] (85,42.14) .. controls (83.48,42.71) and (81.4,42.55) .. (80.43,43.86) .. controls (79.22,45.49) and (80.23,48.05) .. (79.29,49.86) .. controls (78.31,51.73) and (74.76,47.83) .. (73,49) .. controls (70.31,50.79) and (68.91,57.52) .. (66.14,55.86) ;
    \draw [color={rgb, 255:red, 74; green, 144; blue, 226 }  ,draw opacity=1 ][line width=0.75]    (102.33,34.33) .. controls (95.8,28.2) and (83.93,20.9) .. (68.33,28.33) .. controls (52.74,35.77) and (47.85,53.52) .. (50.2,54.9) ;
    \draw [color={rgb, 255:red, 208; green, 2; blue, 27 }  ,draw opacity=1 ][line width=0.75]    (97.2,47.1) .. controls (103.97,38.8) and (109.52,49.57) .. (112,49.9) .. controls (114.48,50.23) and (113.77,40.09) .. (123.8,51.3) ;
    \draw [color={rgb, 255:red, 74; green, 144; blue, 226 }  ,draw opacity=1 ][line width=0.75]    (97.2,47.1) .. controls (100.96,34.93) and (108.44,36.54) .. (113.8,39.7) .. controls (119.16,42.86) and (118.44,39.6) .. (123.8,51.3) ;
    \draw [color={rgb, 255:red, 208; green, 2; blue, 27 }  ,draw opacity=1 ][line width=0.75]    (129.8,47.4) .. controls (125.07,56.25) and (111.29,61.49) .. (96.2,64.6) .. controls (81.1,67.71) and (43,114.6) .. (17.8,63) ;
    \draw [color={rgb, 255:red, 74; green, 144; blue, 226 }  ,draw opacity=1 ][line width=0.75]    (129.8,47.4) .. controls (130.85,53.18) and (109.6,75.27) .. (79.8,74.6) .. controls (50,73.93) and (36.2,92.6) .. (17.8,63) ;
    \draw    (162,16) -- (162,87) ;
    \draw [shift={(162,89)}, rotate = 270] [color={rgb, 255:red, 0; green, 0; blue, 0 }  ][line width=0.75]    (10.93,-3.29) .. controls (6.95,-1.4) and (3.31,-0.3) .. (0,0) .. controls (3.31,0.3) and (6.95,1.4) .. (10.93,3.29)   ;
    \draw [shift={(162,14)}, rotate = 90] [color={rgb, 255:red, 0; green, 0; blue, 0 }  ][line width=0.75]    (10.93,-3.29) .. controls (6.95,-1.4) and (3.31,-0.3) .. (0,0) .. controls (3.31,0.3) and (6.95,1.4) .. (10.93,3.29)   ;
    \draw [color={rgb, 255:red, 208; green, 2; blue, 27 }  ,draw opacity=1 ][line width=0.75]    (102.33,34.33) .. controls (86.6,38.2) and (87.3,27.89) .. (72.2,31) .. controls (57.1,34.11) and (48.67,54.69) .. (50.2,54.9) ;
    \draw [color={rgb, 255:red, 74; green, 144; blue, 226 }  ,draw opacity=1 ][line width=0.75]    (113,29.8) .. controls (106.47,23.67) and (68.33,12.5) .. (52.73,19.93) .. controls (37.14,27.37) and (32.25,45.12) .. (34.6,46.5) ;
    \draw [color={rgb, 255:red, 208; green, 2; blue, 27 }  ,draw opacity=1 ][line width=0.75]    (113,29.8) .. controls (103.8,24.6) and (69.39,18.77) .. (53.8,26.2) .. controls (38.21,33.63) and (32.25,45.12) .. (34.6,46.5) ;
    \draw    (141,94) -- (10,94) ;
    \draw [shift={(8,94)}, rotate = 360] [color={rgb, 255:red, 0; green, 0; blue, 0 }  ][line width=0.75]    (10.93,-3.29) .. controls (6.95,-1.4) and (3.31,-0.3) .. (0,0) .. controls (3.31,0.3) and (6.95,1.4) .. (10.93,3.29)   ;
    \draw [shift={(143,94)}, rotate = 180] [color={rgb, 255:red, 0; green, 0; blue, 0 }  ][line width=0.75]    (10.93,-3.29) .. controls (6.95,-1.4) and (3.31,-0.3) .. (0,0) .. controls (3.31,0.3) and (6.95,1.4) .. (10.93,3.29)   ;

    \draw (168.4,64) node [anchor=north west][inner sep=0.75pt]  [rotate=-270]  {$\ell >\ell _{*}$};
    \draw (64,97.4) node [anchor=north west][inner sep=0.75pt]    {$\Delta$};

    \end{tikzpicture}
  \end{equation}

  In processes involving four applications of rule $7$, at least one of the two $\fp,\fm$ pairs formed must survive for at least $O(\ell^2)$ time in order to increase the support of the operator to $\ell$ part-way through the evolution. But the $\fr,\fb$ pairs released will need to be absorbed, and this is once again unlikely. The same reasoning as above shows that this amplitude is suppressed as $e^{-\mathcal{O}(\ell_*)}$ once summed over all $\ell>\ell_*$.

  We have argued that the first few orders in a series expansion are consistent with the exponential decay of backflow. It may be that this mode of reasoning can be extended order-by-order to diagrams that involve further applications of rule 7, however we were unable to prove rigorously that the fully summed series is on net exponentially suppressed in $\ell_*$.  Our limited numerical evidence (\figref{fig:total}) suggests that the fully summed series of contributions will continue to show exponential decay in $\ell$, but a complete proof is lacking and could be an interesting topic for future investigation.

  
\section{How hard is it to simulate quantum transport classically?}
\label{sec:memory}

We now discuss the implications of our various conjectures for numerical studies of dynamics in many-body systems. 
We will be interested in the following question: 
What resources are required to compute transport properties, such as a diffusion constant, to some fixed precision $\epsilon$? 

A standard approach for calculating the diffusion constant -- used in exact evolution and tensor-network based methods -- is to approximate the true diffusion constant with a finite time approximation. The deviation between the diffusion constant and its finite time approximation is bounded as $|D(t)- D|\leq   C/t^{1/2}$ where $C$ is an $O(1)$ constant \footnote{This follows from the Drude formula and the fact that current auto-correlations decay as $\langle J(t) J(0) \rangle_c \sim t^{-1-d/2}$ \cite{huse_tails}; we focus on $d=1$ in this section.}. Thus we will need to simulate the system for a minimum time $t_c = \mathcal{O}(\epsilon^{-2})$ in order to obtain an estimate for $D$ to within tolerance $\epsilon$.

For exact evolution, we must consider a finite-sized systems, say with $L$ sites. In order to avoid incurring significant finite-size effects, the corresponding Thouless time ought to exceed $t_c$:  This translates to the requirement $L> \mathcal{O}{(\sqrt{t_c})}$, although one can only prove rigorously that finite-size effects are exponentially suppressed in $t_c$ (and thus much smaller than $\epsilon$) if we use the stricter condition $L > v t_c$ for some Lieb-Robinson scale $v$. In either case, if we wish to obtain an $\epsilon$ accurate approximations to the diffusion constant, we will require $L=\mathcal{O}(\mathrm{poly}(\epsilon^{-1})$, which leads to a memory requirement $\exp(\mathcal{O}(\mathrm{poly}(\epsilon^{-1})))$.

In tensor network methods, we need not work with finite systems. However, the entanglement of the time-evolved state/operator is expected to increase linearly in time on theoretical grounds~\cite{jonay2018coarsegrained}. This suggests -- and experience confirms -- that bond dimension needed to control the errors in local observables grows exponentially with time. Using $t_c \sim \epsilon^{-2}$, this would lead to a bond dimension requirement scaling as $\exp[{\mathcal{O}(\epsilon^{-2})}]$. It is possible this bound is too pessimistic: in the small $\epsilon$ limit, it might be possible to leverage the diffusive nature of the transport and truncate the support of our time evolved state/operator to a region of size $\mathcal{O}(\sqrt{t_c})$, which would reduce the bond dimension requirement to $\exp[{\mathcal{O}(\epsilon^{-1})}]$. In either case, we expect that the bond dimension requirement scales as
\begin{equation}\label{eq:err_TEBD}
  \chi_{\mathrm{naive}}=\exp[{\mathcal{O}(\mathrm{poly}(\epsilon^{-1})}],
\end{equation}
which is in-line with the memory requirements for exact evolution.

Modifying the dynamics with DAOE dissipation significantly reduces the memory overhead. This is especially clear in the strong dissipation limit, $\gamma\to\infty,t_\text{D}\to 0$, where we explicitly restrict the accessible Hilbert space to operators with length $\leq\ell_*$; however, it was found in Ref. \onlinecite{me_daoe_1} that even when $\gamma$ and $t_\text{D}$ are both finite, DAOE limits the growth of operator entanglement to a value that remains bounded at all times. The cost is that we introduce a systematic error in the diffusion constant; however, as we argued throughout this paper, this error is bounded by $\exp[{- \mathcal{O}(\ell_*) }]$. Insisting that the systematic error is less than $O(\epsilon)$ requires $\ell_* > \ell_c = \mathcal{O}(|\log(\epsilon)|)$. We still have the requirement that $t>t_c$. For times of this order, the accessible Hilbert space at $\gamma = \infty$ is $4^{\ell_c} \binom{2 v t_c}{\ell_c}$. For small $\epsilon$, this implies that the bond dimension required for a tolerance $\epsilon$ calculation of the diffusion constant scales as
\begin{equation}\label{eq:err_DAOE}
  \chi_{\mathrm{DAOE}}=\exp[{\mathcal{O}(|\log \epsilon |^2)}].
\end{equation}
which is significantly less than either the memory requirement for tensor network simulations in the absence of dissipation in \eqnref{eq:err_TEBD} or that of exact evolution discussed above. 

While it is generally believed that some dynamical properties of quantum many-body systems are not efficiently simulable on classical computers \cite{preskill2012quantum,Aaronson2017,GoogleSupremacy}, our result \eqnref{eq:err_DAOE} suggests that transport properties are (at least in ergodic systems). It would be interesting to find what other problems fall into this latter category, and devise new classical algorithms to tackle them.

\section{Conclusion}\label{sec:conclusion}
We have argued that the `backflow' contributions to hydrodynamic correlations involving operators that act on more than $\ell_*$ different sites are exponentially suppressed in $\ell_*$ (\secref{ssec:first bound}). Moreover, the corrections to correlation functions and transport coefficients due to backflow are themselves arranged in a series in powers of $t^{-1}$; we have made various predictions for the exponents of these contributions, dependent on whether the systems have randomness, and the hydrodynamical correlation function in question (\secref{ss:backflow_hydro_structure}). We have provided numerical evidence for this picture, both in deterministic quantum spin chains (\figref{fig:IsingandLadder}), and in U$(1)$ symmetric random circuit models (\figref{fig:total}). In the latter case, we provided further semi-analytical consistency checks of both our hydrodynamical picture for backflow corrections as well as the assertion that backflow corrections are exponentially suppressed in $\ell_*$ (\secref{sec:U(1)random}). Our conjectured bounds have important consequences for the asymptotic computational resources required for ab initio calculation of transport properties; our results suggest that the DAOE method introduced in Ref. \onlinecite{me_daoe_1} requires significantly smaller numerical resources than existing established methods like ED or TEBD (see e.g. Eqs.~\eqref{eq:err_TEBD} and~\eqref{eq:err_DAOE} in \secref{sec:memory}). 


This work leaves many questions open. It would be worthwhile to generalize the theoretical predictions in this work to finite temperatures/chemical potentials. It is likely that a dramatic modification to these results occurs in higher dimensions at sufficiently low temperatures such that the systems undergo an ordering transition. Moreover it is worth performing a backflow analysis for a variety of truncation methods (including truncating dependent of the spatial support rather than the length of operators); we plan to explore this in forthcoming work. 

In this paper, we focused on one-dimensional systems. Nonetheless, we expect the hydrodynamical framework developed here to apply in higher dimensions; in particular, we expect our arguments in \secref{sec:self-energy} for exponential suppression with $\ell_*$ to generalize quite straightforwardly. We note that it may be practicable to use DAOE in this more general setting, given recent advances in higher dimensional tensor network algorithms. It would be interesting to investigate the growth/shrinkage processes in higher dimensions and with greater detail and control. Forthcoming work by Nahum \emph{et al.} goes in this direction, with a focus on grow-shrink processes in the random circuit without symmetries~\cite{nahumfatop}. 

It would also be interesting to connect our results to existing efforts in the study of complexity theory. Various results are known on the computational resources required to, e.g., calculate the ground state energy density of local Hamiltonians \cite{aharonov_en_density,cubitt_en_density}. What can be said for the calculation of hydrodynamical quantities? Our results arguably indicate that, at least for ergodic systems (which are ubiquitous), surprisingly few resources are required. Finally, there are certainly many conjectures and technical points in the present work which would benefit from further study. For example, can one prove rigorously that backflow is exponentially suppressed in the U(1) circuit averaged model?

\section{Acknowledgements}
We thank Luca Delacretaz, Christopher White, Daniel Parker and Gabriele Pinna for useful discussions. CvK is supported by a UKRI Future Leaders Fellowship MR/T040947/1. The computations described in this paper were performed in part using the University of Birmingham's BlueBEAR HPC service. F.P. acknowledges support from the European Research Council (ERC) under the European Union's Horizon 2020 research and innovation programme (grant agreement No. 771537) and the Deutsche Forschungsgemeinschaft (DFG, German Research Foundation) under Germany's Excellence Strategy EXC-2111-390814868 and TRR 80. T.R. is supported by the Stanford Q-Farm Bloch Postdoctoral Fellowship in Quantum Science and Engineering. TR acknowledges the hospitality of the Aspen Center for Physics, supported by National Science Foundation grant PHY-1607611 and the Kavli Institute for Physics, supported by the National Science Foundation under Grant No. NSF PHY-1748958.

\bibliography{bib} 

\begin{thebibliography}{43}%
\makeatletter
\providecommand \@ifxundefined [1]{%
 \@ifx{#1\undefined}
}%
\providecommand \@ifnum [1]{%
 \ifnum #1\expandafter \@firstoftwo
 \else \expandafter \@secondoftwo
 \fi
}%
\providecommand \@ifx [1]{%
 \ifx #1\expandafter \@firstoftwo
 \else \expandafter \@secondoftwo
 \fi
}%
\providecommand \natexlab [1]{#1}%
\providecommand \enquote  [1]{``#1''}%
\providecommand \bibnamefont  [1]{#1}%
\providecommand \bibfnamefont [1]{#1}%
\providecommand \citenamefont [1]{#1}%
\providecommand \href@noop [0]{\@secondoftwo}%
\providecommand \href [0]{\begingroup \@sanitize@url \@href}%
\providecommand \@href[1]{\@@startlink{#1}\@@href}%
\providecommand \@@href[1]{\endgroup#1\@@endlink}%
\providecommand \@sanitize@url [0]{\catcode `\\12\catcode `\$12\catcode
  `\&12\catcode `\#12\catcode `\^12\catcode `\_12\catcode `\%12\relax}%
\providecommand \@@startlink[1]{}%
\providecommand \@@endlink[0]{}%
\providecommand \url  [0]{\begingroup\@sanitize@url \@url }%
\providecommand \@url [1]{\endgroup\@href {#1}{\urlprefix }}%
\providecommand \urlprefix  [0]{URL }%
\providecommand \Eprint [0]{\href }%
\providecommand \doibase [0]{http://dx.doi.org/}%
\providecommand \selectlanguage [0]{\@gobble}%
\providecommand \bibinfo  [0]{\@secondoftwo}%
\providecommand \bibfield  [0]{\@secondoftwo}%
\providecommand \translation [1]{[#1]}%
\providecommand \BibitemOpen [0]{}%
\providecommand \bibitemStop [0]{}%
\providecommand \bibitemNoStop [0]{.\EOS\space}%
\providecommand \EOS [0]{\spacefactor3000\relax}%
\providecommand \BibitemShut  [1]{\csname bibitem#1\endcsname}%
\let\auto@bib@innerbib\@empty
\bibitem [{\citenamefont {Heitmann}\ \emph {et~al.}(2020)\citenamefont
  {Heitmann}, \citenamefont {Richter}, \citenamefont {Schubert},\ and\
  \citenamefont {Steinigeweg}}]{SteinigewegReview}%
  \BibitemOpen
  \bibfield  {author} {\bibinfo {author} {\bibfnamefont {T.}~\bibnamefont
  {Heitmann}}, \bibinfo {author} {\bibfnamefont {J.}~\bibnamefont {Richter}},
  \bibinfo {author} {\bibfnamefont {D.}~\bibnamefont {Schubert}}, \ and\
  \bibinfo {author} {\bibfnamefont {R.}~\bibnamefont {Steinigeweg}},\ }\href
  {\doibase doi:10.1515/zna-2020-0010} {\bibfield  {journal} {\bibinfo
  {journal} {Zeitschrift f{\"u}r Naturforschung A}\ }\textbf {\bibinfo {volume}
  {75}},\ \bibinfo {pages} {421} (\bibinfo {year} {2020})}\BibitemShut
  {NoStop}%
\bibitem [{\citenamefont {Verstraete}\ \emph {et~al.}(2008)\citenamefont
  {Verstraete}, \citenamefont {Murg},\ and\ \citenamefont
  {Cirac}}]{MurgReview}%
  \BibitemOpen
  \bibfield  {author} {\bibinfo {author} {\bibfnamefont {F.}~\bibnamefont
  {Verstraete}}, \bibinfo {author} {\bibfnamefont {V.}~\bibnamefont {Murg}}, \
  and\ \bibinfo {author} {\bibfnamefont {J.}~\bibnamefont {Cirac}},\ }\href
  {\doibase 10.1080/14789940801912366} {\bibfield  {journal} {\bibinfo
  {journal} {Advances in Physics}\ }\textbf {\bibinfo {volume} {57}},\ \bibinfo
  {pages} {143} (\bibinfo {year} {2008})},\ \Eprint
  {http://arxiv.org/abs/https://doi.org/10.1080/14789940801912366}
  {https://doi.org/10.1080/14789940801912366} \BibitemShut {NoStop}%
\bibitem [{\citenamefont {Schollw\"ock}(2011)}]{SCHOLLWOCK201196}%
  \BibitemOpen
  \bibfield  {author} {\bibinfo {author} {\bibfnamefont {U.}~\bibnamefont
  {Schollw\"ock}},\ }\href {\doibase https://doi.org/10.1016/j.aop.2010.09.012}
  {\bibfield  {journal} {\bibinfo  {journal} {Annals of Physics}\ }\textbf
  {\bibinfo {volume} {326}},\ \bibinfo {pages} {96} (\bibinfo {year} {2011})},\
  \bibinfo {note} {january 2011 Special Issue}\BibitemShut {NoStop}%
\bibitem [{\citenamefont {Paeckel}\ \emph {et~al.}(2019)\citenamefont
  {Paeckel}, \citenamefont {K\"ohler}, \citenamefont {Swoboda}, \citenamefont
  {Manmana}, \citenamefont {Schollw\"ock},\ and\ \citenamefont
  {Hubig}}]{PAECKEL2019167998}%
  \BibitemOpen
  \bibfield  {author} {\bibinfo {author} {\bibfnamefont {S.}~\bibnamefont
  {Paeckel}}, \bibinfo {author} {\bibfnamefont {T.}~\bibnamefont {K\"ohler}},
  \bibinfo {author} {\bibfnamefont {A.}~\bibnamefont {Swoboda}}, \bibinfo
  {author} {\bibfnamefont {S.~R.}\ \bibnamefont {Manmana}}, \bibinfo {author}
  {\bibfnamefont {U.}~\bibnamefont {Schollw\"ock}}, \ and\ \bibinfo {author}
  {\bibfnamefont {C.}~\bibnamefont {Hubig}},\ }\href {\doibase
  https://doi.org/10.1016/j.aop.2019.167998} {\bibfield  {journal} {\bibinfo
  {journal} {Annals of Physics}\ }\textbf {\bibinfo {volume} {411}},\ \bibinfo
  {pages} {167998} (\bibinfo {year} {2019})}\BibitemShut {NoStop}%
\bibitem [{\citenamefont {White}\ \emph {et~al.}(2018)\citenamefont {White},
  \citenamefont {Zaletel}, \citenamefont {Mong},\ and\ \citenamefont
  {Refael}}]{white_zaletel_1}%
  \BibitemOpen
  \bibfield  {author} {\bibinfo {author} {\bibfnamefont {C.~D.}\ \bibnamefont
  {White}}, \bibinfo {author} {\bibfnamefont {M.}~\bibnamefont {Zaletel}},
  \bibinfo {author} {\bibfnamefont {R.~S.~K.}\ \bibnamefont {Mong}}, \ and\
  \bibinfo {author} {\bibfnamefont {G.}~\bibnamefont {Refael}},\ }\href
  {\doibase 10.1103/physrevb.97.035127} {\bibfield  {journal} {\bibinfo
  {journal} {Physical Review B}\ }\textbf {\bibinfo {volume} {97}} (\bibinfo
  {year} {2018}),\ 10.1103/physrevb.97.035127}\BibitemShut {NoStop}%
\bibitem [{\citenamefont {Parker}\ \emph {et~al.}(2019)\citenamefont {Parker},
  \citenamefont {Cao}, \citenamefont {Avdoshkin}, \citenamefont {Scaffidi},\
  and\ \citenamefont {Altman}}]{parkerhypothesis}%
  \BibitemOpen
  \bibfield  {author} {\bibinfo {author} {\bibfnamefont {D.~E.}\ \bibnamefont
  {Parker}}, \bibinfo {author} {\bibfnamefont {X.}~\bibnamefont {Cao}},
  \bibinfo {author} {\bibfnamefont {A.}~\bibnamefont {Avdoshkin}}, \bibinfo
  {author} {\bibfnamefont {T.}~\bibnamefont {Scaffidi}}, \ and\ \bibinfo
  {author} {\bibfnamefont {E.}~\bibnamefont {Altman}},\ }\href {\doibase
  10.1103/PhysRevX.9.041017} {\bibfield  {journal} {\bibinfo  {journal} {Phys.
  Rev. X}\ }\textbf {\bibinfo {volume} {9}},\ \bibinfo {pages} {041017}
  (\bibinfo {year} {2019})}\BibitemShut {NoStop}%
\bibitem [{\citenamefont {Prosen}\ and\ \citenamefont
  {{\v{Z}}nidari{\v{c}}}(2009)}]{marko_method_1}%
  \BibitemOpen
  \bibfield  {author} {\bibinfo {author} {\bibfnamefont {T.}~\bibnamefont
  {Prosen}}\ and\ \bibinfo {author} {\bibfnamefont {M.}~\bibnamefont
  {{\v{Z}}nidari{\v{c}}}},\ }\href {\doibase 10.1088/1742-5468/2009/02/p02035}
  {\bibfield  {journal} {\bibinfo  {journal} {Journal of Statistical Mechanics:
  Theory and Experiment}\ }\textbf {\bibinfo {volume} {2009}},\ \bibinfo
  {pages} {P02035} (\bibinfo {year} {2009})}\BibitemShut {NoStop}%
\bibitem [{\citenamefont {Kvorning}\ \emph {et~al.}(2021)\citenamefont
  {Kvorning}, \citenamefont {Herviou},\ and\ \citenamefont
  {Bardarson}}]{jens_hydro_1}%
  \BibitemOpen
  \bibfield  {author} {\bibinfo {author} {\bibfnamefont {T.~K.}\ \bibnamefont
  {Kvorning}}, \bibinfo {author} {\bibfnamefont {L.}~\bibnamefont {Herviou}}, \
  and\ \bibinfo {author} {\bibfnamefont {J.~H.}\ \bibnamefont {Bardarson}},\
  }\href@noop {} {\enquote {\bibinfo {title} {Time-evolution of local
  information: thermalization dynamics of local observables},}\ } (\bibinfo
  {year} {2021}),\ \Eprint {http://arxiv.org/abs/2105.11206} {arXiv:2105.11206
  [quant-ph]} \BibitemShut {NoStop}%
\bibitem [{\citenamefont {Rakovszky}\ \emph {et~al.}(2020)\citenamefont
  {Rakovszky}, \citenamefont {von Keyserlingk},\ and\ \citenamefont
  {Pollmann}}]{me_daoe_1}%
  \BibitemOpen
  \bibfield  {author} {\bibinfo {author} {\bibfnamefont {T.}~\bibnamefont
  {Rakovszky}}, \bibinfo {author} {\bibfnamefont {C.~W.}\ \bibnamefont {von
  Keyserlingk}}, \ and\ \bibinfo {author} {\bibfnamefont {F.}~\bibnamefont
  {Pollmann}},\ }\href@noop {} {\enquote {\bibinfo {title}
  {Dissipation-assisted operator evolution method for capturing hydrodynamic
  transport},}\ } (\bibinfo {year} {2020}),\ \Eprint
  {http://arxiv.org/abs/2004.05177} {arXiv:2004.05177 [cond-mat.str-el]}
  \BibitemShut {NoStop}%
\bibitem [{\citenamefont {White}(2021)}]{white2021effective}%
  \BibitemOpen
  \bibfield  {author} {\bibinfo {author} {\bibfnamefont {C.~D.}\ \bibnamefont
  {White}},\ }\href@noop {} {\enquote {\bibinfo {title} {Effective dissipation
  rate in a liouvillean graph picture of high-temperature quantum
  hydrodynamics},}\ } (\bibinfo {year} {2021}),\ \Eprint
  {http://arxiv.org/abs/2108.00019} {arXiv:2108.00019 [cond-mat.str-el]}
  \BibitemShut {NoStop}%
\bibitem [{\citenamefont {Wurtz}\ \emph {et~al.}(2018)\citenamefont {Wurtz},
  \citenamefont {Polkovnikov},\ and\ \citenamefont {Sels}}]{WURTZ2018341}%
  \BibitemOpen
  \bibfield  {author} {\bibinfo {author} {\bibfnamefont {J.}~\bibnamefont
  {Wurtz}}, \bibinfo {author} {\bibfnamefont {A.}~\bibnamefont {Polkovnikov}},
  \ and\ \bibinfo {author} {\bibfnamefont {D.}~\bibnamefont {Sels}},\ }\href
  {\doibase https://doi.org/10.1016/j.aop.2018.06.001} {\bibfield  {journal}
  {\bibinfo  {journal} {Annals of Physics}\ }\textbf {\bibinfo {volume}
  {395}},\ \bibinfo {pages} {341} (\bibinfo {year} {2018})}\BibitemShut
  {NoStop}%
\bibitem [{\citenamefont {Reichman}\ and\ \citenamefont
  {Charbonneau}(2005)}]{mct_review}%
  \BibitemOpen
  \bibfield  {author} {\bibinfo {author} {\bibfnamefont {D.~R.}\ \bibnamefont
  {Reichman}}\ and\ \bibinfo {author} {\bibfnamefont {P.}~\bibnamefont
  {Charbonneau}},\ }\href {\doibase 10.1088/1742-5468/2005/05/p05013}
  {\bibfield  {journal} {\bibinfo  {journal} {Journal of Statistical Mechanics:
  Theory and Experiment}\ }\textbf {\bibinfo {volume} {2005}},\ \bibinfo
  {pages} {P05013} (\bibinfo {year} {2005})}\BibitemShut {NoStop}%
\bibitem [{\citenamefont {{Landau}}\ and\ \citenamefont
  {{Lifshitz}}(1987)}]{landau_lifshitz_fluids}%
  \BibitemOpen
  \bibfield  {author} {\bibinfo {author} {\bibfnamefont {L.~D.}\ \bibnamefont
  {{Landau}}}\ and\ \bibinfo {author} {\bibfnamefont {E.~M.}\ \bibnamefont
  {{Lifshitz}}},\ }\href@noop {} {\emph {\bibinfo {title} {{Fluid
  Mechanics}}}}\ (\bibinfo {year} {1987})\BibitemShut {NoStop}%
\bibitem [{\citenamefont {Rakovszky}\ \emph {et~al.}(2018)\citenamefont
  {Rakovszky}, \citenamefont {Pollmann},\ and\ \citenamefont {von
  Keyserlingk}}]{vonKeyserlingk2018_diffusive}%
  \BibitemOpen
  \bibfield  {author} {\bibinfo {author} {\bibfnamefont {T.}~\bibnamefont
  {Rakovszky}}, \bibinfo {author} {\bibfnamefont {F.}~\bibnamefont {Pollmann}},
  \ and\ \bibinfo {author} {\bibfnamefont {C.~W.}\ \bibnamefont {von
  Keyserlingk}},\ }\href {\doibase 10.1103/PhysRevX.8.031058} {\bibfield
  {journal} {\bibinfo  {journal} {Phys. Rev. X}\ }\textbf {\bibinfo {volume}
  {8}},\ \bibinfo {pages} {031058} (\bibinfo {year} {2018})}\BibitemShut
  {NoStop}%
\bibitem [{\citenamefont {Bloembergen}(1949)}]{diffusion_ubiquitous_1}%
  \BibitemOpen
  \bibfield  {author} {\bibinfo {author} {\bibfnamefont {N.}~\bibnamefont
  {Bloembergen}},\ }\href {\doibase
  https://doi.org/10.1016/0031-8914(49)90114-7} {\bibfield  {journal} {\bibinfo
   {journal} {Physica}\ }\textbf {\bibinfo {volume} {15}},\ \bibinfo {pages}
  {386 } (\bibinfo {year} {1949})}\BibitemShut {NoStop}%
\bibitem [{\citenamefont {Gennes}(1958)}]{diffusion_ubiquitous_2}%
  \BibitemOpen
  \bibfield  {author} {\bibinfo {author} {\bibfnamefont {P.~D.}\ \bibnamefont
  {Gennes}},\ }\href {\doibase https://doi.org/10.1016/0022-3697(58)90120-3}
  {\bibfield  {journal} {\bibinfo  {journal} {Journal of Physics and Chemistry
  of Solids}\ }\textbf {\bibinfo {volume} {4}},\ \bibinfo {pages} {223 }
  (\bibinfo {year} {1958})}\BibitemShut {NoStop}%
\bibitem [{\citenamefont {Kadanoff}\ and\ \citenamefont
  {Martin}(1963)}]{diffusion_ubiquitous_3}%
  \BibitemOpen
  \bibfield  {author} {\bibinfo {author} {\bibfnamefont {L.~P.}\ \bibnamefont
  {Kadanoff}}\ and\ \bibinfo {author} {\bibfnamefont {P.~C.}\ \bibnamefont
  {Martin}},\ }\href {\doibase https://doi.org/10.1016/0003-4916(63)90078-2}
  {\bibfield  {journal} {\bibinfo  {journal} {Annals of Physics}\ }\textbf
  {\bibinfo {volume} {24}},\ \bibinfo {pages} {419 } (\bibinfo {year}
  {1963})}\BibitemShut {NoStop}%
\bibitem [{\citenamefont {Jonay}\ \emph {et~al.}(2018)\citenamefont {Jonay},
  \citenamefont {Huse},\ and\ \citenamefont {Nahum}}]{jonay2018coarsegrained}%
  \BibitemOpen
  \bibfield  {author} {\bibinfo {author} {\bibfnamefont {C.}~\bibnamefont
  {Jonay}}, \bibinfo {author} {\bibfnamefont {D.~A.}\ \bibnamefont {Huse}}, \
  and\ \bibinfo {author} {\bibfnamefont {A.}~\bibnamefont {Nahum}},\
  }\href@noop {} {\enquote {\bibinfo {title} {Coarse-grained dynamics of
  operator and state entanglement},}\ } (\bibinfo {year} {2018}),\ \Eprint
  {http://arxiv.org/abs/1803.00089} {arXiv:1803.00089 [cond-mat.stat-mech]}
  \BibitemShut {NoStop}%
\bibitem [{\citenamefont {Forster}(2018)}]{forster}%
  \BibitemOpen
  \bibfield  {author} {\bibinfo {author} {\bibfnamefont {D.}~\bibnamefont
  {Forster}},\ }\href@noop {} {\emph {\bibinfo {title} {Hydrodynamic
  fluctuations, broken symmetry, and correlation functions}}}\ (\bibinfo
  {publisher} {CRC Press},\ \bibinfo {year} {2018})\BibitemShut {NoStop}%
\bibitem [{\citenamefont {Mukerjee}\ \emph {et~al.}(2006)\citenamefont
  {Mukerjee}, \citenamefont {Oganesyan},\ and\ \citenamefont
  {Huse}}]{huse_tails}%
  \BibitemOpen
  \bibfield  {author} {\bibinfo {author} {\bibfnamefont {S.}~\bibnamefont
  {Mukerjee}}, \bibinfo {author} {\bibfnamefont {V.}~\bibnamefont {Oganesyan}},
  \ and\ \bibinfo {author} {\bibfnamefont {D.}~\bibnamefont {Huse}},\ }\href
  {\doibase 10.1103/physrevb.73.035113} {\bibfield  {journal} {\bibinfo
  {journal} {Physical Review B}\ }\textbf {\bibinfo {volume} {73}} (\bibinfo
  {year} {2006}),\ 10.1103/physrevb.73.035113}\BibitemShut {NoStop}%
\bibitem [{\citenamefont {Khemani}\ \emph {et~al.}(2018)\citenamefont
  {Khemani}, \citenamefont {Vishwanath},\ and\ \citenamefont
  {Huse}}]{Khemani_2018}%
  \BibitemOpen
  \bibfield  {author} {\bibinfo {author} {\bibfnamefont {V.}~\bibnamefont
  {Khemani}}, \bibinfo {author} {\bibfnamefont {A.}~\bibnamefont {Vishwanath}},
  \ and\ \bibinfo {author} {\bibfnamefont {D.~A.}\ \bibnamefont {Huse}},\
  }\href {\doibase 10.1103/physrevx.8.031057} {\bibfield  {journal} {\bibinfo
  {journal} {Physical Review X}\ }\textbf {\bibinfo {volume} {8}} (\bibinfo
  {year} {2018}),\ 10.1103/physrevx.8.031057}\BibitemShut {NoStop}%
\bibitem [{\citenamefont {Lux}\ \emph {et~al.}(2014)\citenamefont {Lux},
  \citenamefont {M\"uller}, \citenamefont {Mitra},\ and\ \citenamefont
  {Rosch}}]{RoschTails}%
  \BibitemOpen
  \bibfield  {author} {\bibinfo {author} {\bibfnamefont {J.}~\bibnamefont
  {Lux}}, \bibinfo {author} {\bibfnamefont {J.}~\bibnamefont {M\"uller}},
  \bibinfo {author} {\bibfnamefont {A.}~\bibnamefont {Mitra}}, \ and\ \bibinfo
  {author} {\bibfnamefont {A.}~\bibnamefont {Rosch}},\ }\href {\doibase
  10.1103/physreva.89.053608} {\bibfield  {journal} {\bibinfo  {journal}
  {Physical Review A}\ }\textbf {\bibinfo {volume} {89}} (\bibinfo {year}
  {2014}),\ 10.1103/physreva.89.053608}\BibitemShut {NoStop}%
\bibitem [{\citenamefont {Chen-Lin}\ \emph {et~al.}(2019)\citenamefont
  {Chen-Lin}, \citenamefont {Delacr\'etaz},\ and\ \citenamefont
  {Hartnoll}}]{delacretaz2019}%
  \BibitemOpen
  \bibfield  {author} {\bibinfo {author} {\bibfnamefont {X.}~\bibnamefont
  {Chen-Lin}}, \bibinfo {author} {\bibfnamefont {L.~V.}\ \bibnamefont
  {Delacr\'etaz}}, \ and\ \bibinfo {author} {\bibfnamefont {S.~A.}\
  \bibnamefont {Hartnoll}},\ }\href {\doibase 10.1103/physrevlett.122.091602}
  {\bibfield  {journal} {\bibinfo  {journal} {Physical Review Letters}\
  }\textbf {\bibinfo {volume} {122}} (\bibinfo {year} {2019}),\
  10.1103/physrevlett.122.091602}\BibitemShut {NoStop}%
\bibitem [{\citenamefont {Glorioso}\ \emph {et~al.}(2021)\citenamefont
  {Glorioso}, \citenamefont {Delacr\'etaz}, \citenamefont {Chen}, \citenamefont
  {Nandkishore},\ and\ \citenamefont {Lucas}}]{delacretazsu2}%
  \BibitemOpen
  \bibfield  {author} {\bibinfo {author} {\bibfnamefont {P.}~\bibnamefont
  {Glorioso}}, \bibinfo {author} {\bibfnamefont {L.~V.}\ \bibnamefont
  {Delacr\'etaz}}, \bibinfo {author} {\bibfnamefont {X.}~\bibnamefont {Chen}},
  \bibinfo {author} {\bibfnamefont {R.~M.}\ \bibnamefont {Nandkishore}}, \ and\
  \bibinfo {author} {\bibfnamefont {A.}~\bibnamefont {Lucas}},\ }\href
  {\doibase 10.21468/SciPostPhys.10.1.015} {\bibfield  {journal} {\bibinfo
  {journal} {SciPost Phys.}\ }\textbf {\bibinfo {volume} {10}},\ \bibinfo
  {pages} {15} (\bibinfo {year} {2021})}\BibitemShut {NoStop}%
\bibitem [{\citenamefont {Delacr\'etaz}(2020)}]{delacretazOPE}%
  \BibitemOpen
  \bibfield  {author} {\bibinfo {author} {\bibfnamefont {L.}~\bibnamefont
  {Delacr\'etaz}},\ }\href {\doibase 10.21468/scipostphys.9.3.034} {\bibfield
  {journal} {\bibinfo  {journal} {SciPost Physics}\ }\textbf {\bibinfo {volume}
  {9}} (\bibinfo {year} {2020}),\ 10.21468/scipostphys.9.3.034}\BibitemShut
  {NoStop}%
\bibitem [{\citenamefont {Nahum}\ \emph {et~al.}(2018)\citenamefont {Nahum},
  \citenamefont {Vijay},\ and\ \citenamefont {Haah}}]{nahum_no_symm}%
  \BibitemOpen
  \bibfield  {author} {\bibinfo {author} {\bibfnamefont {A.}~\bibnamefont
  {Nahum}}, \bibinfo {author} {\bibfnamefont {S.}~\bibnamefont {Vijay}}, \ and\
  \bibinfo {author} {\bibfnamefont {J.}~\bibnamefont {Haah}},\ }\href {\doibase
  10.1103/PhysRevX.8.021014} {\bibfield  {journal} {\bibinfo  {journal} {Phys.
  Rev. X}\ }\textbf {\bibinfo {volume} {8}},\ \bibinfo {pages} {021014}
  (\bibinfo {year} {2018})}\BibitemShut {NoStop}%
\bibitem [{\citenamefont {von Keyserlingk}\ \emph {et~al.}(2018)\citenamefont
  {von Keyserlingk}, \citenamefont {Rakovszky}, \citenamefont {Pollmann},\ and\
  \citenamefont {Sondhi}}]{vonKeyserlingk2017_no_symm}%
  \BibitemOpen
  \bibfield  {author} {\bibinfo {author} {\bibfnamefont {C.~W.}\ \bibnamefont
  {von Keyserlingk}}, \bibinfo {author} {\bibfnamefont {T.}~\bibnamefont
  {Rakovszky}}, \bibinfo {author} {\bibfnamefont {F.}~\bibnamefont {Pollmann}},
  \ and\ \bibinfo {author} {\bibfnamefont {S.~L.}\ \bibnamefont {Sondhi}},\
  }\href {\doibase 10.1103/PhysRevX.8.021013} {\bibfield  {journal} {\bibinfo
  {journal} {Phys. Rev. X}\ }\textbf {\bibinfo {volume} {8}},\ \bibinfo {pages}
  {021013} (\bibinfo {year} {2018})}\BibitemShut {NoStop}%
\bibitem [{\citenamefont {Nahum}\ \emph {et~al.}(ming)\citenamefont {Nahum},
  \citenamefont {Roy}, \citenamefont {Vijay},\ and\ \citenamefont
  {Zhou}}]{nahumfatop}%
  \BibitemOpen
  \bibfield  {author} {\bibinfo {author} {\bibfnamefont {A.}~\bibnamefont
  {Nahum}}, \bibinfo {author} {\bibfnamefont {S.}~\bibnamefont {Roy}}, \bibinfo
  {author} {\bibfnamefont {S.}~\bibnamefont {Vijay}}, \ and\ \bibinfo {author}
  {\bibfnamefont {T.}~\bibnamefont {Zhou}},\ }\href@noop {} {\enquote {\bibinfo
  {title} {Chaos and the time-ordered two-point function},}\ } (\bibinfo {year}
  {forthcoming})\BibitemShut {NoStop}%
\bibitem [{\citenamefont {Ho}\ and\ \citenamefont {Abanin}(2017)}]{abaninho}%
  \BibitemOpen
  \bibfield  {author} {\bibinfo {author} {\bibfnamefont {W.~W.}\ \bibnamefont
  {Ho}}\ and\ \bibinfo {author} {\bibfnamefont {D.~A.}\ \bibnamefont
  {Abanin}},\ }\href {\doibase 10.1103/PhysRevB.95.094302} {\bibfield
  {journal} {\bibinfo  {journal} {Phys. Rev. B}\ }\textbf {\bibinfo {volume}
  {95}},\ \bibinfo {pages} {094302} (\bibinfo {year} {2017})}\BibitemShut
  {NoStop}%
\bibitem [{\citenamefont {Leviatan}\ \emph {et~al.}(2017)\citenamefont
  {Leviatan}, \citenamefont {Pollmann}, \citenamefont {Bardarson},
  \citenamefont {Huse},\ and\ \citenamefont {Altman}}]{Leviatan2017}%
  \BibitemOpen
  \bibfield  {author} {\bibinfo {author} {\bibfnamefont {E.}~\bibnamefont
  {Leviatan}}, \bibinfo {author} {\bibfnamefont {F.}~\bibnamefont {Pollmann}},
  \bibinfo {author} {\bibfnamefont {J.~H.}\ \bibnamefont {Bardarson}}, \bibinfo
  {author} {\bibfnamefont {D.~A.}\ \bibnamefont {Huse}}, \ and\ \bibinfo
  {author} {\bibfnamefont {E.}~\bibnamefont {Altman}},\ }\href@noop {}
  {\enquote {\bibinfo {title} {Quantum thermalization dynamics with
  matrix-product states},}\ } (\bibinfo {year} {2017}),\ \Eprint
  {http://arxiv.org/abs/arXiv:1702.08894} {arXiv:1702.08894} \BibitemShut
  {NoStop}%
\bibitem [{\citenamefont {Steinigeweg}\ \emph {et~al.}(2014)\citenamefont
  {Steinigeweg}, \citenamefont {Heidrich-Meisner}, \citenamefont {Gemmer},
  \citenamefont {Michielsen},\ and\ \citenamefont
  {De~Raedt}}]{Steinigeweg2014_2}%
  \BibitemOpen
  \bibfield  {author} {\bibinfo {author} {\bibfnamefont {R.}~\bibnamefont
  {Steinigeweg}}, \bibinfo {author} {\bibfnamefont {F.}~\bibnamefont
  {Heidrich-Meisner}}, \bibinfo {author} {\bibfnamefont {J.}~\bibnamefont
  {Gemmer}}, \bibinfo {author} {\bibfnamefont {K.}~\bibnamefont {Michielsen}},
  \ and\ \bibinfo {author} {\bibfnamefont {H.}~\bibnamefont {De~Raedt}},\
  }\href {\doibase 10.1103/PhysRevB.90.094417} {\bibfield  {journal} {\bibinfo
  {journal} {Phys. Rev. B}\ }\textbf {\bibinfo {volume} {90}},\ \bibinfo
  {pages} {094417} (\bibinfo {year} {2014})}\BibitemShut {NoStop}%
\bibitem [{\citenamefont {Kloss}\ \emph {et~al.}(2018)\citenamefont {Kloss},
  \citenamefont {Lev},\ and\ \citenamefont {Reichman}}]{Kloss2018}%
  \BibitemOpen
  \bibfield  {author} {\bibinfo {author} {\bibfnamefont {B.}~\bibnamefont
  {Kloss}}, \bibinfo {author} {\bibfnamefont {Y.~B.}\ \bibnamefont {Lev}}, \
  and\ \bibinfo {author} {\bibfnamefont {D.}~\bibnamefont {Reichman}},\ }\href
  {\doibase 10.1103/PhysRevB.97.024307} {\bibfield  {journal} {\bibinfo
  {journal} {Phys. Rev. B}\ }\textbf {\bibinfo {volume} {97}},\ \bibinfo
  {pages} {024307} (\bibinfo {year} {2018})}\BibitemShut {NoStop}%
\bibitem [{\citenamefont {Karthik}\ \emph {et~al.}(2007)\citenamefont
  {Karthik}, \citenamefont {Sharma},\ and\ \citenamefont
  {Lakshminarayan}}]{Karthik2007}%
  \BibitemOpen
  \bibfield  {author} {\bibinfo {author} {\bibfnamefont {J.}~\bibnamefont
  {Karthik}}, \bibinfo {author} {\bibfnamefont {A.}~\bibnamefont {Sharma}}, \
  and\ \bibinfo {author} {\bibfnamefont {A.}~\bibnamefont {Lakshminarayan}},\
  }\href {\doibase 10.1103/PhysRevA.75.022304} {\bibfield  {journal} {\bibinfo
  {journal} {Phys. Rev. A}\ }\textbf {\bibinfo {volume} {75}},\ \bibinfo
  {pages} {022304} (\bibinfo {year} {2007})}\BibitemShut {NoStop}%
\bibitem [{\citenamefont {Kim}\ and\ \citenamefont {Huse}(2013)}]{KimHuse2013}%
  \BibitemOpen
  \bibfield  {author} {\bibinfo {author} {\bibfnamefont {H.}~\bibnamefont
  {Kim}}\ and\ \bibinfo {author} {\bibfnamefont {D.~A.}\ \bibnamefont {Huse}},\
  }\href {\doibase 10.1103/PhysRevLett.111.127205} {\bibfield  {journal}
  {\bibinfo  {journal} {Phys. Rev. Lett.}\ }\textbf {\bibinfo {volume} {111}},\
  \bibinfo {pages} {127205} (\bibinfo {year} {2013})}\BibitemShut {NoStop}%
\bibitem [{\citenamefont {Vidal}(2003)}]{VidalTEBD}%
  \BibitemOpen
  \bibfield  {author} {\bibinfo {author} {\bibfnamefont {G.}~\bibnamefont
  {Vidal}},\ }\href {\doibase 10.1103/PhysRevLett.91.147902} {\bibfield
  {journal} {\bibinfo  {journal} {Phys. Rev. Lett.}\ }\textbf {\bibinfo
  {volume} {91}},\ \bibinfo {pages} {147902} (\bibinfo {year}
  {2003})}\BibitemShut {NoStop}%
\bibitem [{\citenamefont {Karrasch}\ \emph {et~al.}(2015)\citenamefont
  {Karrasch}, \citenamefont {Kennes},\ and\ \citenamefont
  {Heidrich-Meisner}}]{Karrasch2015}%
  \BibitemOpen
  \bibfield  {author} {\bibinfo {author} {\bibfnamefont {C.}~\bibnamefont
  {Karrasch}}, \bibinfo {author} {\bibfnamefont {D.~M.}\ \bibnamefont
  {Kennes}}, \ and\ \bibinfo {author} {\bibfnamefont {F.}~\bibnamefont
  {Heidrich-Meisner}},\ }\href {\doibase 10.1103/PhysRevB.91.115130} {\bibfield
   {journal} {\bibinfo  {journal} {Phys. Rev. B}\ }\textbf {\bibinfo {volume}
  {91}},\ \bibinfo {pages} {115130} (\bibinfo {year} {2015})}\BibitemShut
  {NoStop}%
\bibitem [{\citenamefont {Rakovszky}\ \emph {et~al.}(2019)\citenamefont
  {Rakovszky}, \citenamefont {Pollmann},\ and\ \citenamefont {von
  Keyserlingk}}]{Rakovszky_2019}%
  \BibitemOpen
  \bibfield  {author} {\bibinfo {author} {\bibfnamefont {T.}~\bibnamefont
  {Rakovszky}}, \bibinfo {author} {\bibfnamefont {F.}~\bibnamefont {Pollmann}},
  \ and\ \bibinfo {author} {\bibfnamefont {C.}~\bibnamefont {von
  Keyserlingk}},\ }\href {\doibase 10.1103/physrevlett.122.250602} {\bibfield
  {journal} {\bibinfo  {journal} {Physical Review Letters}\ }\textbf {\bibinfo
  {volume} {122}} (\bibinfo {year} {2019}),\
  10.1103/physrevlett.122.250602}\BibitemShut {NoStop}%
\bibitem [{\citenamefont {Huang}(2020)}]{Huang2020}%
  \BibitemOpen
  \bibfield  {author} {\bibinfo {author} {\bibfnamefont {Y.}~\bibnamefont
  {Huang}},\ }\href {\doibase 10.1088/2633-1357/abd1e2} {\ \textbf {\bibinfo
  {volume} {1}},\ \bibinfo {pages} {035205} (\bibinfo {year}
  {2020})}\BibitemShut {NoStop}%
\bibitem [{\citenamefont {Preskill}(2012)}]{preskill2012quantum}%
  \BibitemOpen
  \bibfield  {author} {\bibinfo {author} {\bibfnamefont {J.}~\bibnamefont
  {Preskill}},\ }\href@noop {} {\bibfield  {journal} {\bibinfo  {journal}
  {arXiv preprint arXiv:1203.5813}\ } (\bibinfo {year} {2012})}\BibitemShut
  {NoStop}%
\bibitem [{\citenamefont {Aaronson}\ and\ \citenamefont
  {Chen}(2017)}]{Aaronson2017}%
  \BibitemOpen
  \bibfield  {author} {\bibinfo {author} {\bibfnamefont {S.}~\bibnamefont
  {Aaronson}}\ and\ \bibinfo {author} {\bibfnamefont {L.}~\bibnamefont
  {Chen}},\ }in\ \href {\doibase 10.4230/LIPIcs.CCC.2017.22} {\emph {\bibinfo
  {booktitle} {32nd Computational Complexity Conference (CCC 2017)}}},\
  \bibinfo {series} {Leibniz International Proceedings in Informatics
  (LIPIcs)}, Vol.~\bibinfo {volume} {79},\ \bibinfo {editor} {edited by\
  \bibinfo {editor} {\bibfnamefont {R.}~\bibnamefont {O'Donnell}}}\ (\bibinfo
  {publisher} {Schloss Dagstuhl--Leibniz-Zentrum fuer Informatik},\ \bibinfo
  {address} {Dagstuhl, Germany},\ \bibinfo {year} {2017})\ pp.\ \bibinfo
  {pages} {22:1--22:67}\BibitemShut {NoStop}%
\bibitem [{\citenamefont {Arute}\ \emph {et~al.}(2019)\citenamefont {Arute},
  \citenamefont {Arya}, \citenamefont {Babbush}, \citenamefont {Bacon},
  \citenamefont {Bardin}, \citenamefont {Barends}, \citenamefont {Biswas},
  \citenamefont {Boixo}, \citenamefont {Brandao}, \citenamefont {Buell},
  \citenamefont {Burkett}, \citenamefont {Chen}, \citenamefont {Chen},
  \citenamefont {Chiaro}, \citenamefont {Collins}, \citenamefont {Courtney},
  \citenamefont {Dunsworth}, \citenamefont {Farhi}, \citenamefont {Foxen},
  \citenamefont {Fowler}, \citenamefont {Gidney}, \citenamefont {Giustina},
  \citenamefont {Graff}, \citenamefont {Guerin}, \citenamefont {Habegger},
  \citenamefont {Harrigan}, \citenamefont {Hartmann}, \citenamefont {Ho},
  \citenamefont {Hoffmann}, \citenamefont {Huang}, \citenamefont {Humble},
  \citenamefont {Isakov}, \citenamefont {Jeffrey}, \citenamefont {Jiang},
  \citenamefont {Kafri}, \citenamefont {Kechedzhi}, \citenamefont {Kelly},
  \citenamefont {Klimov}, \citenamefont {Knysh}, \citenamefont {Korotkov},
  \citenamefont {Kostritsa}, \citenamefont {Landhuis}, \citenamefont
  {Lindmark}, \citenamefont {Lucero}, \citenamefont {Lyakh}, \citenamefont
  {Mandr{\`{a}}}, \citenamefont {McClean}, \citenamefont {McEwen},
  \citenamefont {Megrant}, \citenamefont {Mi}, \citenamefont {Michielsen},
  \citenamefont {Mohseni}, \citenamefont {Mutus}, \citenamefont {Naaman},
  \citenamefont {Neeley}, \citenamefont {Neill}, \citenamefont {Niu},
  \citenamefont {Ostby}, \citenamefont {Petukhov}, \citenamefont {Platt},
  \citenamefont {Quintana}, \citenamefont {Rieffel}, \citenamefont {Roushan},
  \citenamefont {Rubin}, \citenamefont {Sank}, \citenamefont {Satzinger},
  \citenamefont {Smelyanskiy}, \citenamefont {Sung}, \citenamefont
  {Trevithick}, \citenamefont {Vainsencher}, \citenamefont {Villalonga},
  \citenamefont {White}, \citenamefont {Yao}, \citenamefont {Yeh},
  \citenamefont {Zalcman}, \citenamefont {Neven},\ and\ \citenamefont
  {Martinis}}]{GoogleSupremacy}%
  \BibitemOpen
  \bibfield  {author} {\bibinfo {author} {\bibfnamefont {F.}~\bibnamefont
  {Arute}}, \bibinfo {author} {\bibfnamefont {K.}~\bibnamefont {Arya}},
  \bibinfo {author} {\bibfnamefont {R.}~\bibnamefont {Babbush}}, \bibinfo
  {author} {\bibfnamefont {D.}~\bibnamefont {Bacon}}, \bibinfo {author}
  {\bibfnamefont {J.~C.}\ \bibnamefont {Bardin}}, \bibinfo {author}
  {\bibfnamefont {R.}~\bibnamefont {Barends}}, \bibinfo {author} {\bibfnamefont
  {R.}~\bibnamefont {Biswas}}, \bibinfo {author} {\bibfnamefont
  {S.}~\bibnamefont {Boixo}}, \bibinfo {author} {\bibfnamefont {F.~G. S.~L.}\
  \bibnamefont {Brandao}}, \bibinfo {author} {\bibfnamefont {D.~A.}\
  \bibnamefont {Buell}}, \bibinfo {author} {\bibfnamefont {B.}~\bibnamefont
  {Burkett}}, \bibinfo {author} {\bibfnamefont {Y.}~\bibnamefont {Chen}},
  \bibinfo {author} {\bibfnamefont {Z.}~\bibnamefont {Chen}}, \bibinfo {author}
  {\bibfnamefont {B.}~\bibnamefont {Chiaro}}, \bibinfo {author} {\bibfnamefont
  {R.}~\bibnamefont {Collins}}, \bibinfo {author} {\bibfnamefont
  {W.}~\bibnamefont {Courtney}}, \bibinfo {author} {\bibfnamefont
  {A.}~\bibnamefont {Dunsworth}}, \bibinfo {author} {\bibfnamefont
  {E.}~\bibnamefont {Farhi}}, \bibinfo {author} {\bibfnamefont
  {B.}~\bibnamefont {Foxen}}, \bibinfo {author} {\bibfnamefont
  {A.}~\bibnamefont {Fowler}}, \bibinfo {author} {\bibfnamefont
  {C.}~\bibnamefont {Gidney}}, \bibinfo {author} {\bibfnamefont
  {M.}~\bibnamefont {Giustina}}, \bibinfo {author} {\bibfnamefont
  {R.}~\bibnamefont {Graff}}, \bibinfo {author} {\bibfnamefont
  {K.}~\bibnamefont {Guerin}}, \bibinfo {author} {\bibfnamefont
  {S.}~\bibnamefont {Habegger}}, \bibinfo {author} {\bibfnamefont {M.~P.}\
  \bibnamefont {Harrigan}}, \bibinfo {author} {\bibfnamefont {M.~J.}\
  \bibnamefont {Hartmann}}, \bibinfo {author} {\bibfnamefont {A.}~\bibnamefont
  {Ho}}, \bibinfo {author} {\bibfnamefont {M.}~\bibnamefont {Hoffmann}},
  \bibinfo {author} {\bibfnamefont {T.}~\bibnamefont {Huang}}, \bibinfo
  {author} {\bibfnamefont {T.~S.}\ \bibnamefont {Humble}}, \bibinfo {author}
  {\bibfnamefont {S.~V.}\ \bibnamefont {Isakov}}, \bibinfo {author}
  {\bibfnamefont {E.}~\bibnamefont {Jeffrey}}, \bibinfo {author} {\bibfnamefont
  {Z.}~\bibnamefont {Jiang}}, \bibinfo {author} {\bibfnamefont
  {D.}~\bibnamefont {Kafri}}, \bibinfo {author} {\bibfnamefont
  {K.}~\bibnamefont {Kechedzhi}}, \bibinfo {author} {\bibfnamefont
  {J.}~\bibnamefont {Kelly}}, \bibinfo {author} {\bibfnamefont {P.~V.}\
  \bibnamefont {Klimov}}, \bibinfo {author} {\bibfnamefont {S.}~\bibnamefont
  {Knysh}}, \bibinfo {author} {\bibfnamefont {A.}~\bibnamefont {Korotkov}},
  \bibinfo {author} {\bibfnamefont {F.}~\bibnamefont {Kostritsa}}, \bibinfo
  {author} {\bibfnamefont {D.}~\bibnamefont {Landhuis}}, \bibinfo {author}
  {\bibfnamefont {M.}~\bibnamefont {Lindmark}}, \bibinfo {author}
  {\bibfnamefont {E.}~\bibnamefont {Lucero}}, \bibinfo {author} {\bibfnamefont
  {D.}~\bibnamefont {Lyakh}}, \bibinfo {author} {\bibfnamefont
  {S.}~\bibnamefont {Mandr{\`{a}}}}, \bibinfo {author} {\bibfnamefont {J.~R.}\
  \bibnamefont {McClean}}, \bibinfo {author} {\bibfnamefont {M.}~\bibnamefont
  {McEwen}}, \bibinfo {author} {\bibfnamefont {A.}~\bibnamefont {Megrant}},
  \bibinfo {author} {\bibfnamefont {X.}~\bibnamefont {Mi}}, \bibinfo {author}
  {\bibfnamefont {K.}~\bibnamefont {Michielsen}}, \bibinfo {author}
  {\bibfnamefont {M.}~\bibnamefont {Mohseni}}, \bibinfo {author} {\bibfnamefont
  {J.}~\bibnamefont {Mutus}}, \bibinfo {author} {\bibfnamefont
  {O.}~\bibnamefont {Naaman}}, \bibinfo {author} {\bibfnamefont
  {M.}~\bibnamefont {Neeley}}, \bibinfo {author} {\bibfnamefont
  {C.}~\bibnamefont {Neill}}, \bibinfo {author} {\bibfnamefont {M.~Y.}\
  \bibnamefont {Niu}}, \bibinfo {author} {\bibfnamefont {E.}~\bibnamefont
  {Ostby}}, \bibinfo {author} {\bibfnamefont {A.}~\bibnamefont {Petukhov}},
  \bibinfo {author} {\bibfnamefont {J.~C.}\ \bibnamefont {Platt}}, \bibinfo
  {author} {\bibfnamefont {C.}~\bibnamefont {Quintana}}, \bibinfo {author}
  {\bibfnamefont {E.~G.}\ \bibnamefont {Rieffel}}, \bibinfo {author}
  {\bibfnamefont {P.}~\bibnamefont {Roushan}}, \bibinfo {author} {\bibfnamefont
  {N.~C.}\ \bibnamefont {Rubin}}, \bibinfo {author} {\bibfnamefont
  {D.}~\bibnamefont {Sank}}, \bibinfo {author} {\bibfnamefont {K.~J.}\
  \bibnamefont {Satzinger}}, \bibinfo {author} {\bibfnamefont {V.}~\bibnamefont
  {Smelyanskiy}}, \bibinfo {author} {\bibfnamefont {K.~J.}\ \bibnamefont
  {Sung}}, \bibinfo {author} {\bibfnamefont {M.~D.}\ \bibnamefont
  {Trevithick}}, \bibinfo {author} {\bibfnamefont {A.}~\bibnamefont
  {Vainsencher}}, \bibinfo {author} {\bibfnamefont {B.}~\bibnamefont
  {Villalonga}}, \bibinfo {author} {\bibfnamefont {T.}~\bibnamefont {White}},
  \bibinfo {author} {\bibfnamefont {Z.~J.}\ \bibnamefont {Yao}}, \bibinfo
  {author} {\bibfnamefont {P.}~\bibnamefont {Yeh}}, \bibinfo {author}
  {\bibfnamefont {A.}~\bibnamefont {Zalcman}}, \bibinfo {author} {\bibfnamefont
  {H.}~\bibnamefont {Neven}}, \ and\ \bibinfo {author} {\bibfnamefont {J.~M.}\
  \bibnamefont {Martinis}},\ }\href {\doibase 10.1038/s41586-019-1666-5} {\
  \textbf {\bibinfo {volume} {574}},\ \bibinfo {pages} {505} (\bibinfo {year}
  {2019})}\BibitemShut {NoStop}%
\bibitem [{\citenamefont {Aharonov}\ and\ \citenamefont
  {Irani}(2021)}]{aharonov_en_density}%
  \BibitemOpen
  \bibfield  {author} {\bibinfo {author} {\bibfnamefont {D.}~\bibnamefont
  {Aharonov}}\ and\ \bibinfo {author} {\bibfnamefont {S.}~\bibnamefont
  {Irani}},\ }\href@noop {} {\enquote {\bibinfo {title} {Hamiltonian complexity
  in the thermodynamic limit},}\ } (\bibinfo {year} {2021}),\ \Eprint
  {http://arxiv.org/abs/2107.06201} {arXiv:2107.06201 [quant-ph]} \BibitemShut
  {NoStop}%
\bibitem [{\citenamefont {Watson}\ and\ \citenamefont
  {Cubitt}(2021)}]{cubitt_en_density}%
  \BibitemOpen
  \bibfield  {author} {\bibinfo {author} {\bibfnamefont {J.~D.}\ \bibnamefont
  {Watson}}\ and\ \bibinfo {author} {\bibfnamefont {T.~S.}\ \bibnamefont
  {Cubitt}},\ }\href@noop {} {\enquote {\bibinfo {title} {Computational
  complexity of the ground state energy density problem},}\ } (\bibinfo {year}
  {2021}),\ \Eprint {http://arxiv.org/abs/2107.05060} {arXiv:2107.05060
  [quant-ph]} \BibitemShut {NoStop}%
\end{thebibliography}%
\end{document}